\documentclass[aps,twocolumn,showpacs,byrevtex,prl,reprint, nofootinbib]{revtex4-1}
\usepackage{xspace}
\usepackage{epsfig}
\usepackage{epstopdf}
\usepackage{dcolumn}
\usepackage{amssymb}   
\usepackage{amsmath}
\usepackage{mathrsfs}
\usepackage{bm}
\usepackage{xcolor}
\usepackage[normalem]{ulem}
\usepackage{graphicx}
\usepackage{subfigure}
\usepackage{siunitx}
\usepackage{pstricks}
\usepackage{multirow}
\usepackage{pst-node}
\usepackage{rotating}
\usepackage{times}
\usepackage{soul}
\usepackage{ulem}
\usepackage{balance}
\usepackage{lipsum}
\usepackage[english]{babel}
\addto{\captionsenglish}{%

}
\usepackage[colorlinks,linkcolor=blue, citecolor=blue]{hyperref}
\usepackage{lineno}
\usepackage[T1]{fontenc}
\usepackage{lmodern}
\usepackage{bm}
\newcommand{\BESIIIorcid}[1]{\href{https://orcid.org/#1}{\hspace*{0.1em}\raisebox{-0.45ex}{\includegraphics[width=1em]{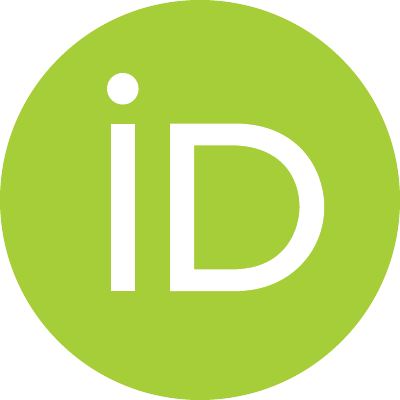}}}}

\begin{document}
\normalsize
\parskip=5pt plus 1pt minus 1pt

\title{Measurement of Born cross sections for $e^+e^-\to p\bar p$ at $\sqrt{s} =3.510-4.946$ GeV}
\author{BESIII Collaboration}
\thanks{Full author list given at the end of the paper.}
\begin{abstract}
We report a measurement of the Born cross section and the effective form factor for the $e^+e^-\to p\bar{p}$ reaction at 47 center-of-mass energies between 3.510 and 4.946 GeV. 
The measurement is performed using the energy-scan technique and is based on data corresponding to an integrated luminosity of 26 fb\(^{-1}\) collected with the BESIII detector at the BEPCII collider. For the first time, the moduli of the electromagnetic form factor ratio $|G_{E}/G_{M}|$ and of the magnetic form factor $|G_{M}|$ are determined with high precision by analyzing the distribution of the polar angle of the proton at a large timelike momentum transfer.
 These  results provide essential insights into the nature of charmonium(-like) states above the open-charm threshold and the dynamics underlying the proton electromagnetic form factors.

\end{abstract}
\maketitle
\section{Introduction}
Over the past decades, a series of charmonium(-like) states have been observed since the discovery of the $J/\psi$ meson in 1974. Below the open-charm threshold, the mass spectrum of the observed charmonium states is in good agreement with theoretical calculations based on the potential quark model~\cite{Barnes:2005pb}, providing strong support for the validity and predictive power of the theoretical framework.
Above this threshold, the potential quark model predicts the existence of six vector charmonium states between the open-charm threshold and $4.7\,{\rm GeV}$, namely $\psi(1D)$, $\psi(3S)$, $\psi(2D)$, $\psi(4S)$, $\psi(3D)$ and $\psi(5S)$~\cite{ParticleDataGroup:2024cfk}. 
However, the experimentally observed vector states in this energy region are found to be more abundant than theoretically predicted.
The decays of the three states, $\psi(4040)$, $\psi(4160)$, and $\psi(4415)$, observed from the inclusive hadronic cross sections, are dominated by open-charm processes~\cite{BES:2001ckj}. In contrast, other states, such as $Y(4260)$, $Y(4360)$, and $Y(4660)$, exhibit strong couplings to hidden-charm final states~\cite{ParticleDataGroup:2024cfk}. The first $Y$ state, $Y(4260)$, was discovered in the $\it{B}$ factories in its decay to $\pi^+\pi^-J/\psi$, produced in $e^+e^-$ annihilation with initial stat radiation (ISR), $i.e.$, through the reaction $e^+e^-\to \gamma_{\rm ISR}\pi^+\pi^-J/\psi$~\cite{BaBar:2005hhc, Belle:2007dxy}. 

The overpopulation of structures in this energy region, together with 
discrepancies between the properties predicted by the potential model and the experimental measurements, makes them potential candidates for exotic states
~\cite{
Close:2005iz,WangQ:2014cvms,Briceno:2015rlt,
Chen:2016qju, Olsen:2017bmm, Guo:2017jvc, Chen:2022asf, Wang:2025dur,Bai:2026atm}.
This situation reflects our limited understanding of the strong interaction, particularly its nonperturbative aspects. To achieve a more comprehensive understanding, more experimental information is needed.
The baryon pair production in charmonium(-like) states or directly in $e^+e^-$ annihilation exhibits a straightforward topology in terms of the final states and provides a clear depiction of the underlying interaction mechanism, which is assumed to be dominated by three-gluon or one-photon processes. In particular, the study of two-body baryonic decays above open-charm threshold in $e^+e^-$ collisions provides a new avenue for understanding the nature of these states~\cite{Wang:2019mhs,Qian:2021neg}.
Additionally, studies of the electromagnetic form factors (EMFFs) 
can shed light on the internal structure of baryons. Several experimental studies of baryon anti-baryon ($B\bar{B}$) pair production have been performed by the Belle experiment~\cite{Belle:2008xmh} and BESIII experiment~\cite{
BESIII:2019cuv, BESIII:2020ktn,BESIII:2021aer,BESIII:2021ccp,BESIII:2021cvv,Wang:2022zyc,BESIII:2022mfx,BESIII:2022lsz,Liu:2023xhg,Liu:2023xhg, BESIII:2023lkg,BESIII:2023rse,BESIII:2023euh,
BESIII:2024umc,BESIII:2024ogz,BESIII:2024ues,BESIII:2024dmr,BESIII:2024gql,
BESIII:2025yzk,Zhang:2025qmo,BESIII:2025lbj,BESIII:2025dke,BESIII:2026ala,Zhang:2026qjt,BESIII:2025dke,BESIII:2026ala,Zhang:2026qjt,BESIII:2026hgj,BESIII:2026hie,BESIII:2026oyx}. 
Apart from two pieces of evidence of $\psi(3770)\to\Lambda\bar\Lambda$ and $\Xi^-\bar\Xi^+$~\cite{BESIII:2021ccp, BESIII:2023rse}, as well as two possible observations of $\psi(3770)\to\Sigma^-\bar\Sigma^+$ and $K^-\bar{\Xi}^+\Lambda$~\cite{BESIII:2026hie, BESIII:2026oyx},
no significant indication for $B\bar{B}$ decay of other vector charmonium(-like) states has been found. Thus, more precise measurements of exclusive cross sections for $B\bar{B}$ final states above the open-charm threshold are crucial.

In this paper, we report a measurement of the Born cross section and the effective form factor for the reaction of $e^+e^-\to p\bar p$ using data corresponding to a total integrated luminosity of \SI{26}{fb^{-1}} collected at center-of-mass (c.m.) energies ($\sqrt{s}$) between 3.510 and \SI{4.946}{GeV} with the BESIII detector~\cite{BESIII:2009fln} at the BEPCII collider~\cite{Yu:2016cof}. In addition, the moduli of the EMFFs ratio and $|G_M|$ are extracted by analyzing the polar angle distribution with high precision, and for timelike squared momentum transfer larger than previous BESIII measurements.

\section{BESIII Detector and Monte Carlo simulation}
The BESIII detector~\cite{BESIII:2009fln} records symmetric $e^+e^-$ collisions 
provided by the BEPCII storage ring~\cite{Yu:2016cof} in the center-of-mass energy range from 1.84 to 4.95~GeV,
with a peak luminosity of $1.1 \times 10^{33}\;\text{cm}^{-2}\text{s}^{-1}$ 
achieved at $\sqrt{s} = 3.773\;\text{GeV}$. 
Large data samples have been collected in this energy region~\cite{BESIII:2020nme,Lu:2020imt,Zhang:2022bdc}. The cylindrical core of the BESIII detector covers 93\% of the full solid angle and consists of a helium-based
multilayer drift chamber~(MDC), a time-of-flight system~(TOF), and a CsI(Tl) electromagnetic calorimeter~(EMC),
which are all enclosed in a superconducting solenoidal magnet providing a 1.0~T magnetic field.
The solenoid is surrounded by an octagonal flux-return yoke made of steel, interleaved with resistive-plate-counter muon-identification modules.
The charged-particle momentum resolution at $1~{\rm GeV}/c$ is $0.5\%$, and the ${\rm d}E/{\rm d}x$ resolution is $6\%$ for electrons from Bhabha scattering. The EMC measures photon energies with a
resolution of $2.5\%$ ($5\%$) at $1$~GeV in the barrel (end cap) region. The time resolution in the plastic scintillator TOF barrel region is 68~ps, while that in the end cap region was 110~ps. The end cap TOF
system was upgraded in 2015 using multigap resistive plate chamber technology, providing a time resolution of
60~ps, which benefits 82\% of the data used in this analysis~\cite{Li:2017jpg,Guo:2017sjt,Cao:2020ibk}.

To evaluate detection efficiencies and estimate backgrounds, simulated
data samples are produced using {\sc geant4}-based Monte Carlo (MC)
software~\cite{GEANT4:2002zbu}, which incorporates the geometric description
of the BESIII detector~\cite{Huang:2022wuo} and the detector
response. The simulation of the  $e^+e^-\to p\bar p$ process models
the beam energy spread in the $e^+e^-$ reaction, employing {\sc kkmc}~\cite{Jadach:2000ir}. To determine the detection efficiencies of the reaction $e^+e^-\to p\bar p$, a sample of $10^5$ MC events is generated for each energy point according to the angular distribution, which can be written as a function of the angle $\theta$ between the direction of the nucleon or the antinucleon and the beam as 
$1+\eta \rm cos^{2}\theta$, where $\eta$ is the angular distribution parameter at the energy point of measurement. To examine potential background channels, an inclusive MC sample produced at $\sqrt{s} =3.773$ GeV is analyzed. This includes the production of $D\bar{D}$ pairs (including quantum coherence for the neutral $D$ channels), the non-$D\bar{D}$ decays of the $\psi(3770)$, the ISR production of the $J/\psi$ and $\psi(3686)$ states, and the continuum processes incorporated in {\sc kkmc}~\cite{Jadach:2000ir}. 

\section{Event selection}
The final state for the $e^+e^-\to p \bar p$ reaction consists of a proton-antiproton pair, which is produced back-to-back in the c.m. system. The two candidate charged tracks reconstructed in the MDC are required to satisfy $V_r < 1.0 \rm $ cm 
and $|V_z|< 10.0$ cm, where $V_r$ and $V_z$ are the distances of closest approach of the reconstructed track to the interaction point, projected in the plane transverse to the beam and along the beam direction, respectively. 
The net charge of the two tracks is required to be zero.
The polar angle $\theta$ of the track is required to be within the region with $|\rm cos\theta|<0.8$. Using information from the barrel TOF, likelihoods $\mathcal{L}_h$ $(h = p, K, \pi)$ for different particle hypotheses are calculated. Proton and antiproton candidates must satisfy $\mathcal{L}_{p,\bar p}>0.001$, as well as $\mathcal{L}_{p,\bar p}>\mathcal{L}_K$ and $\mathcal{L}_{p,\bar p}>\mathcal{L}_{\pi}$. Further, we require the opening angle of the two charged tracks to satisfy $\rm \theta_{opening}>3.1$~rad in the $e^+e^-$ c.m.~system and the $E/p<0.5$ for positive tracks, where $E$ denotes the deposited energy in the EMC and $p$ represents the track momentum.
To remove the background from $e^+e^-\to\mu^+\mu^-$ events, the hit depth of tracks in the muon counter is required to be less than \SI{40}{cm}. Finally, for antiproton candidates, the absolute value of the difference between the measured and expected momentum is required to be within three times the momentum resolution.

\begin{figure*}[!htpb]
    \centering
    \includegraphics[width=0.32\linewidth]{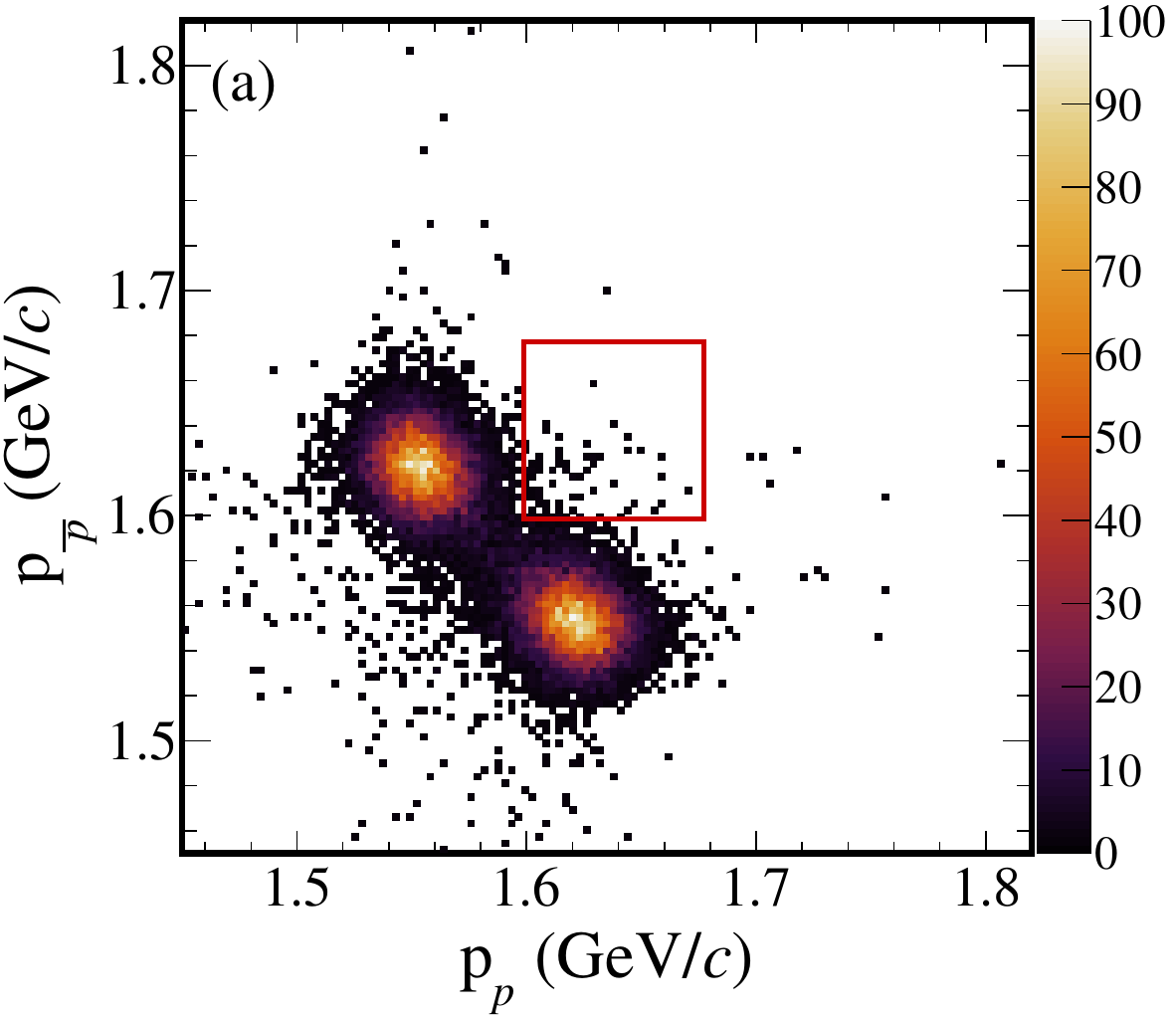}
    \includegraphics[width=0.32\linewidth]{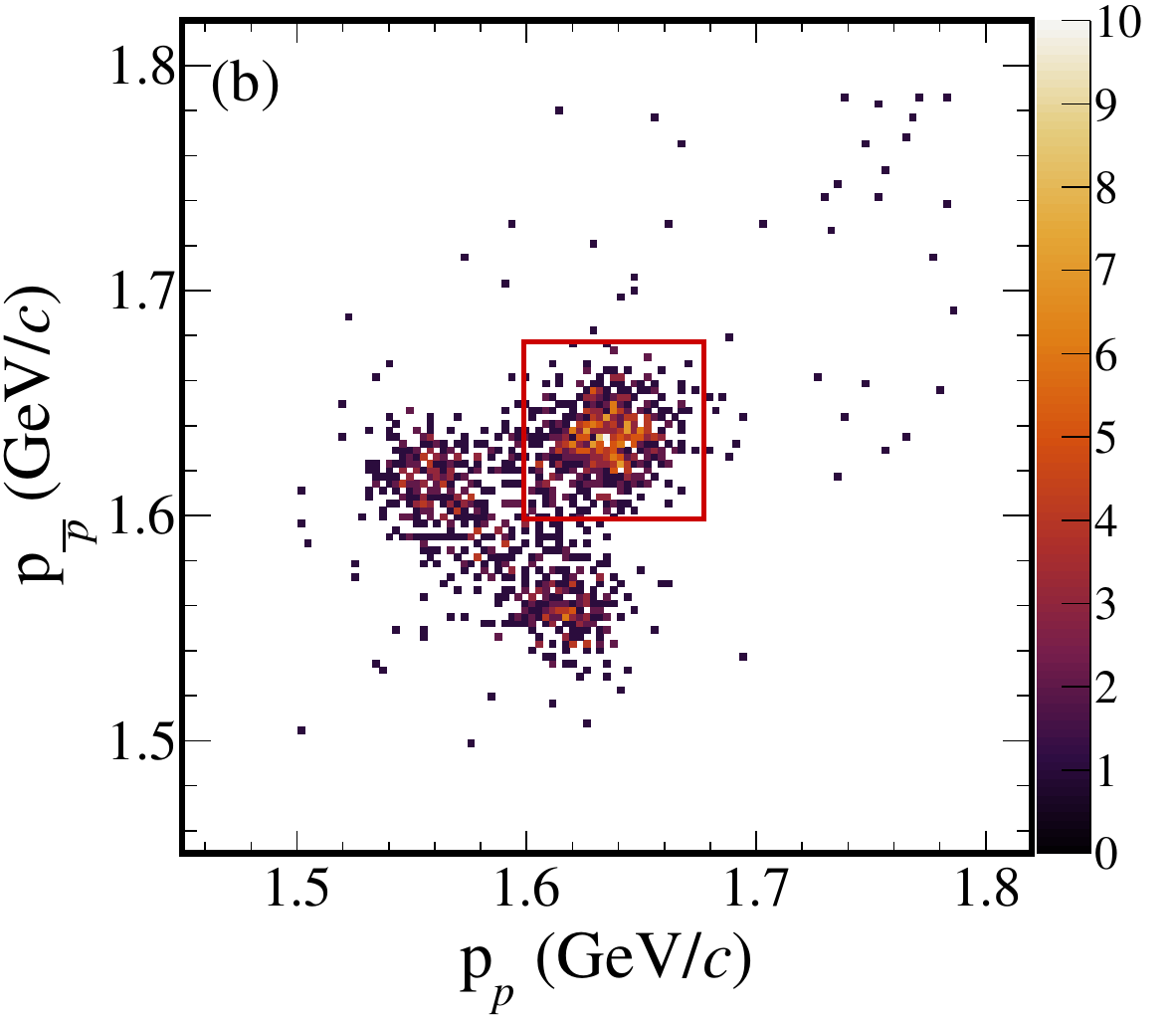}
    \includegraphics[width=0.32\linewidth]{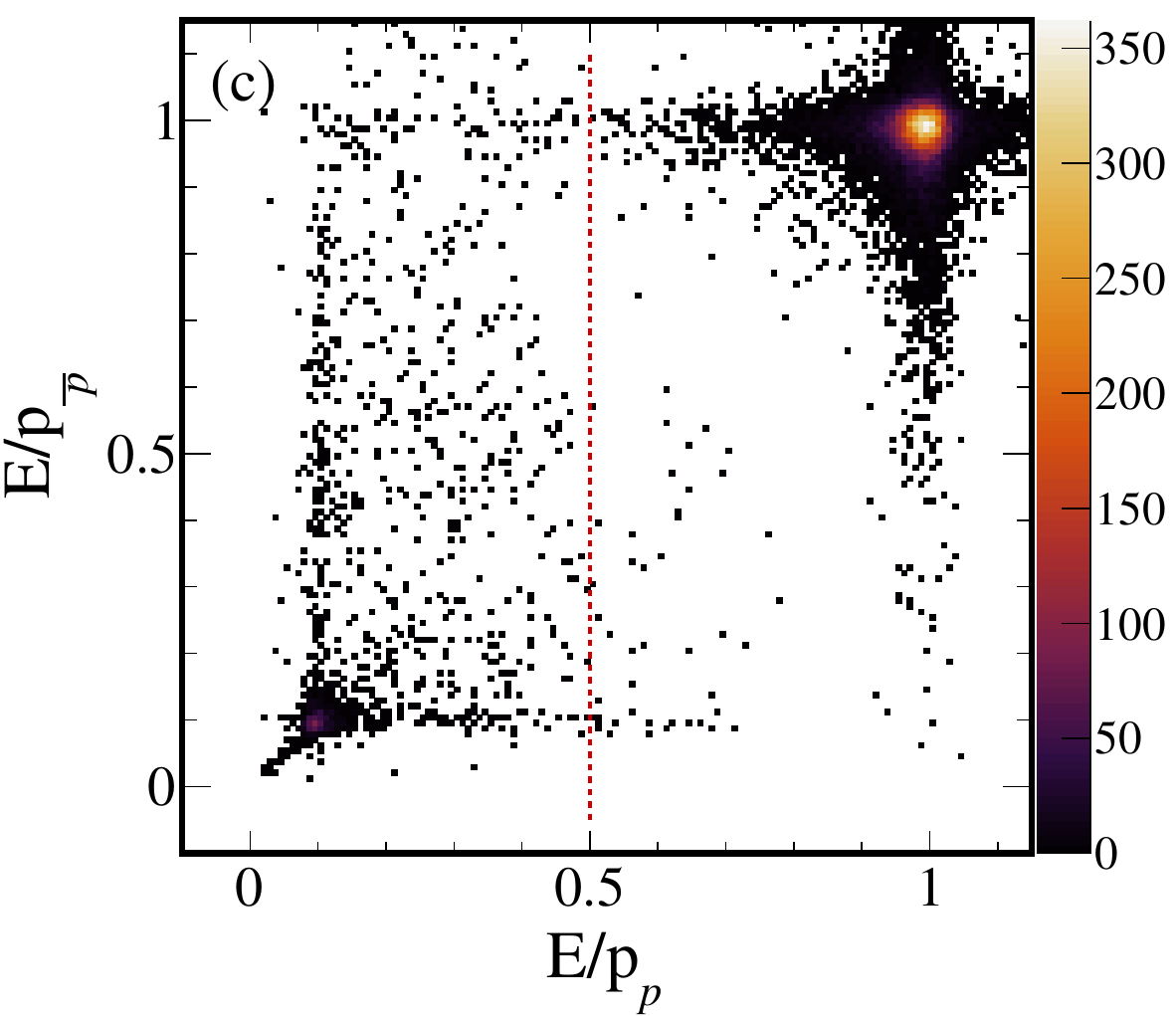}
    \caption{Two-dimensional distributions of ISR background events from the MC simulated sample (a) and the data sample at $\sqrt{s}=3.773 \;\rm GeV$ (b) and the distribution of Bhabha events from data at $\sqrt{s}=4.178\;\rm GeV$ (c). The red boxes define the signal regions, expected to be within three times the proton momentum resolution and the red dashed line represent the threshold value of the $E/p$.}
    \label{bkg}
\end{figure*}

Background from ISR to the lower lying $\psi(3686)$ resonance, which is not taken into account in the ISR correction procedure, is estimated with a MC-simulated sample. As shown in Fig.~\ref{bkg}, this process gives rise to a peaking structure adjacent to the signal region. However, the number of expected background events from this process is
negligible within the signal region according to the normalization in this analysis. 
Nevertheless, its background shape must be taken into account in the signal yields extraction, as described in the next section. Another background is from Bhabha scattering events, where an electron is misidentified as a proton. This background is also negligible with a requirement of $E/p<0.5$ for the positive charge tracks. Figure~\ref{bkg} shows the two-dimensional distributions for backgrounds from ISR and Bhabha events.
After applying the above event selection criteria and analyzing the inclusive MC sample at $\sqrt{s} = 3.773$ GeV, the number of surviving background events is found to be very small. Moreover, these events are distributed quite uniformly in the signal region.

\section{Born cross section measurement}
\subsection{Extraction of signal yields}
\begin{figure*}[tpb!]
    \centering
    \includegraphics[width=.245\linewidth]{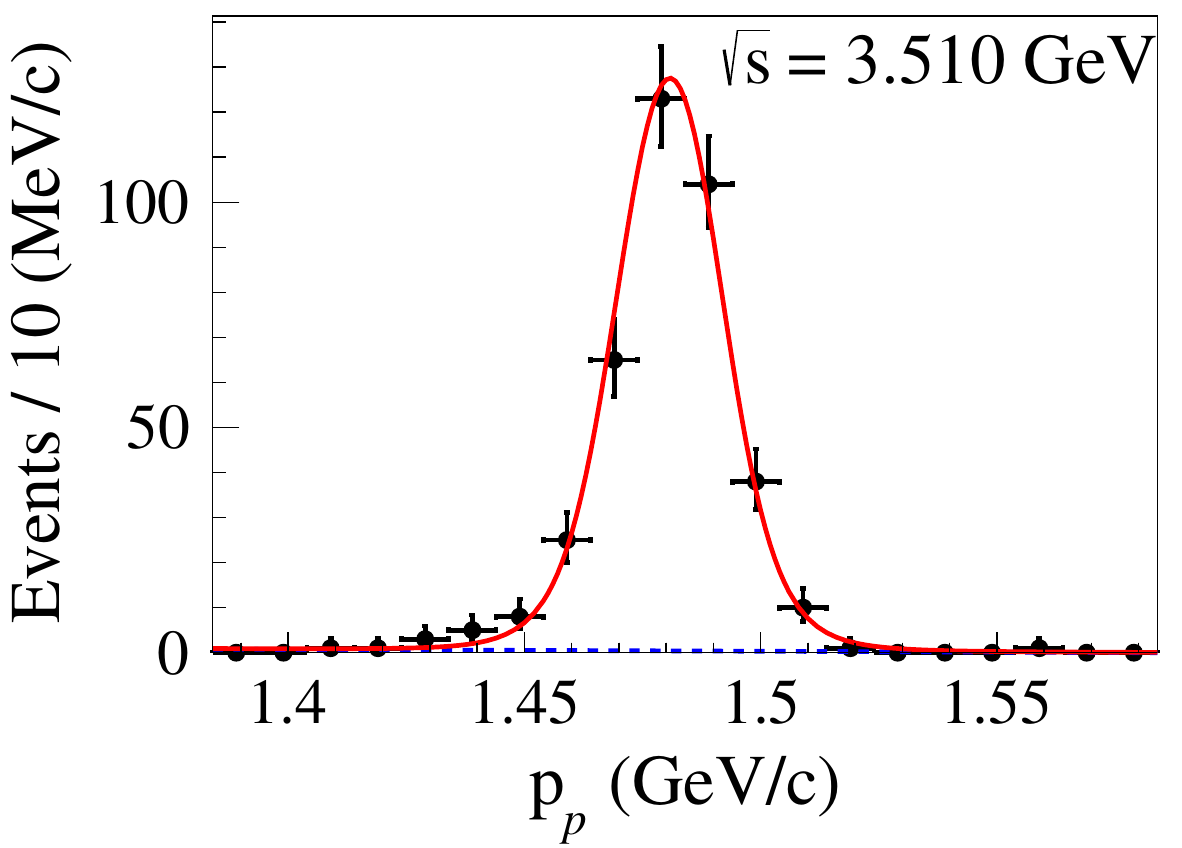}
    \includegraphics[width=.245\linewidth]{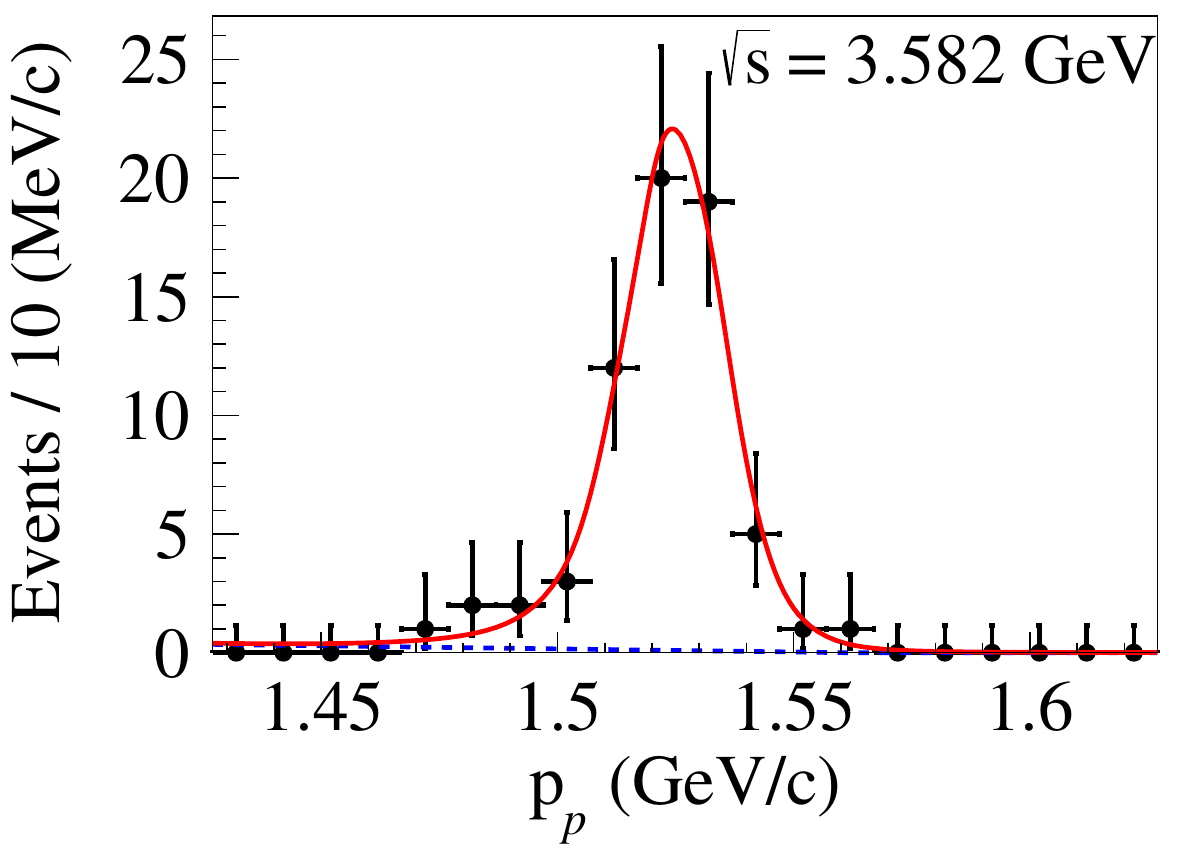}
    \includegraphics[width=.245\linewidth]{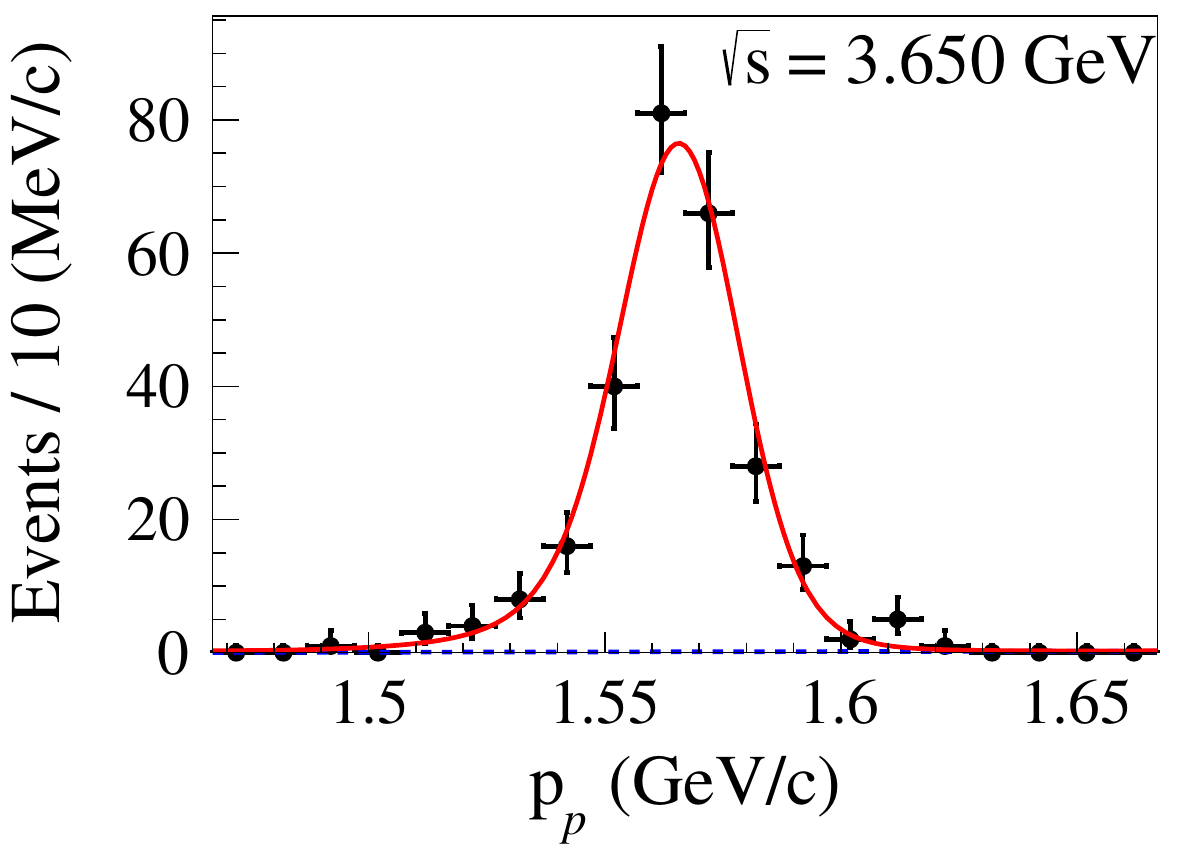}
    \includegraphics[width=.245\linewidth]{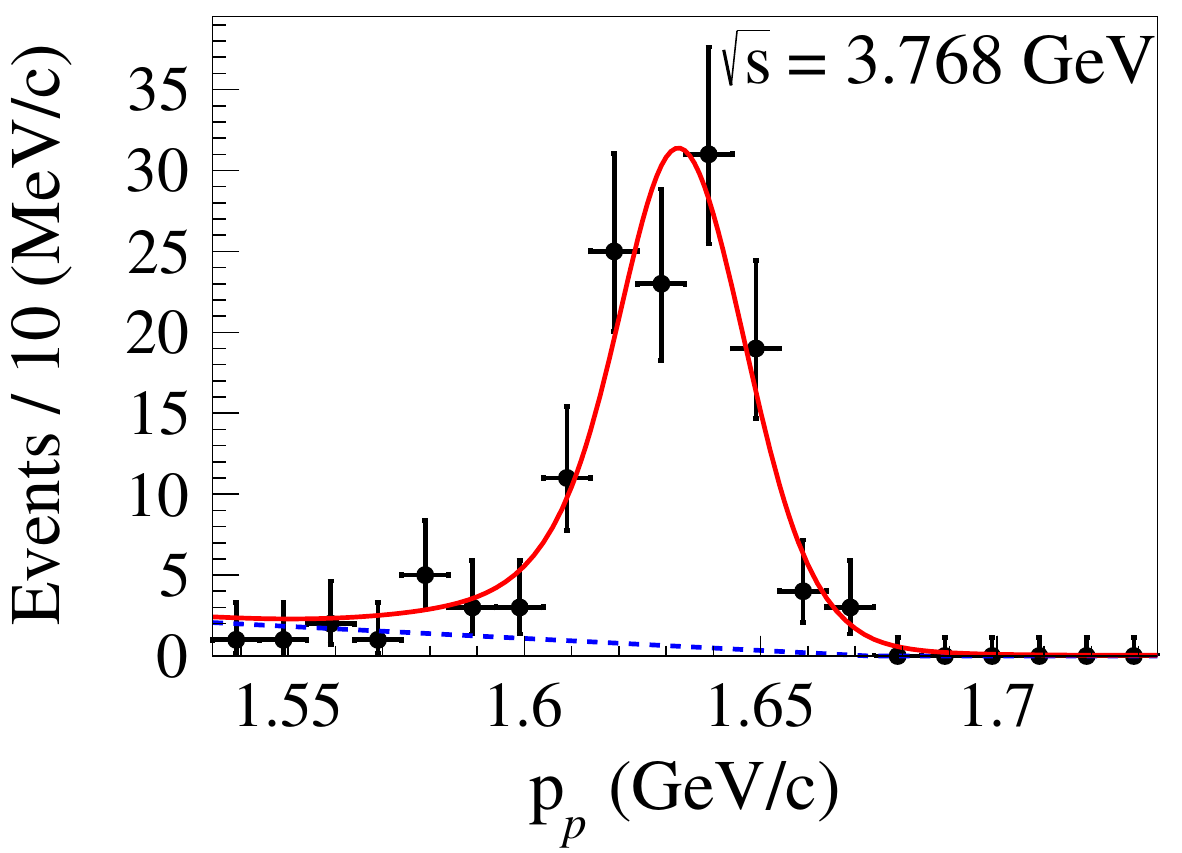}\\
    \includegraphics[width=.245\linewidth]{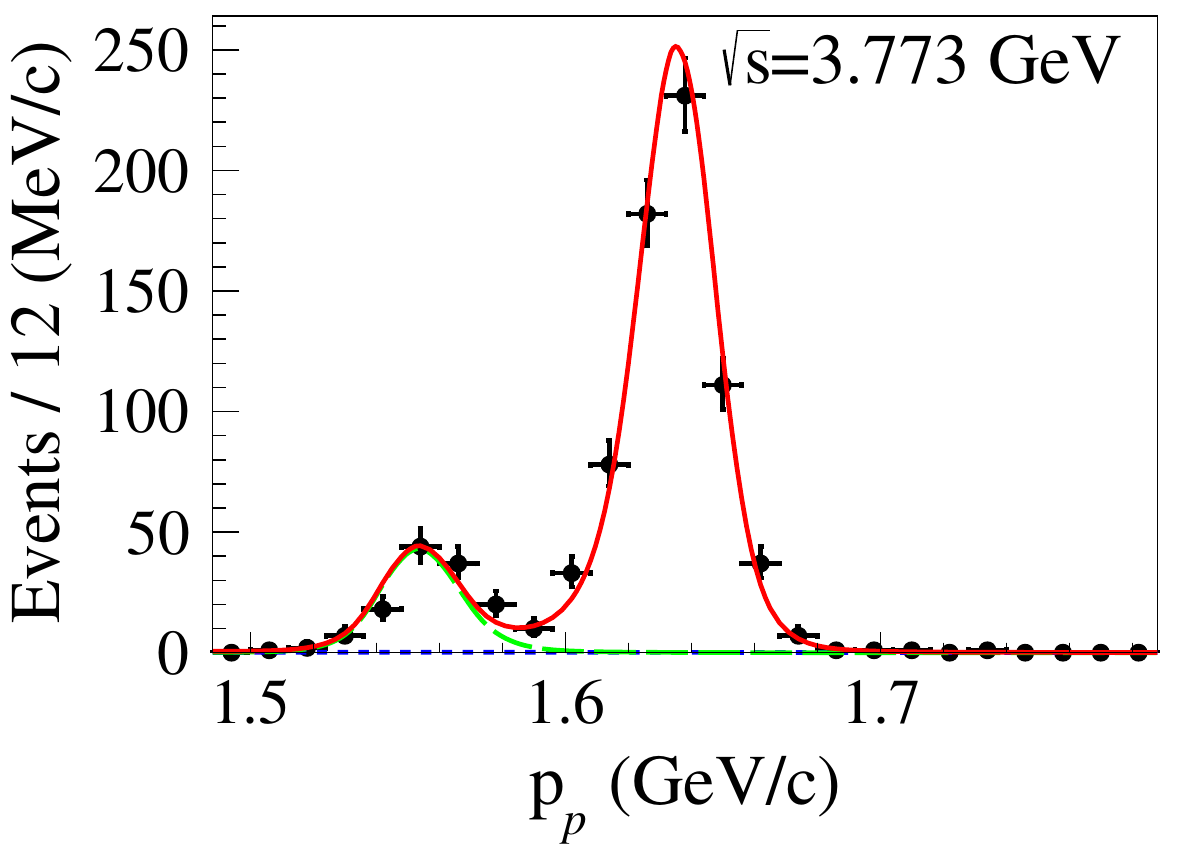}
    \includegraphics[width=.245\linewidth]{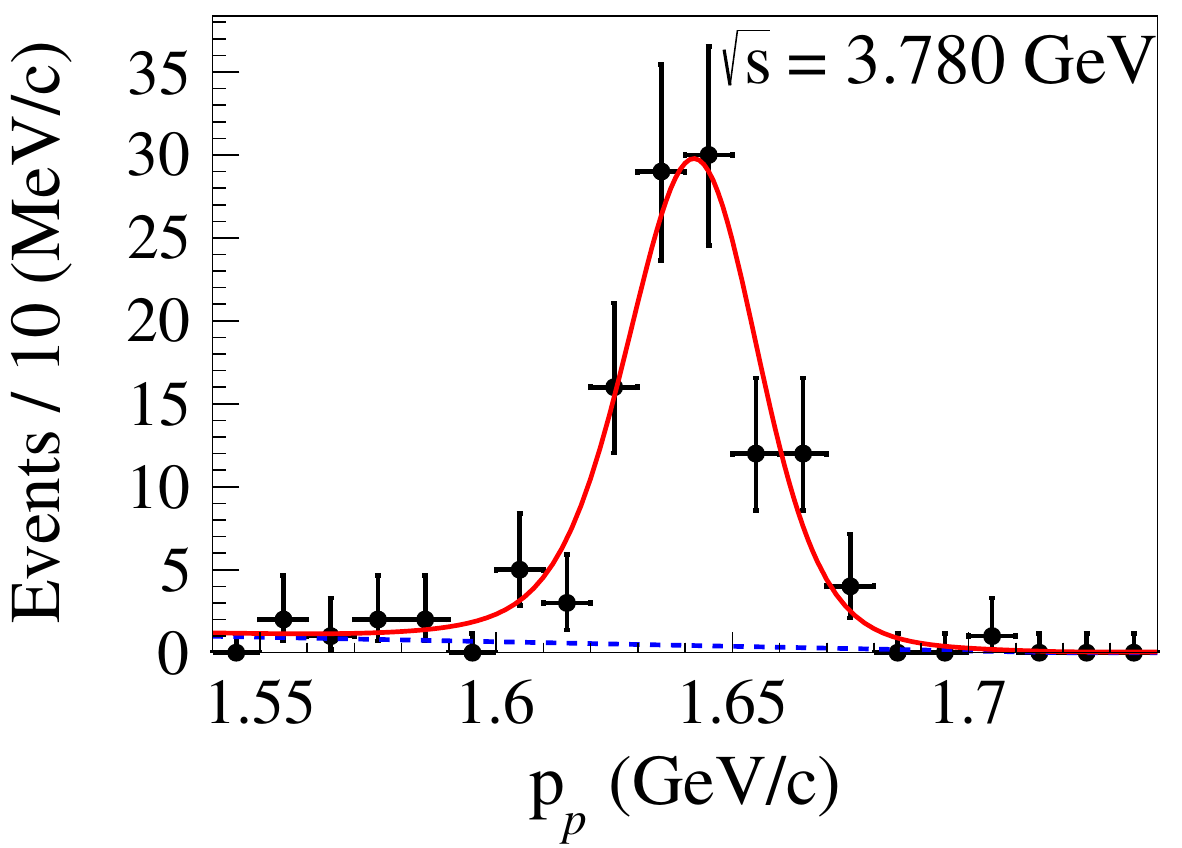}
    \includegraphics[width=.245\linewidth]{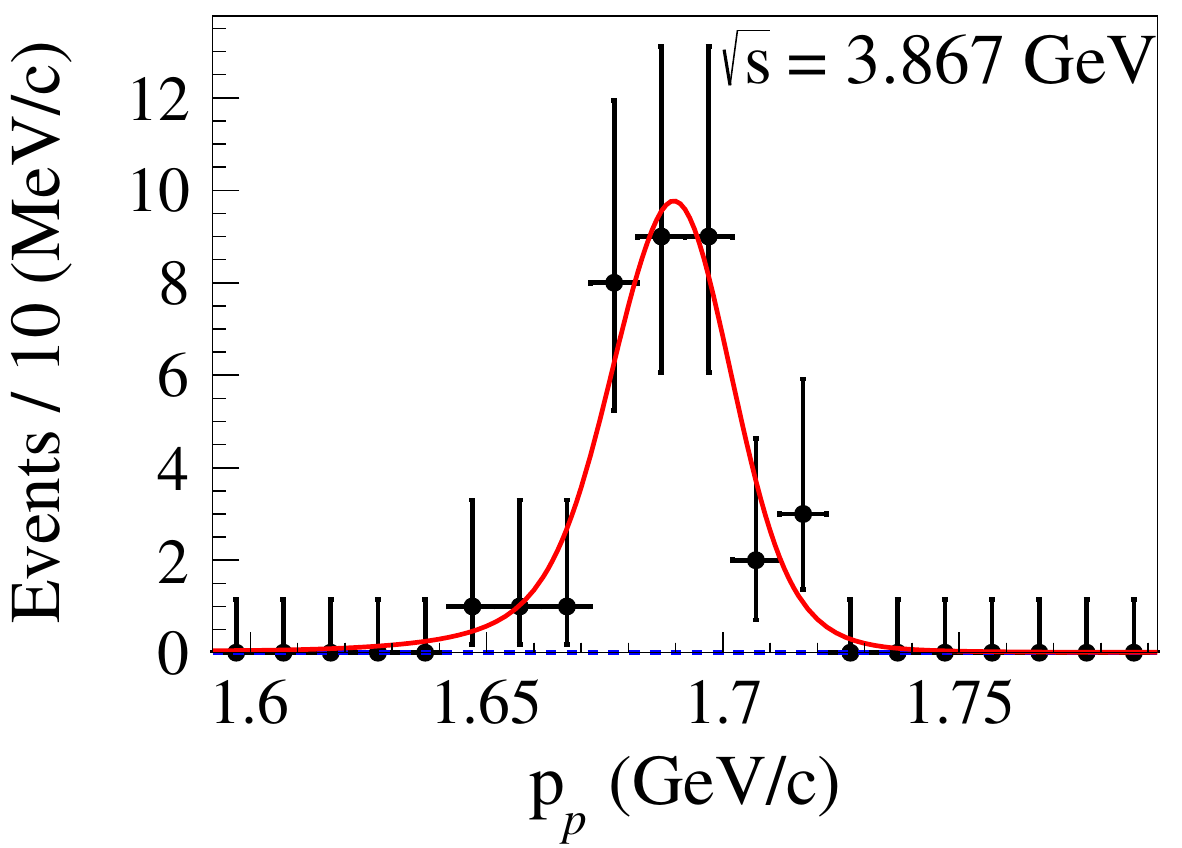}
    \includegraphics[width=.245\linewidth]{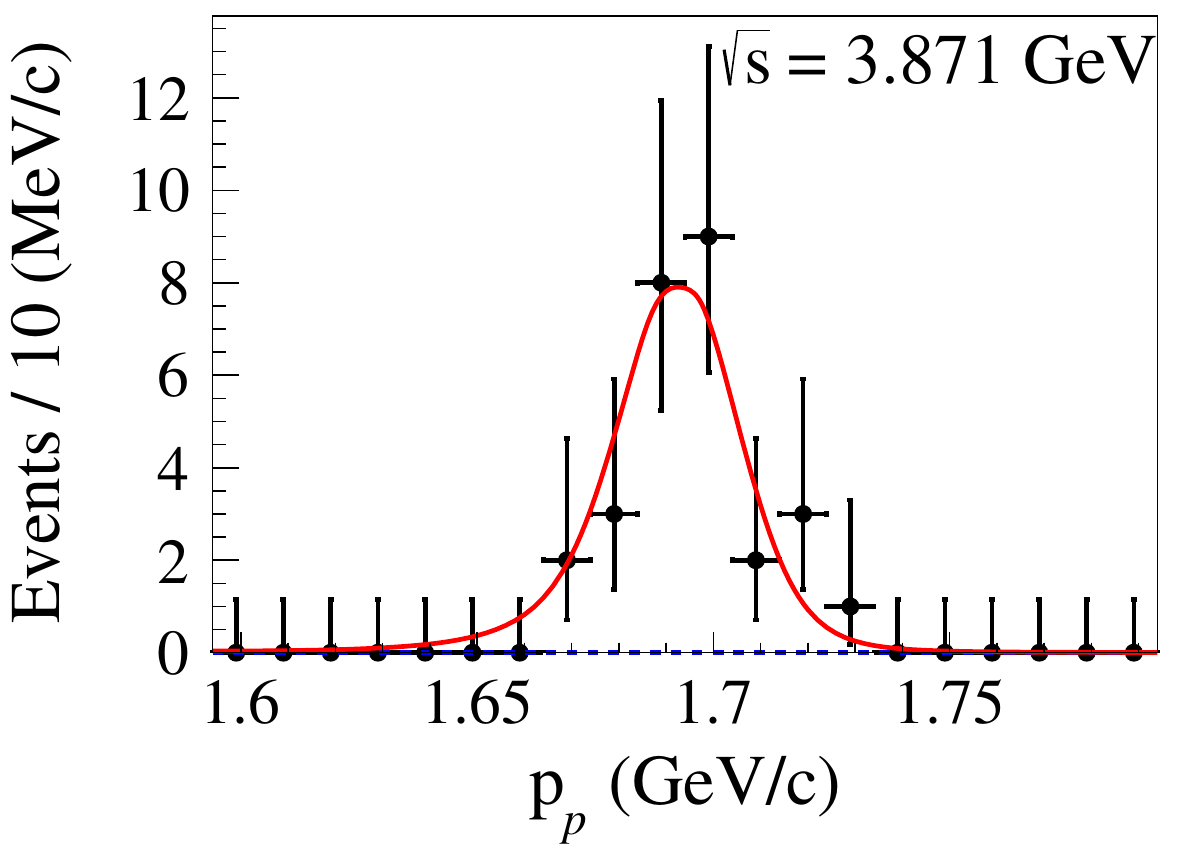}\\
    \includegraphics[width=.245\linewidth]{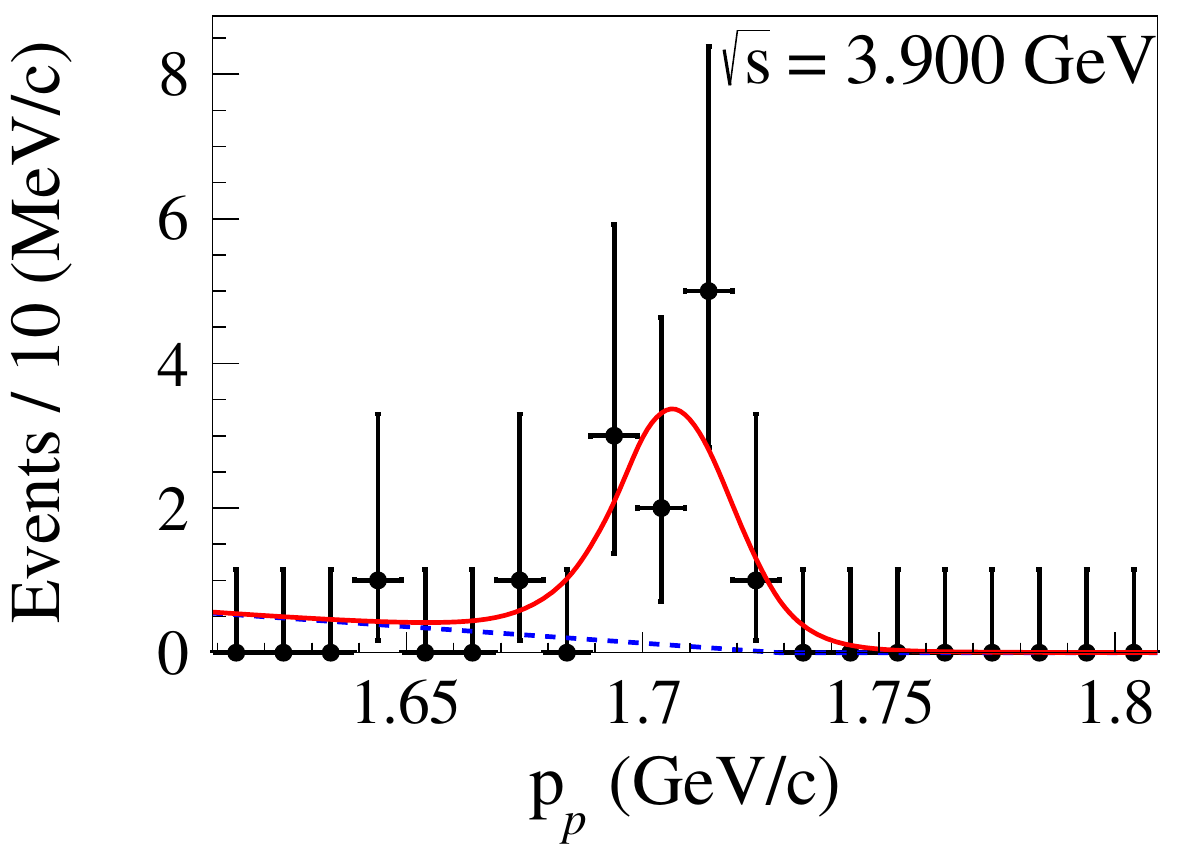}
    \includegraphics[width=.245\linewidth]{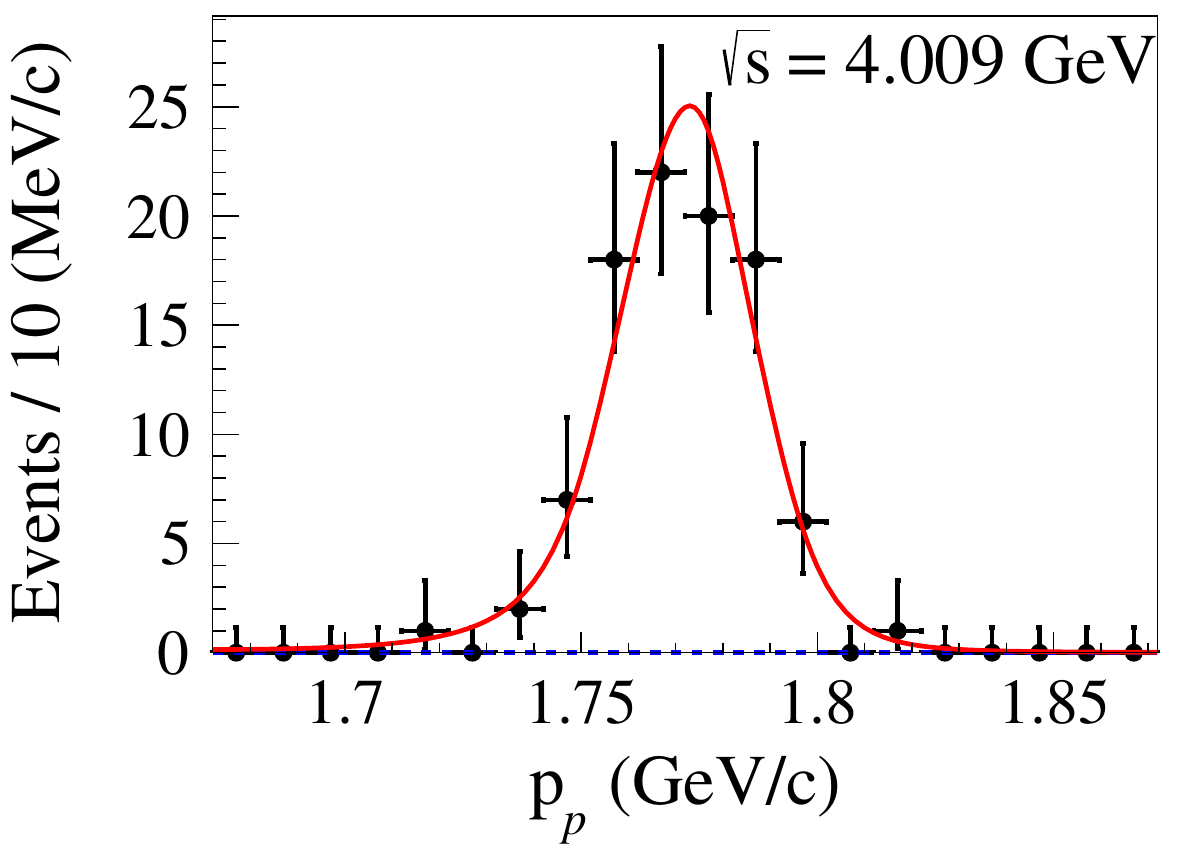}
    \includegraphics[width=.245\linewidth]{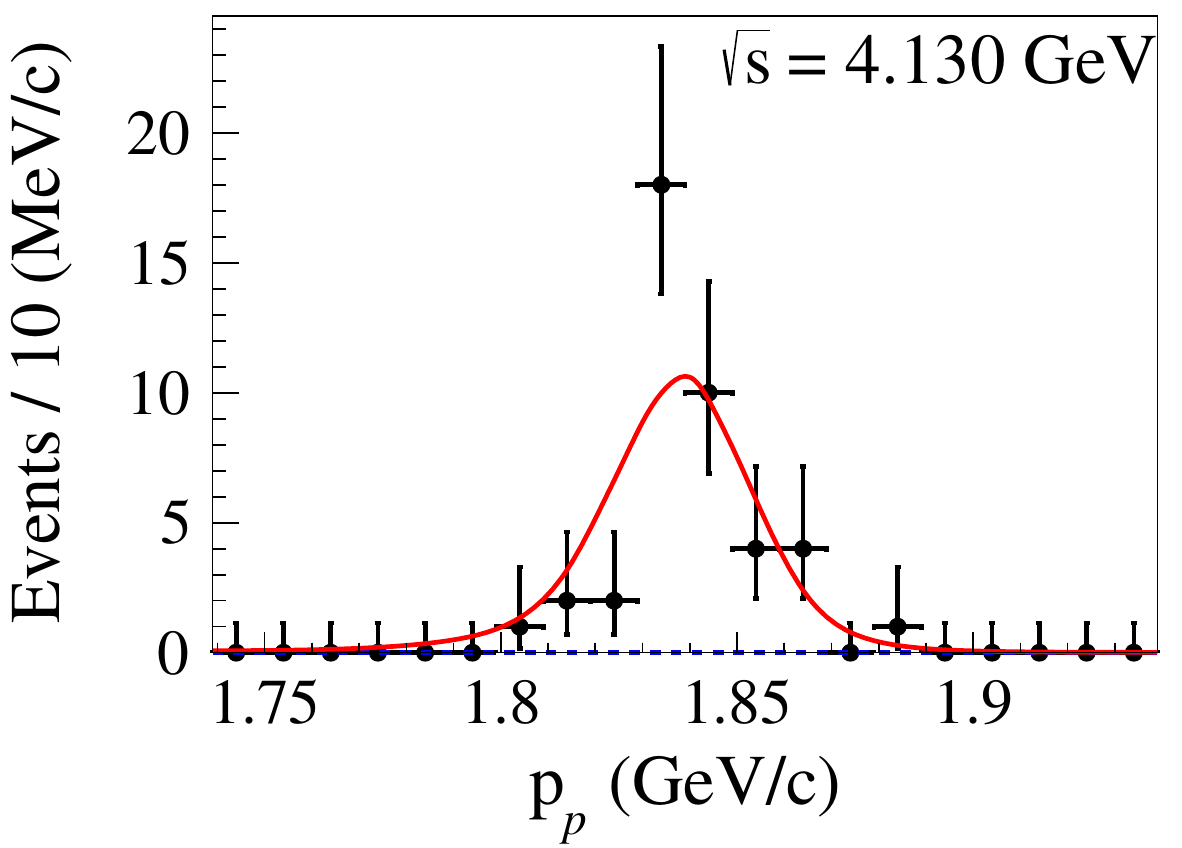}
    \includegraphics[width=.245\linewidth]{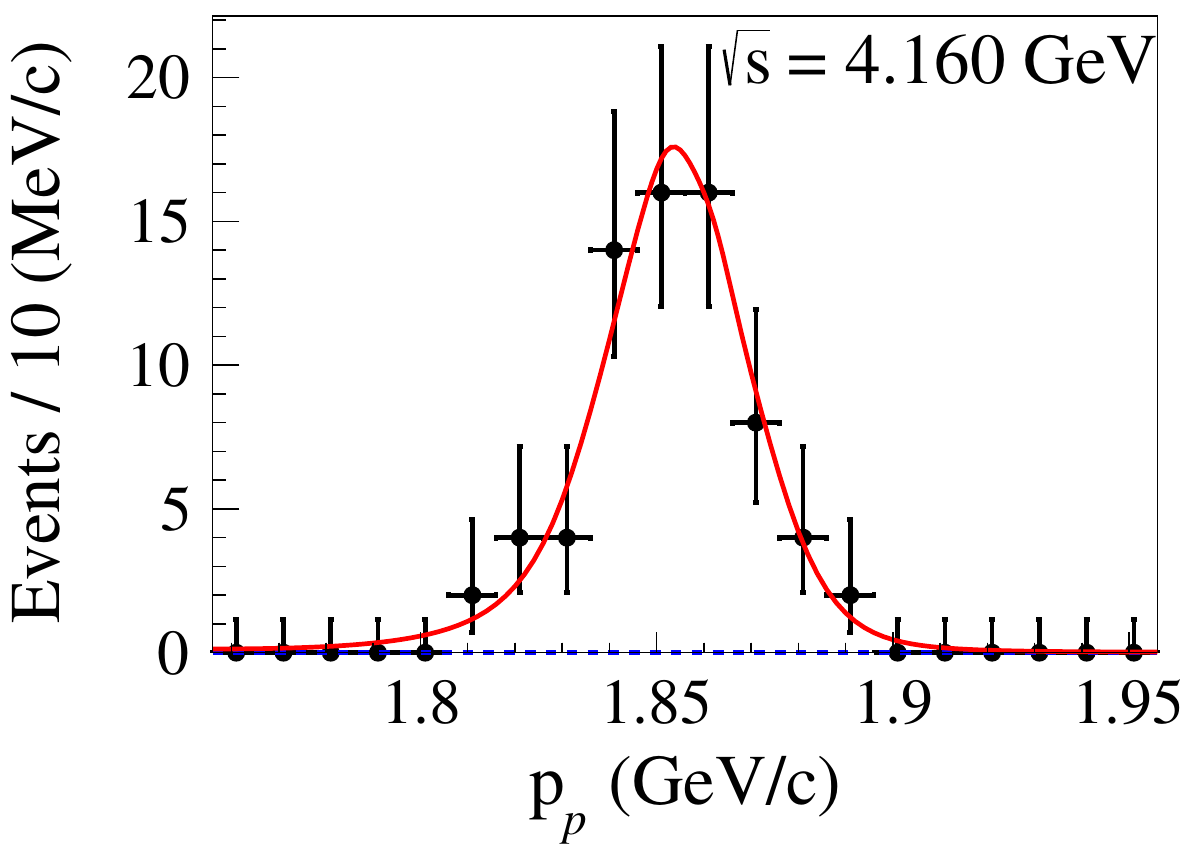}\\
    \includegraphics[width=.245\linewidth]{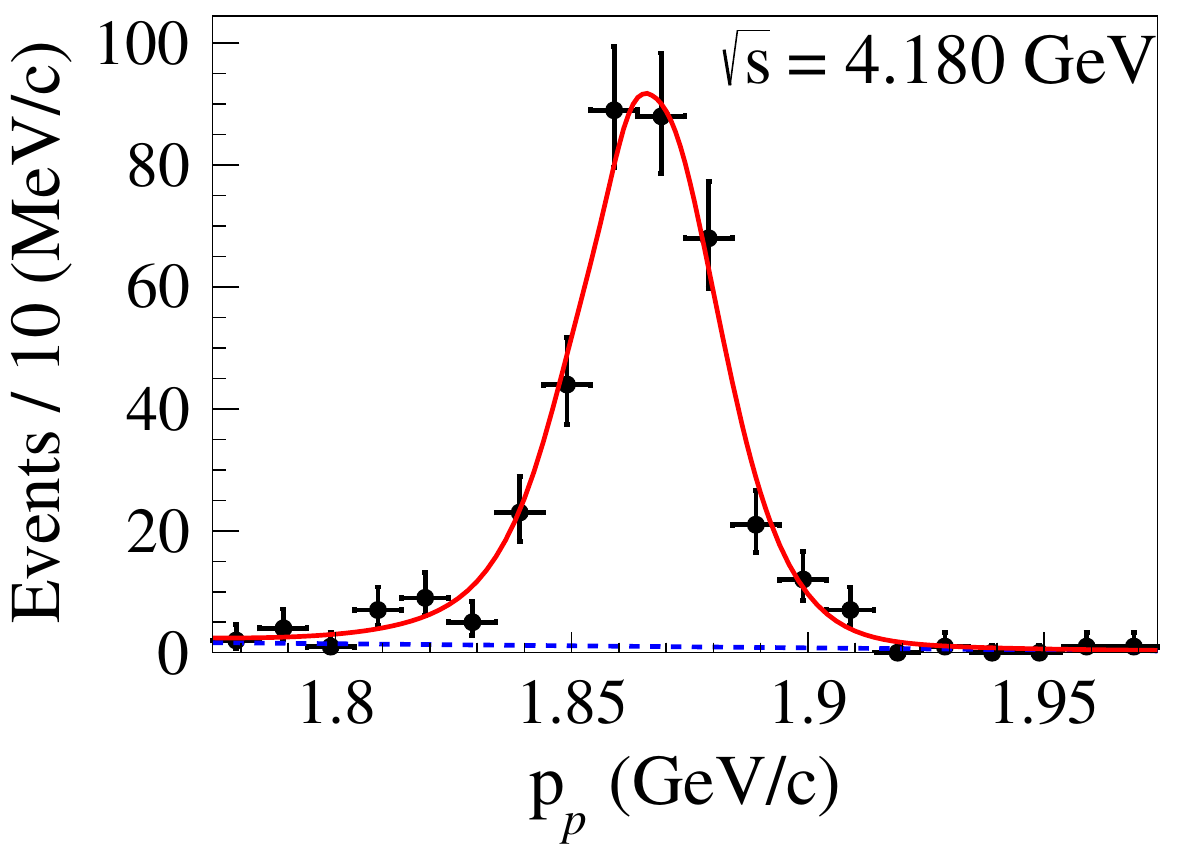}
    \includegraphics[width=.245\linewidth]{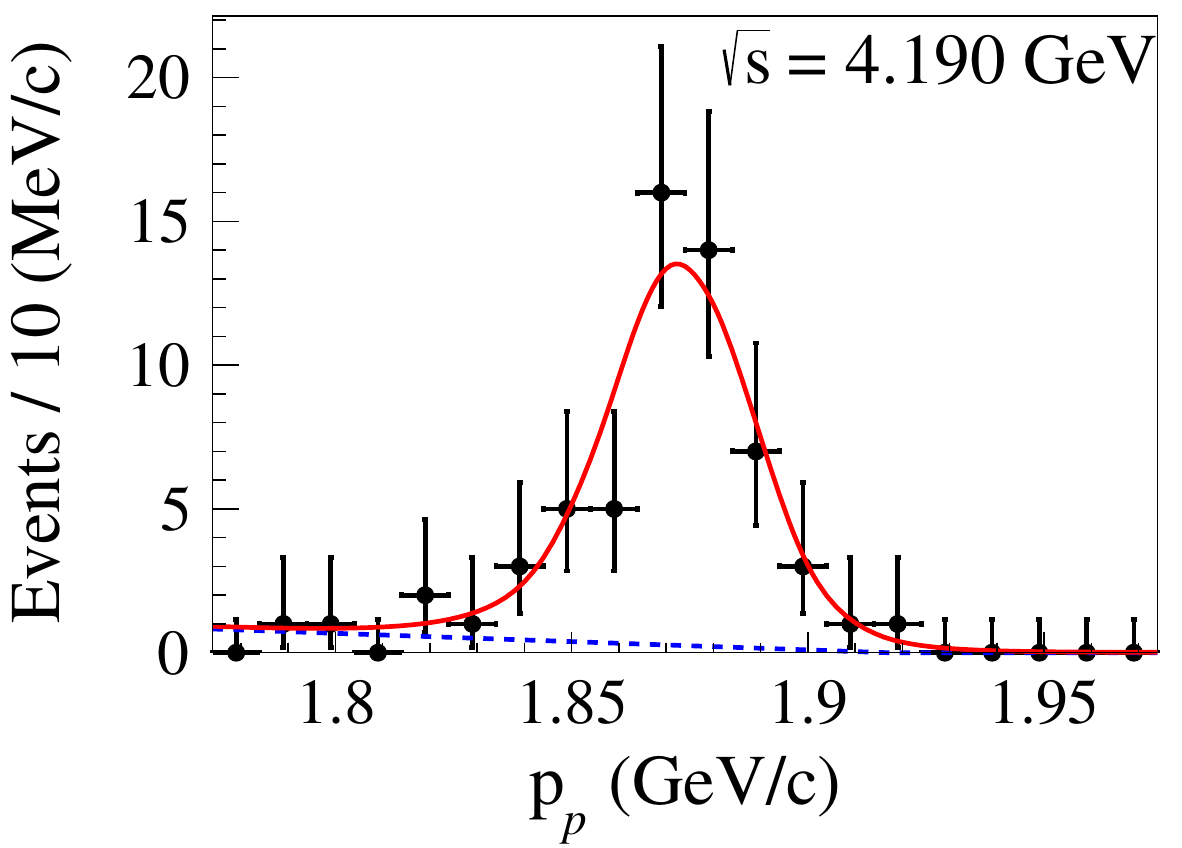}
    \includegraphics[width=.245\linewidth]{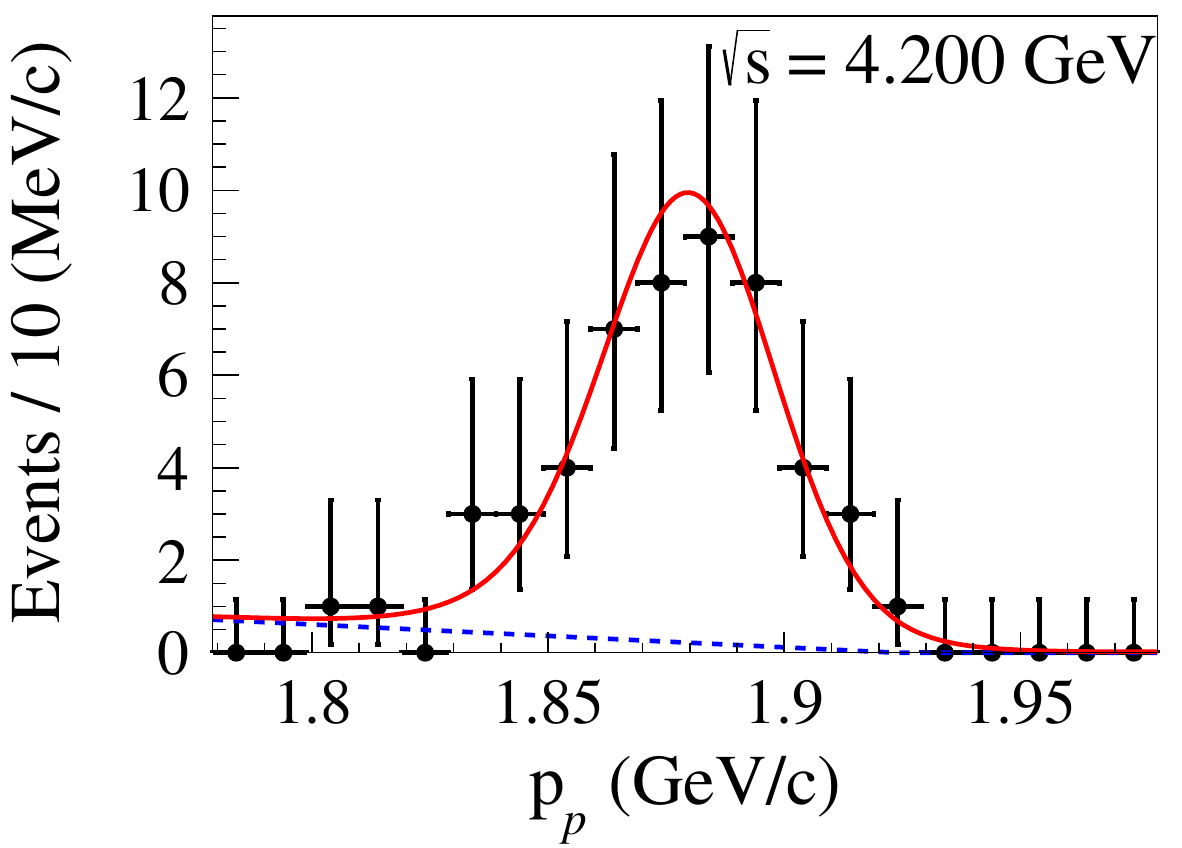}
    \includegraphics[width=.245\linewidth]{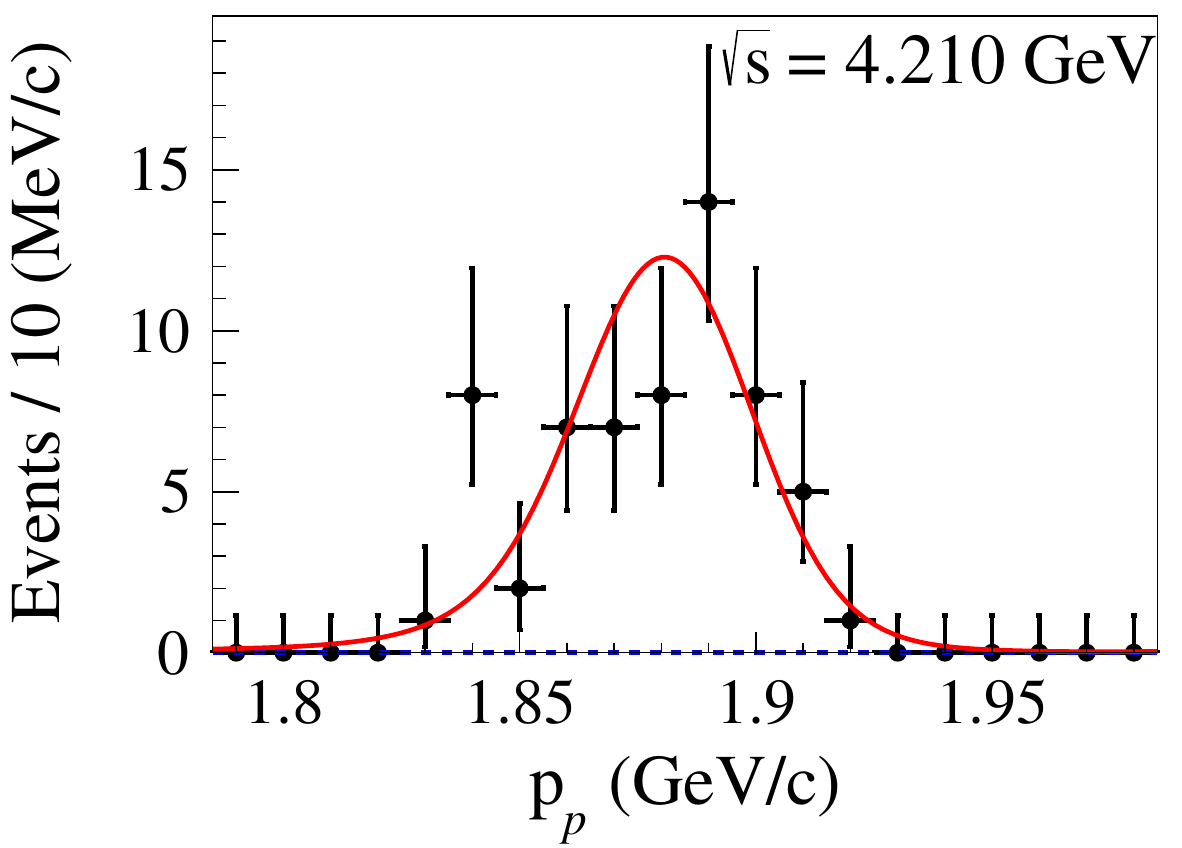}\\
    \includegraphics[width=.245\linewidth]{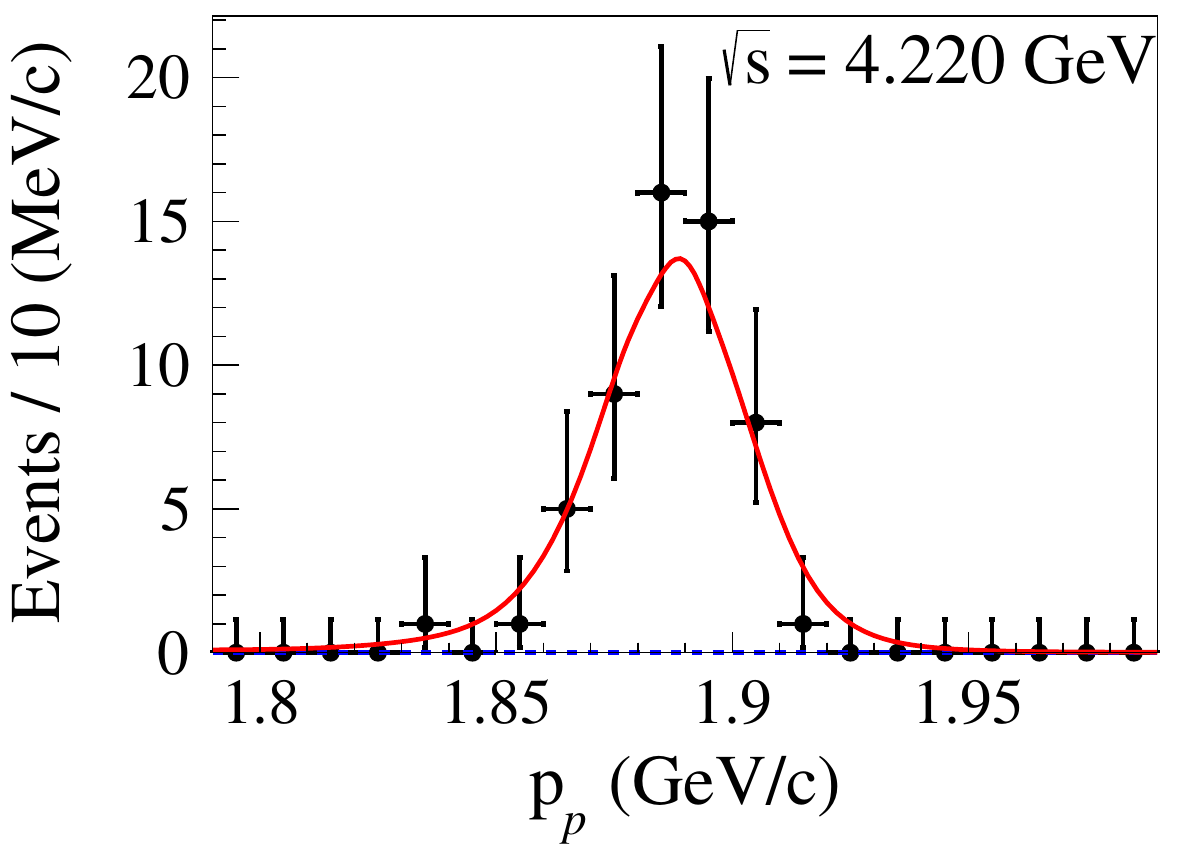}
    \includegraphics[width=.245\linewidth]{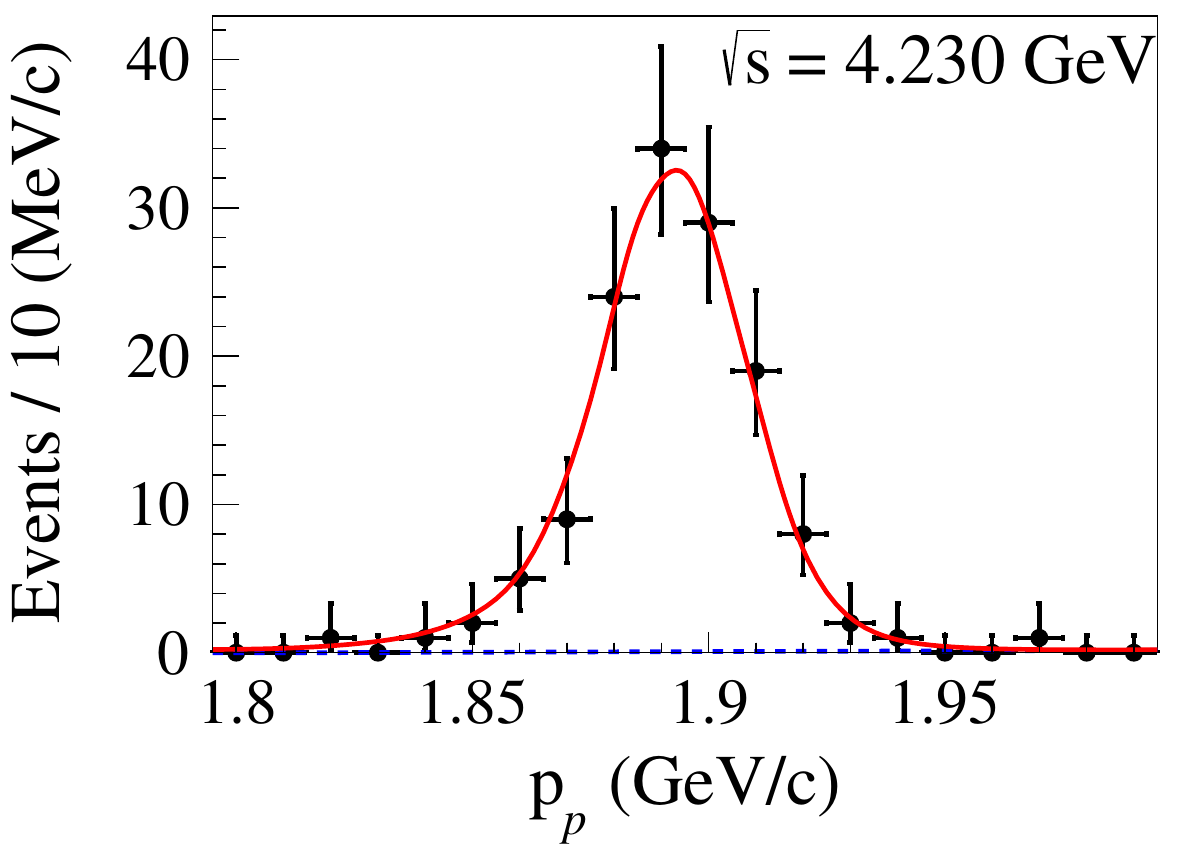}
    \includegraphics[width=.245\linewidth]{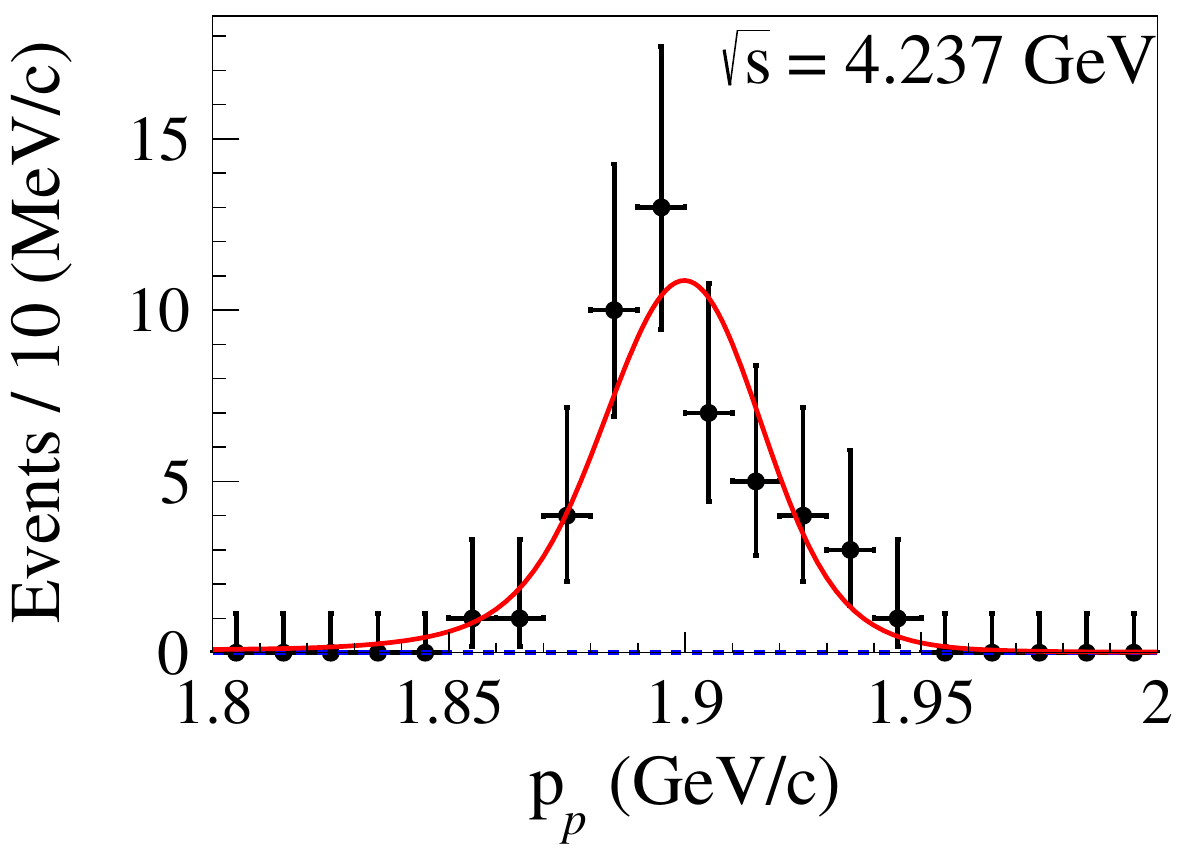}
    \includegraphics[width=.245\linewidth]{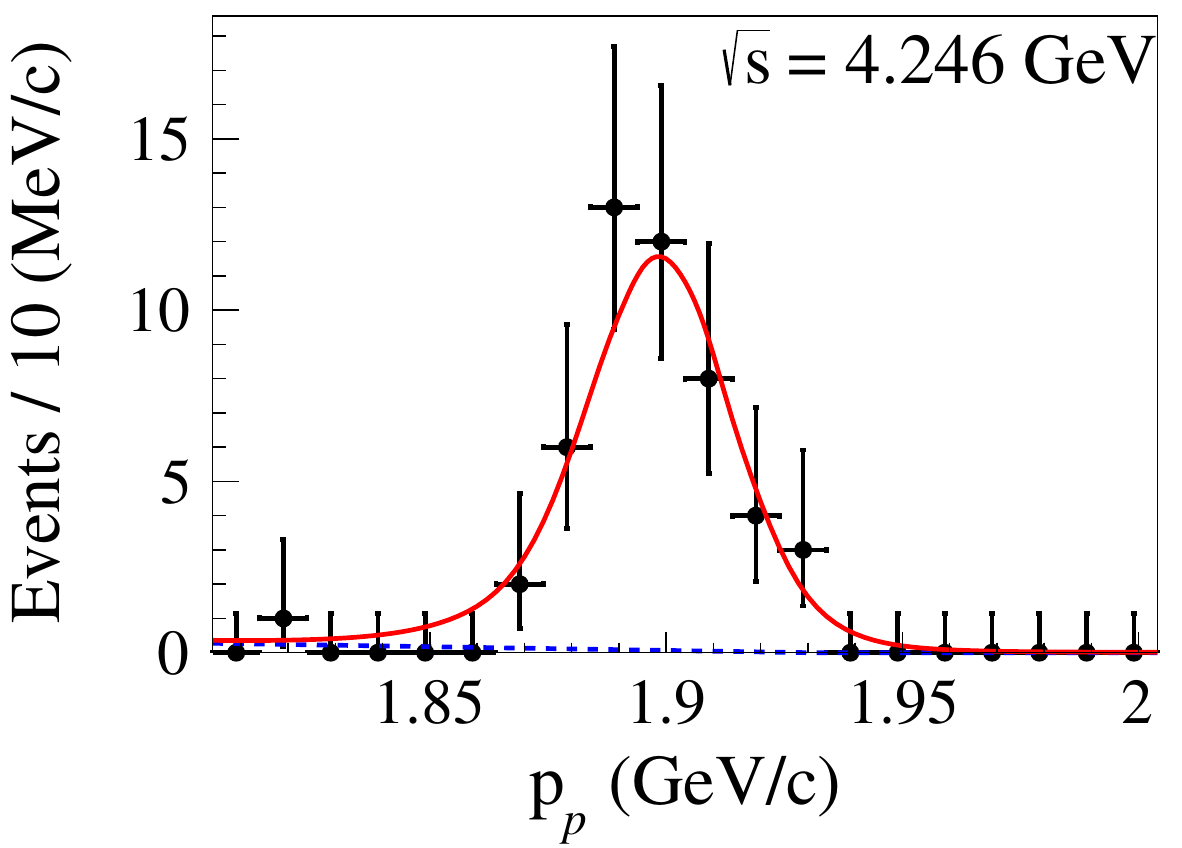}\\
    \includegraphics[width=.245\linewidth]{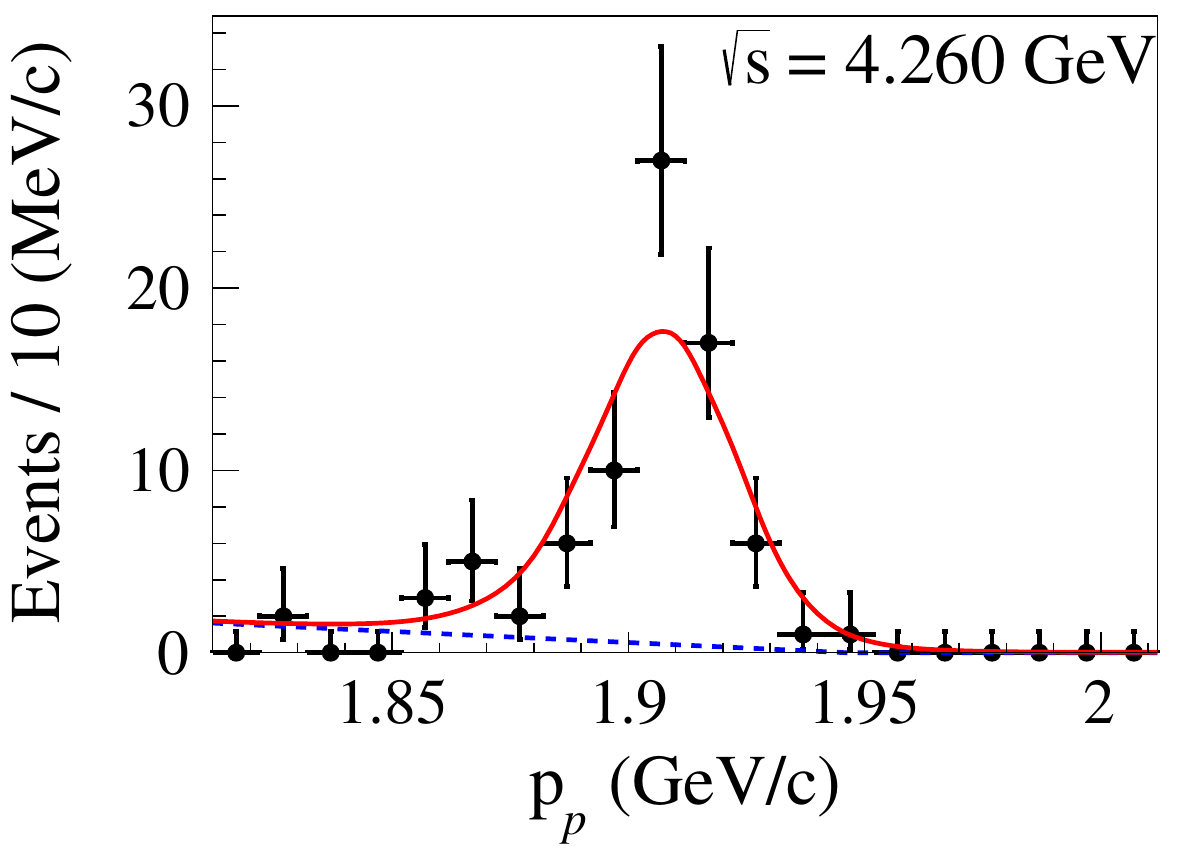}
    \includegraphics[width=.245\linewidth]{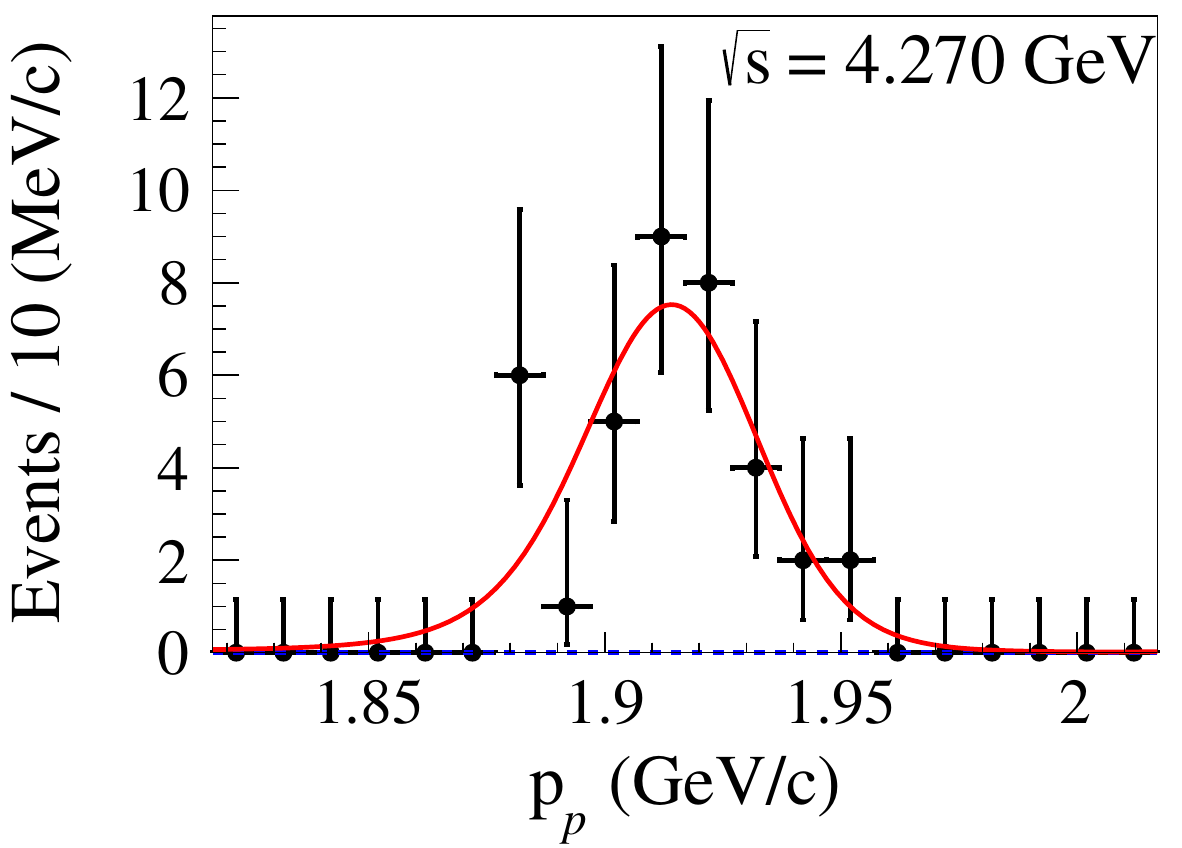}
    \includegraphics[width=.245\linewidth]{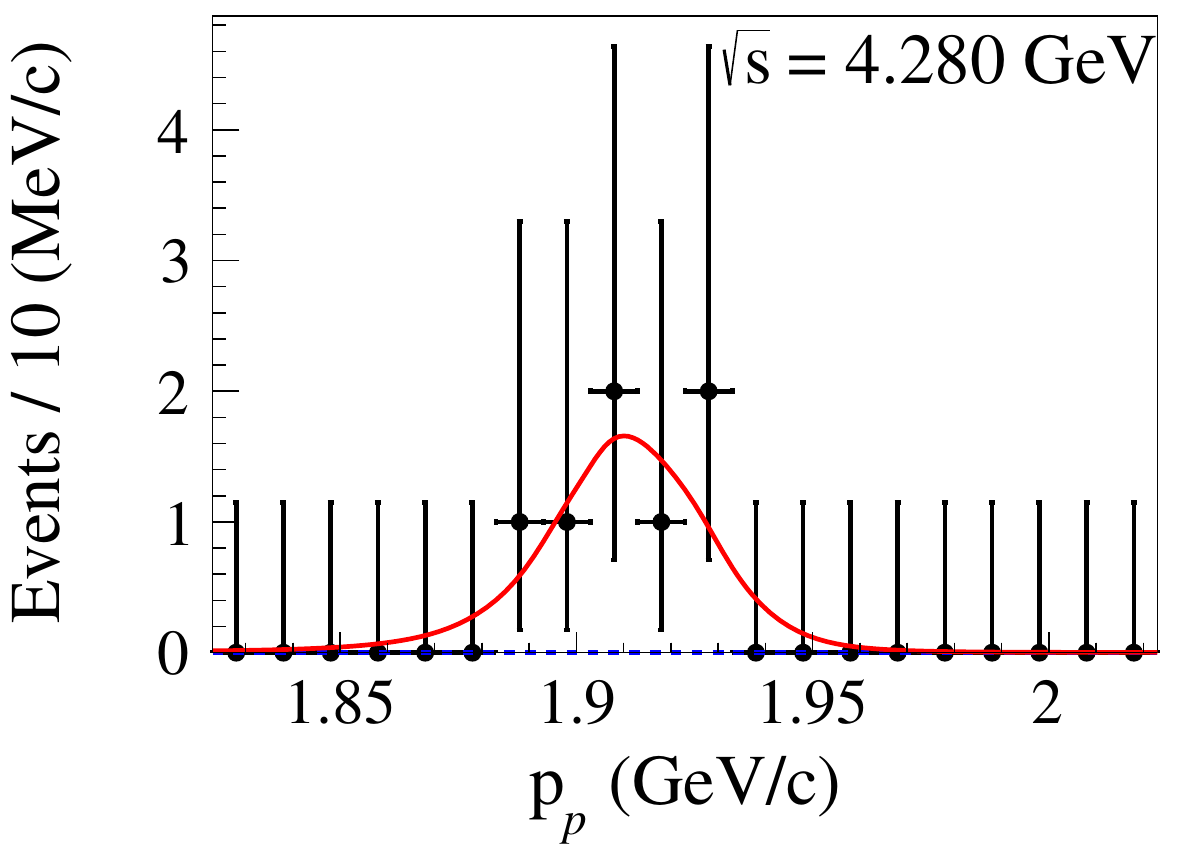}
    \includegraphics[width=.245\linewidth]{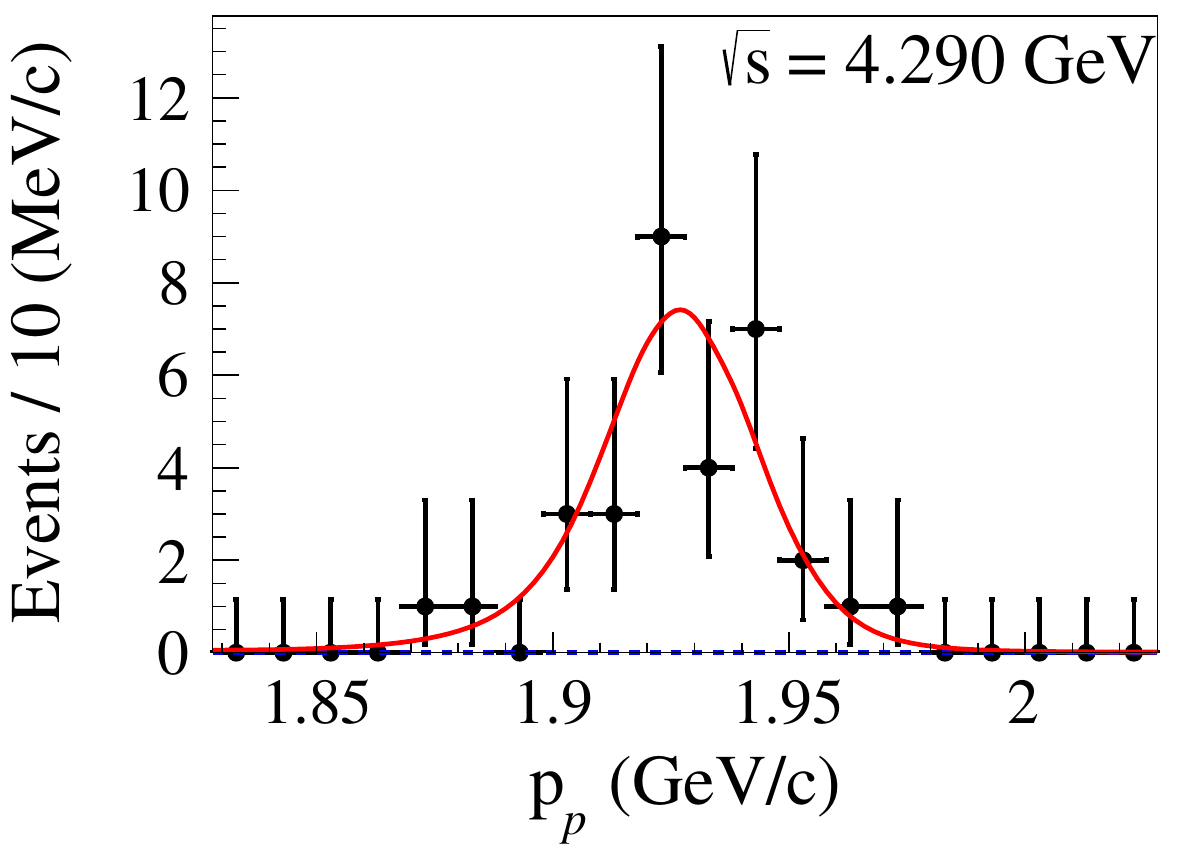}\\
    \includegraphics[width=.245\linewidth]{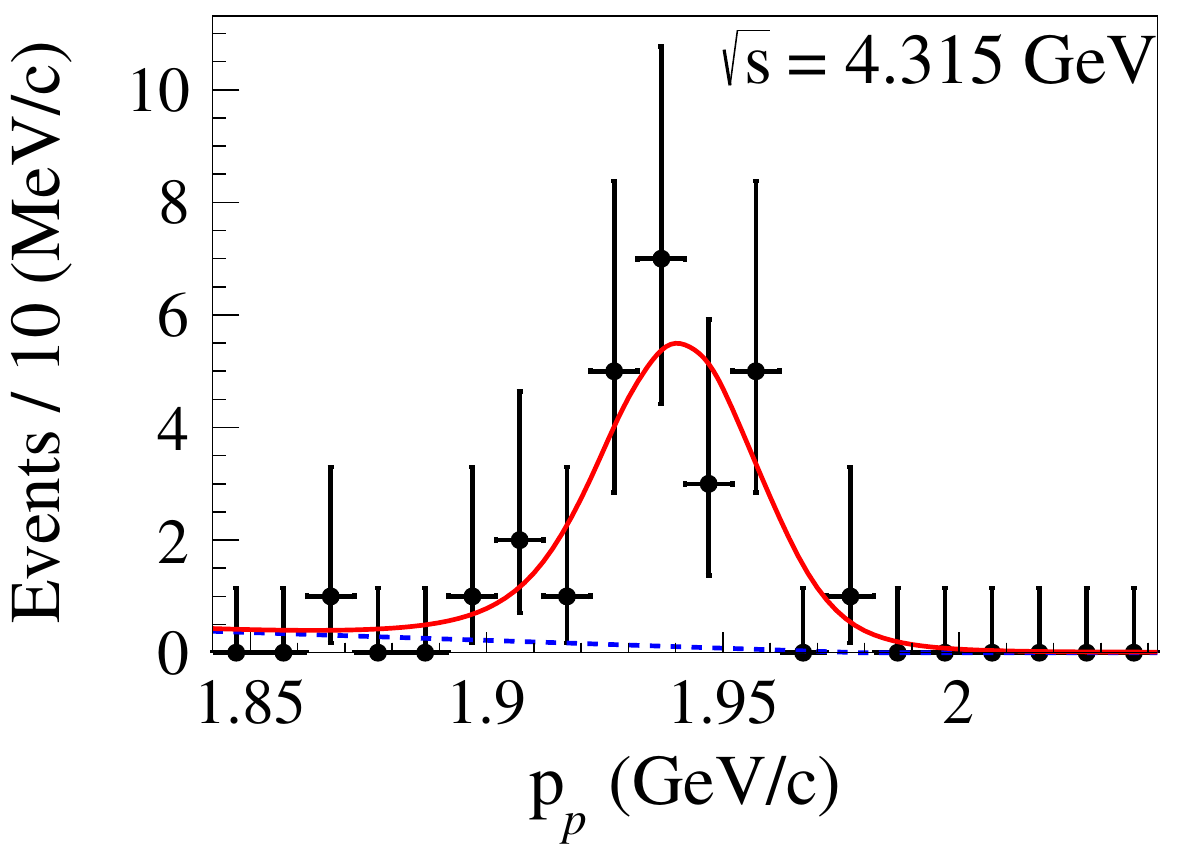}
    \includegraphics[width=.245\linewidth]{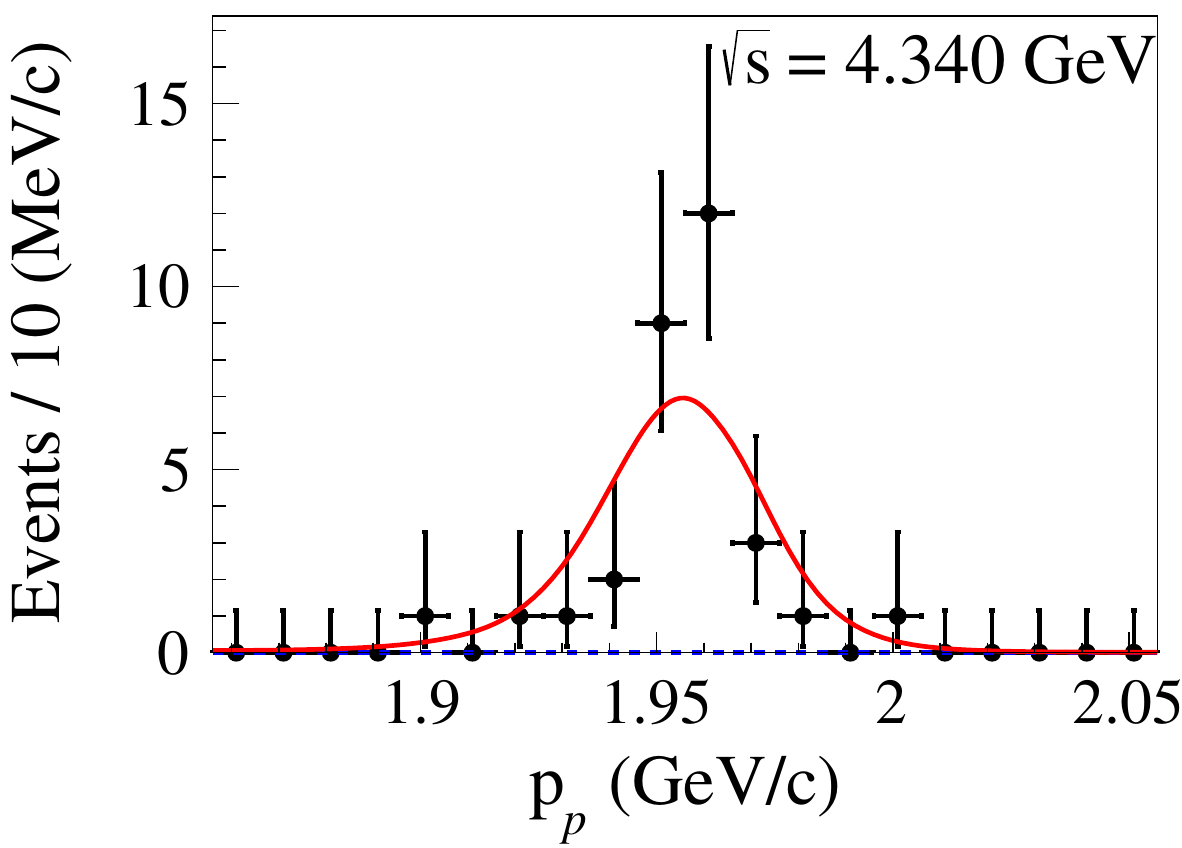}
    \includegraphics[width=.245\linewidth]{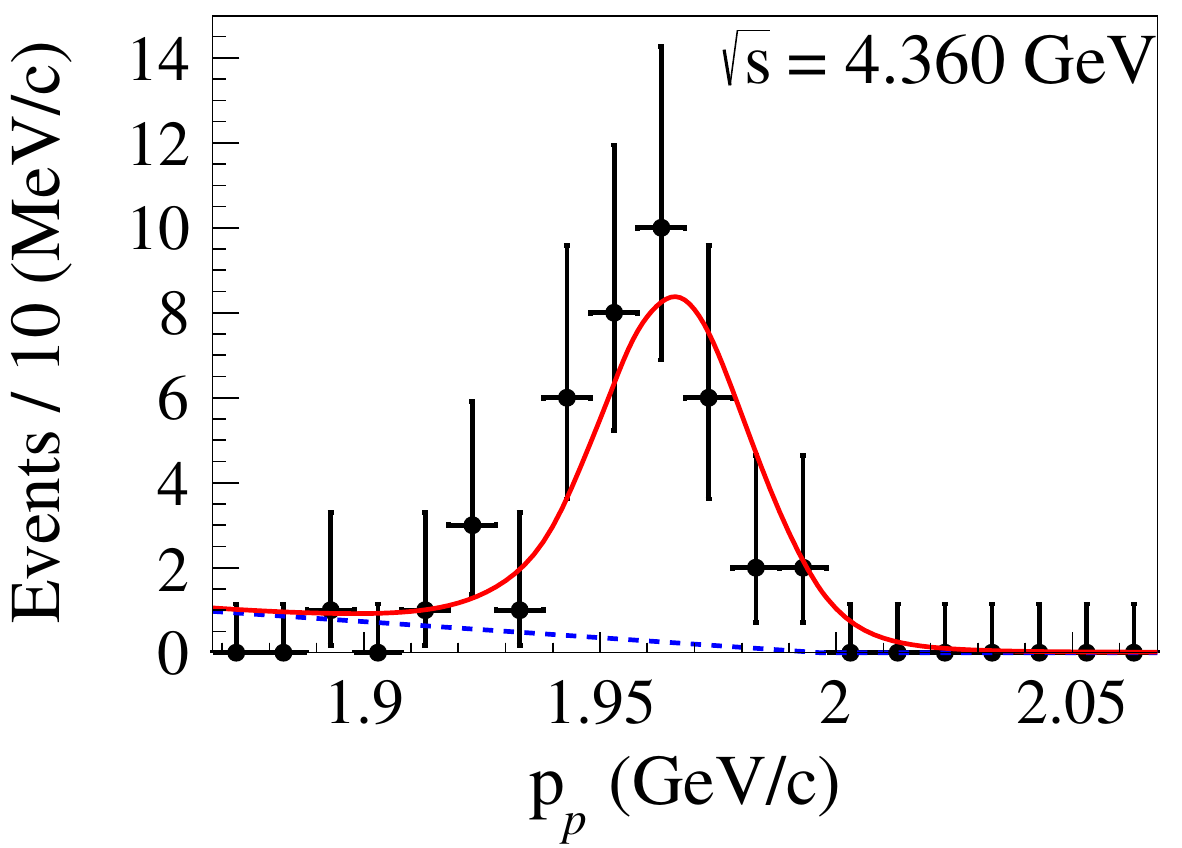}
    \includegraphics[width=.245\linewidth]{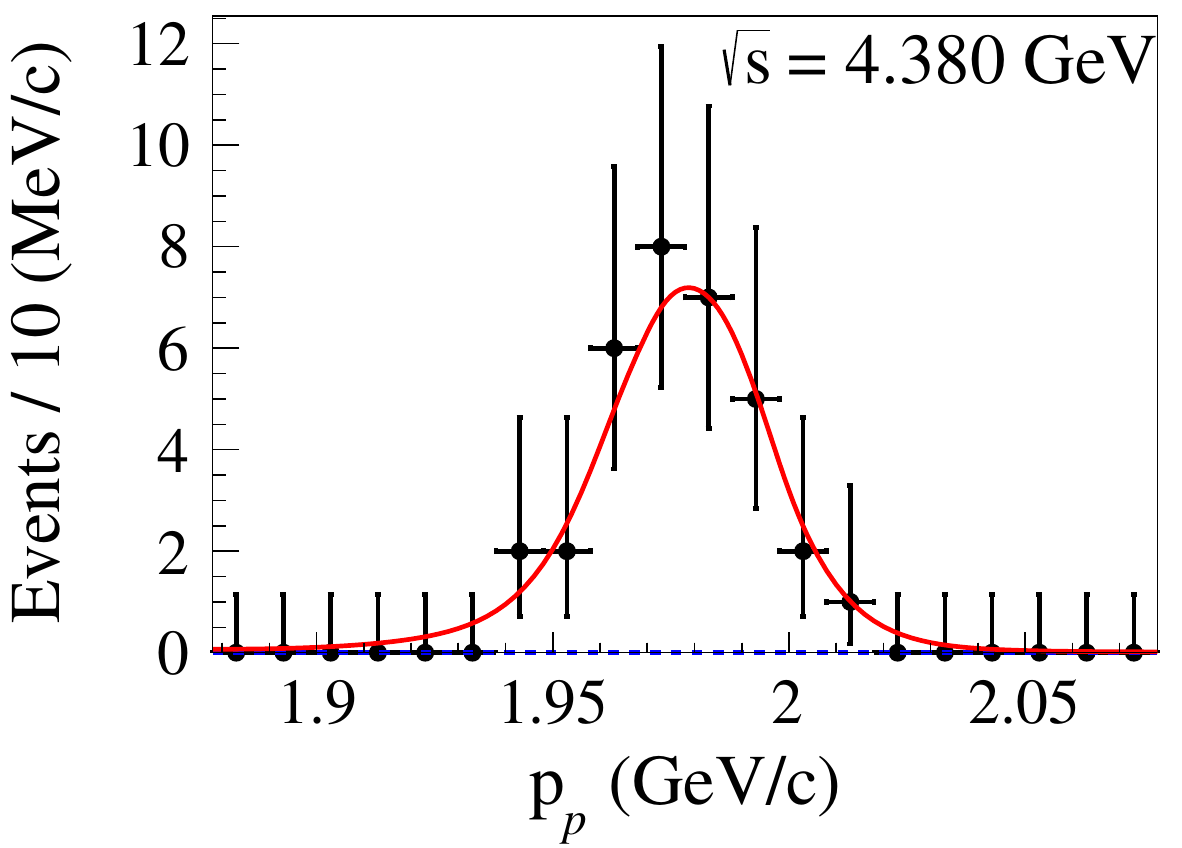}\\
    \caption{Fits to the proton momentum distributions at $\sqrt{s}=3.510-4.380$ GeV. The dots with error bars are data. The red solid lines represent the total fit, and the blue dashed lines stand for the backgrounds. The green dashed line in the fit at $\sqrt{s}=3.773$ GeV represents the background from ISR to the lower lying $\psi(3686)$ resonance.}
    \label{extract_signal_yield}
\end{figure*}

\begin{figure*}[tbp!]
    \centering
    \includegraphics[width=.245\linewidth]{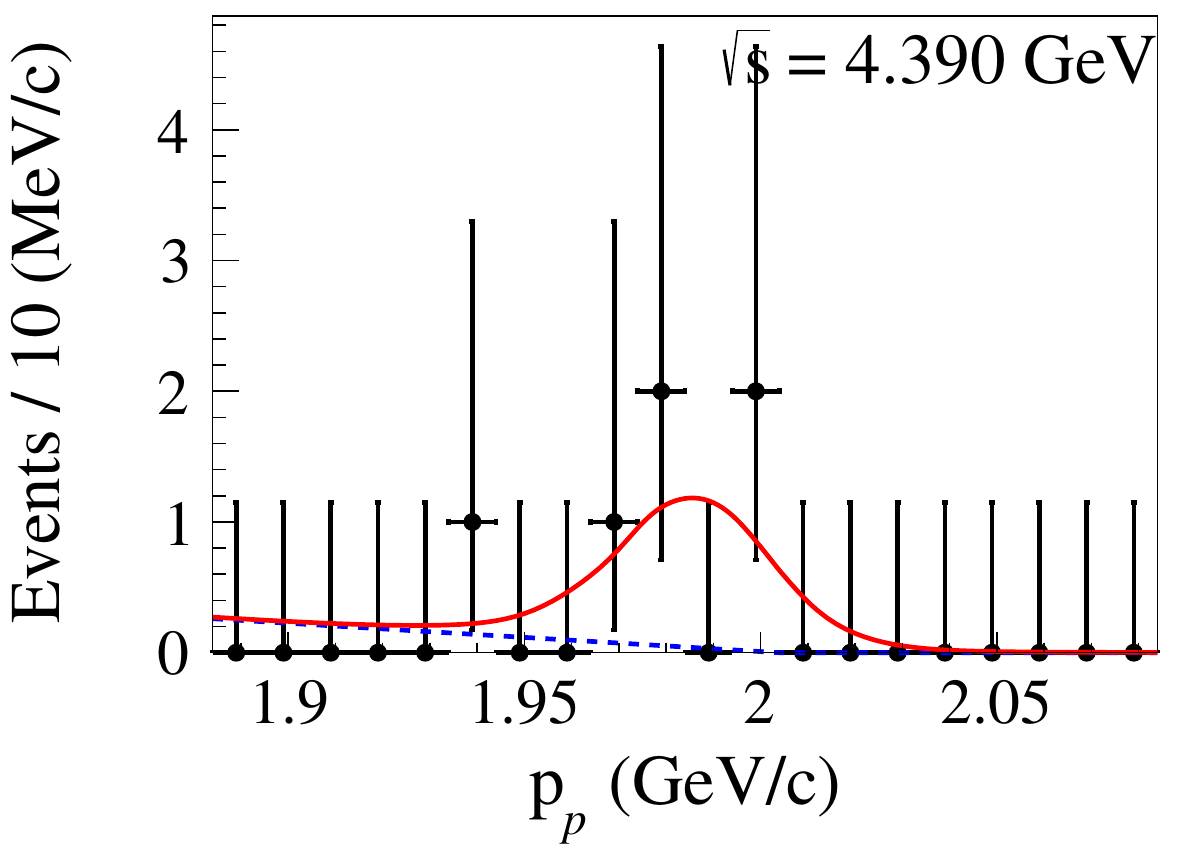}
    \includegraphics[width=.245\linewidth]{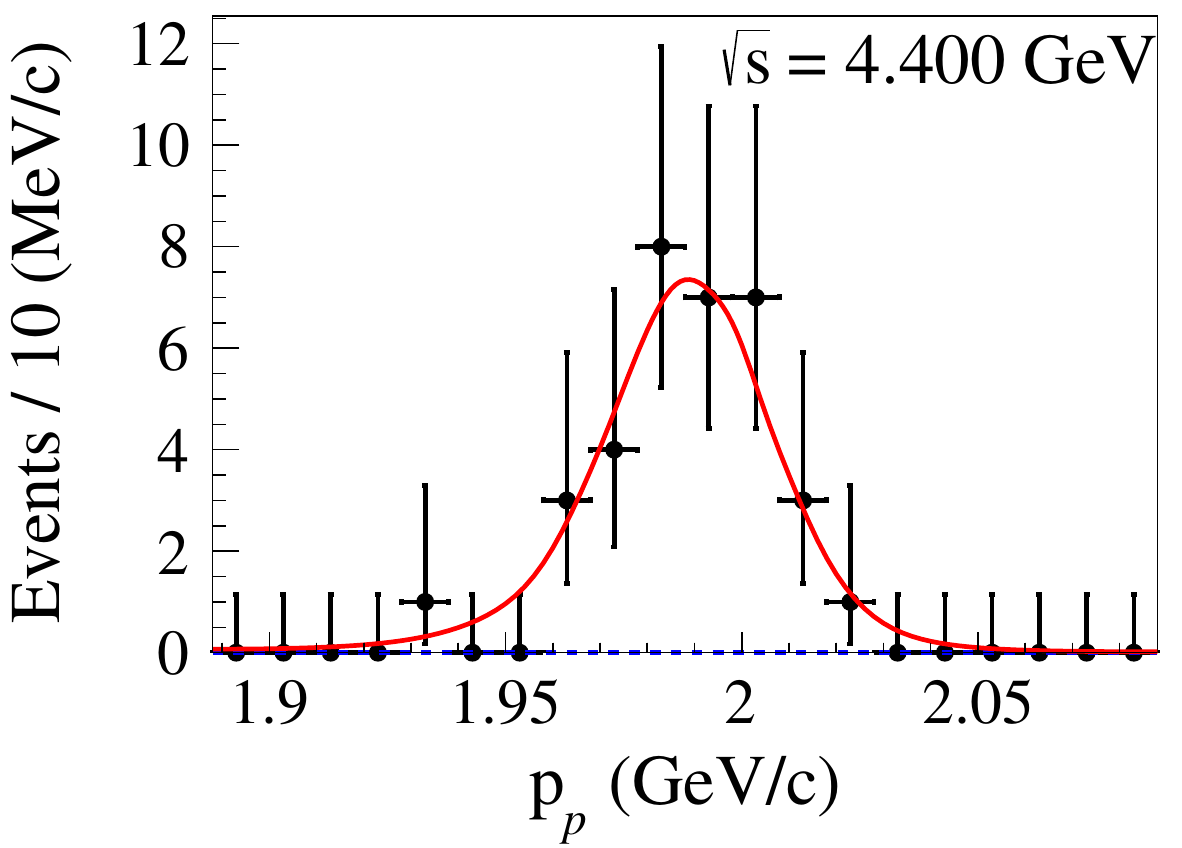}
    \includegraphics[width=.245\linewidth]{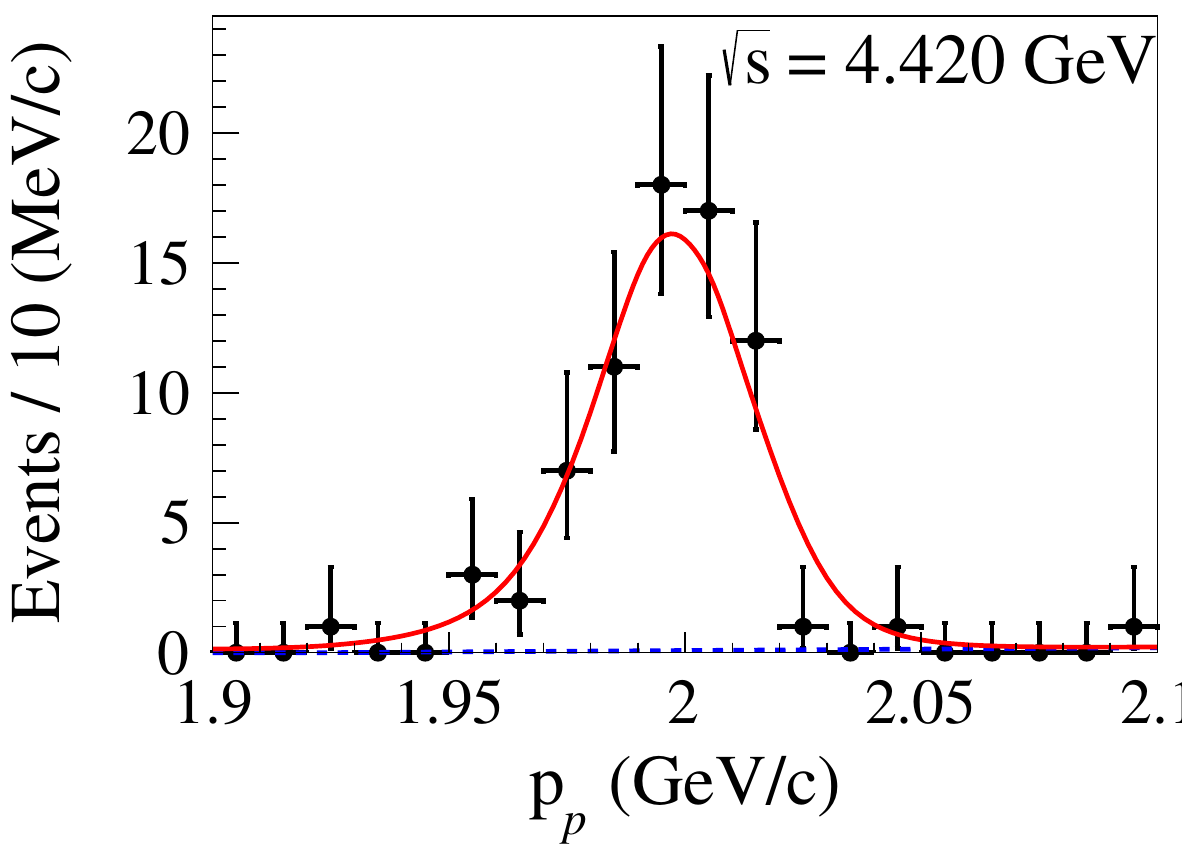}
    \includegraphics[width=.245\linewidth]{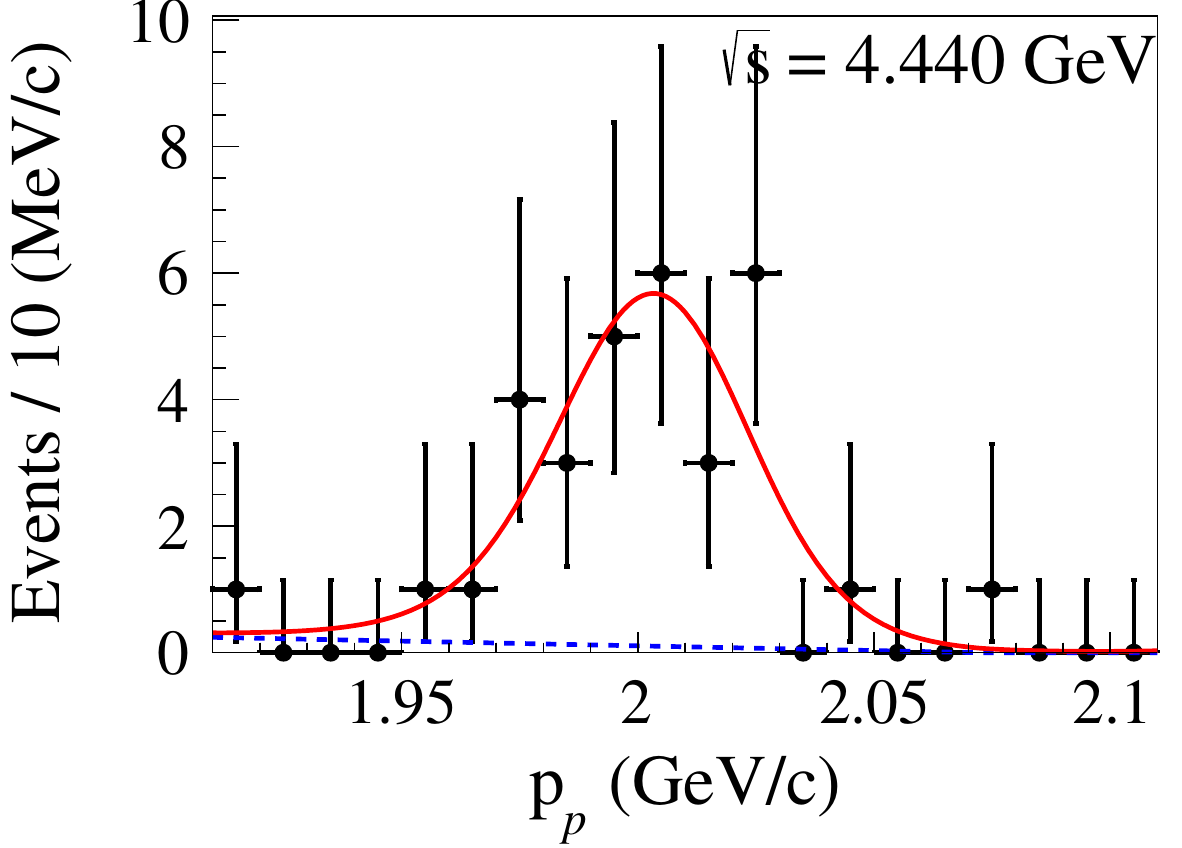}\\
    \includegraphics[width=.245\linewidth]{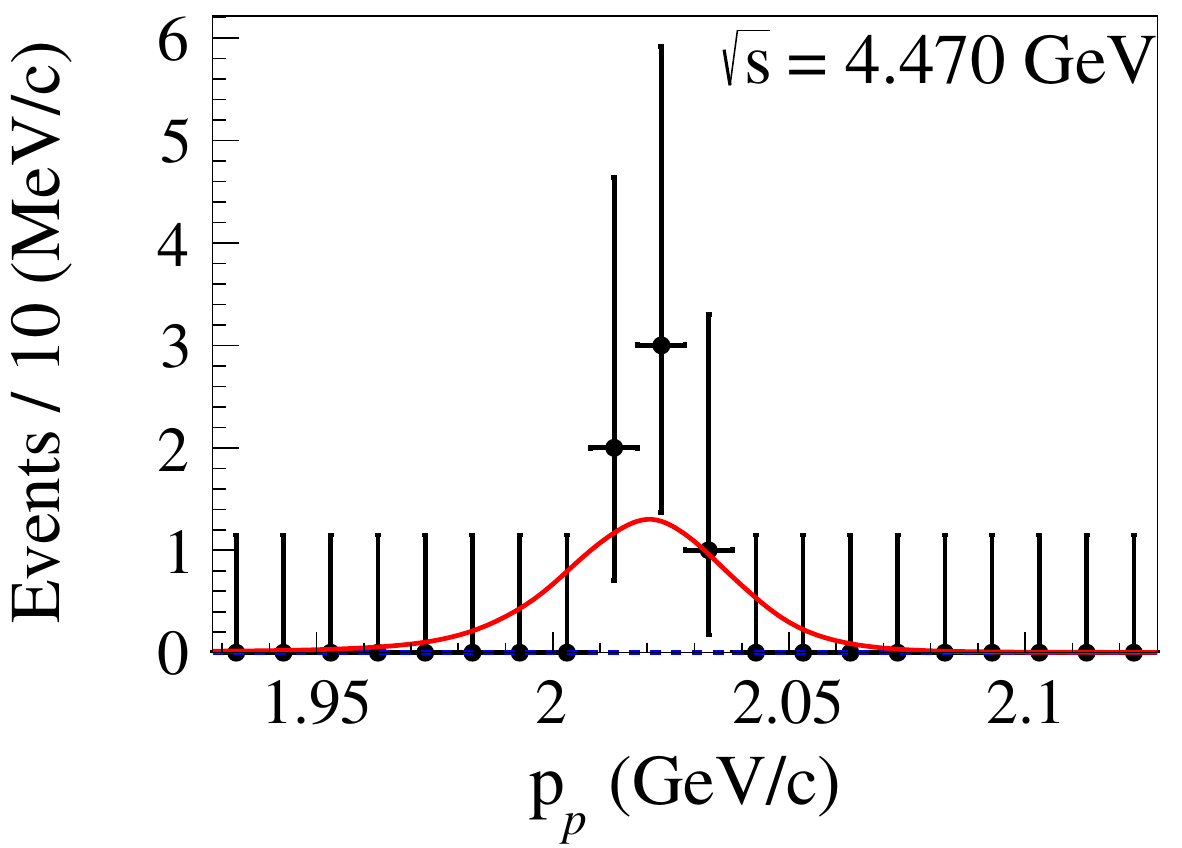}
    \includegraphics[width=.245\linewidth]{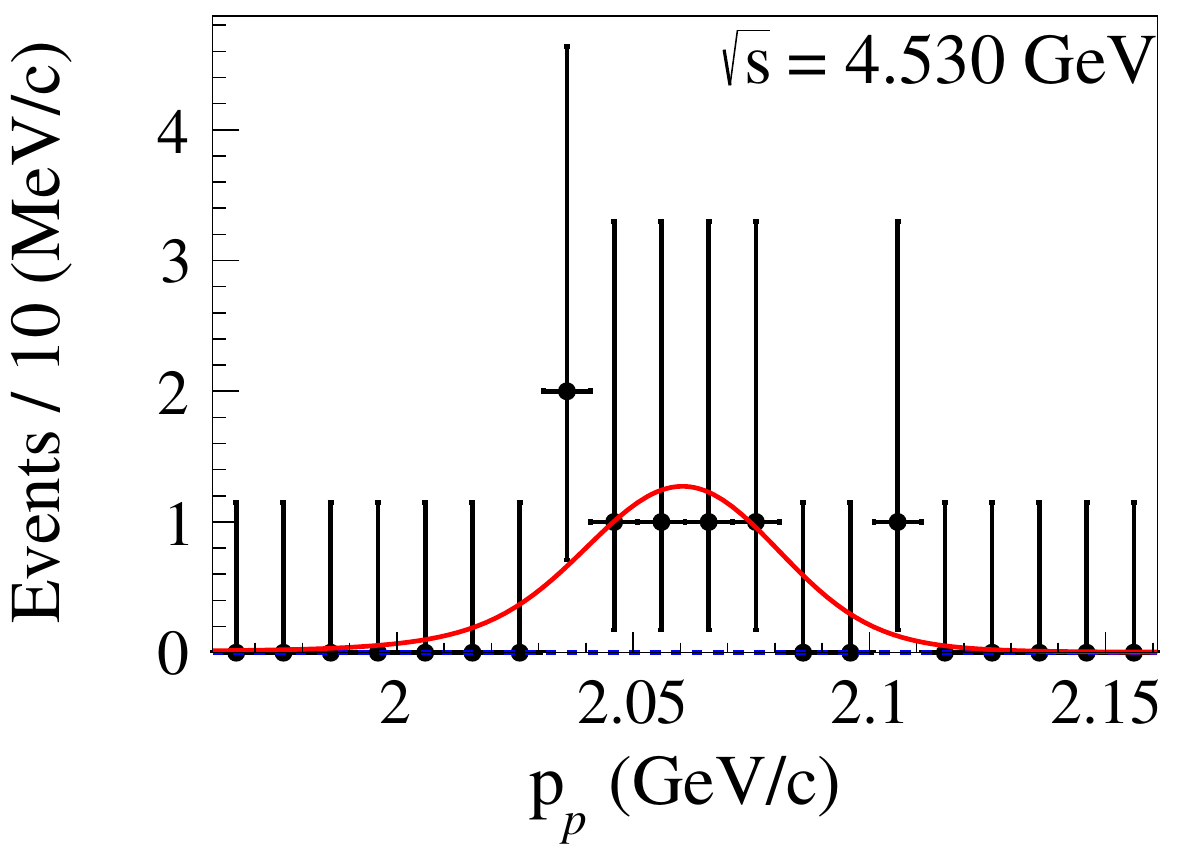}
    \includegraphics[width=.245\linewidth]{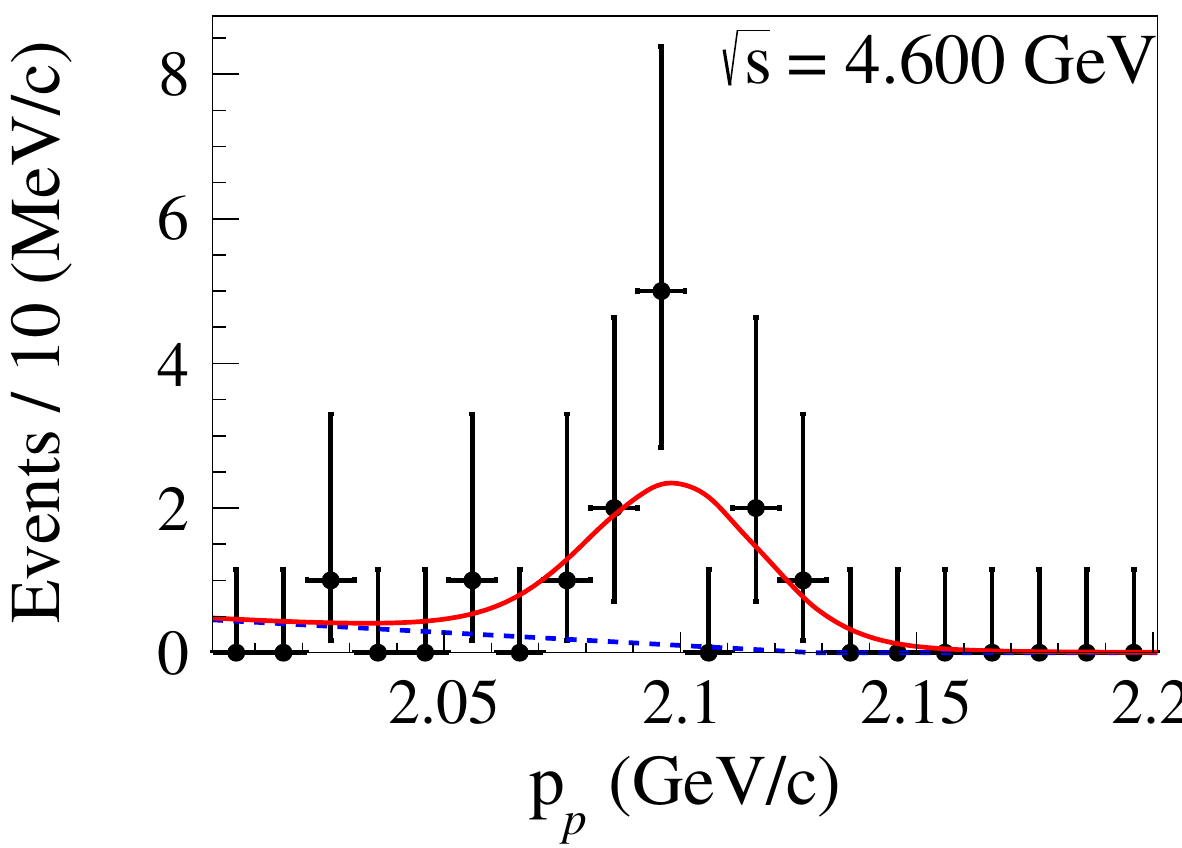}
    \includegraphics[width=.245\linewidth]{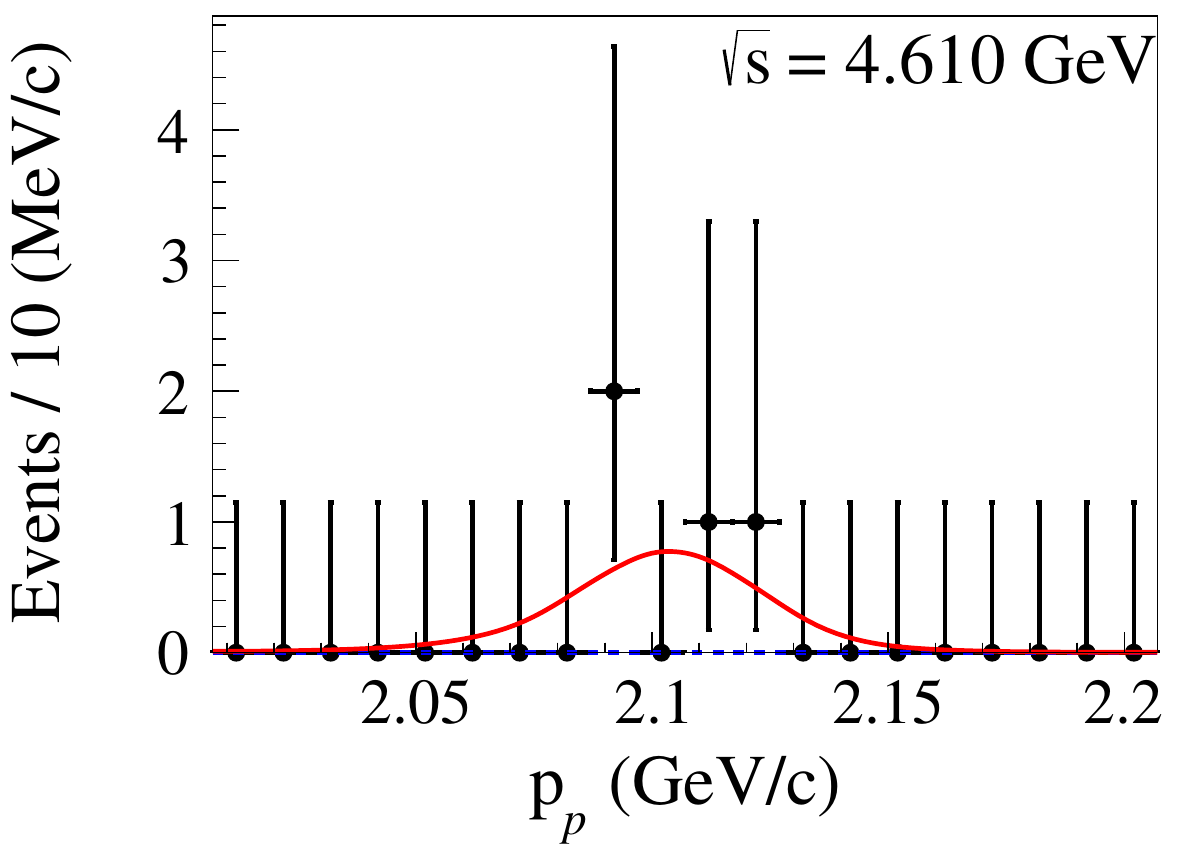}\\
    \includegraphics[width=.245\linewidth]{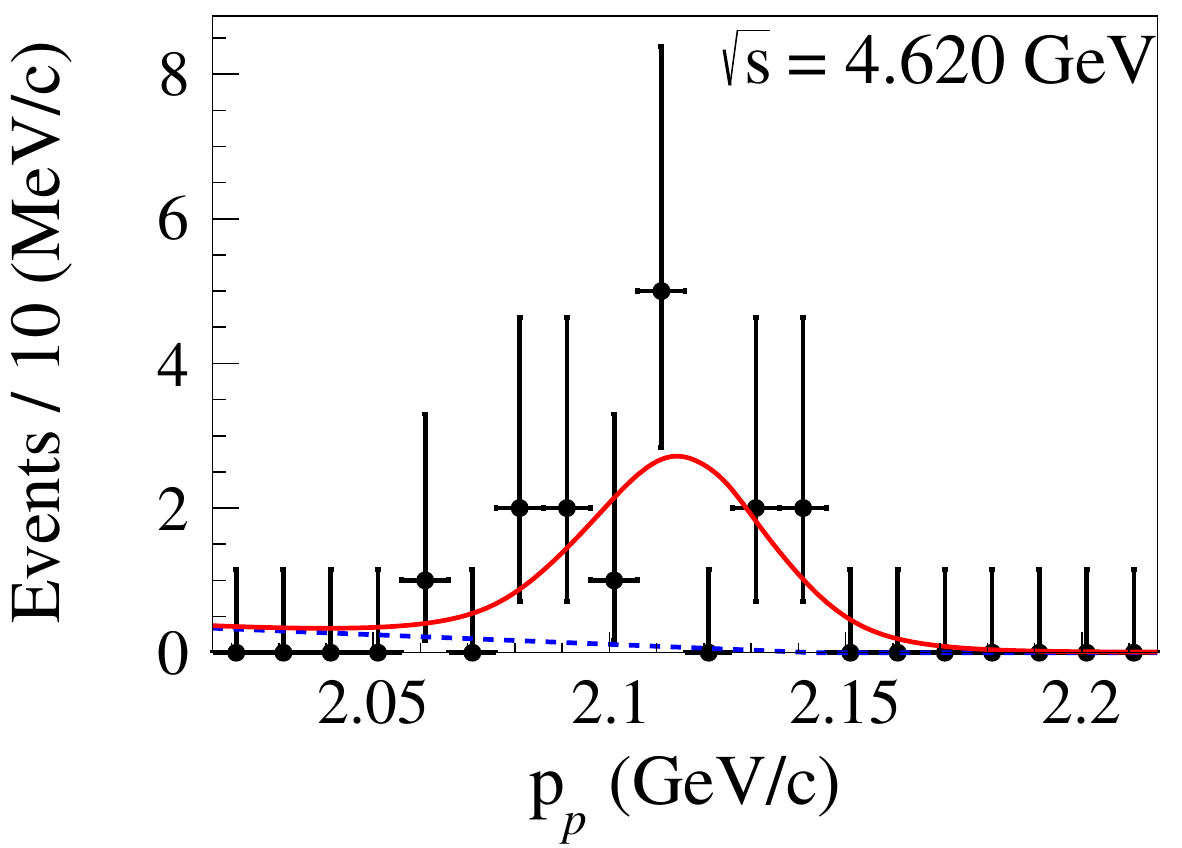}
    \includegraphics[width=.245\linewidth]{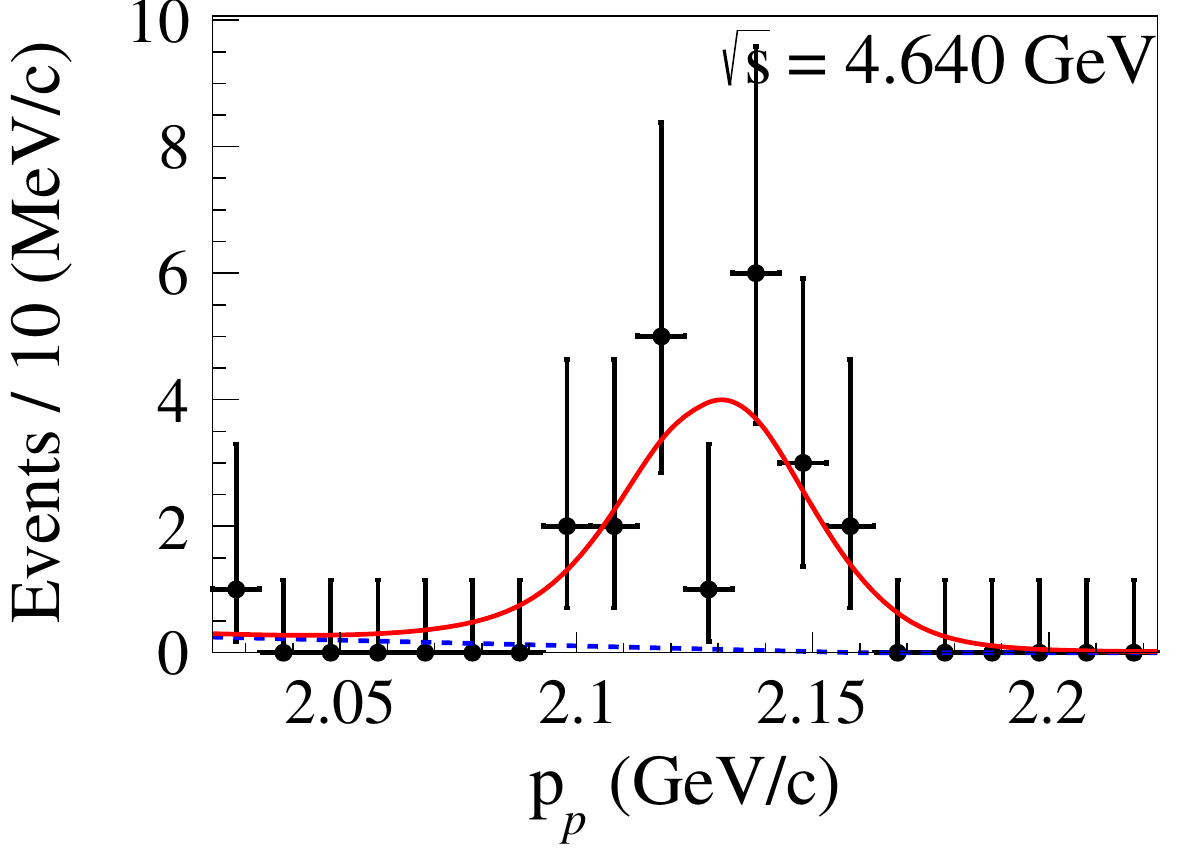}
    \includegraphics[width=.245\linewidth]{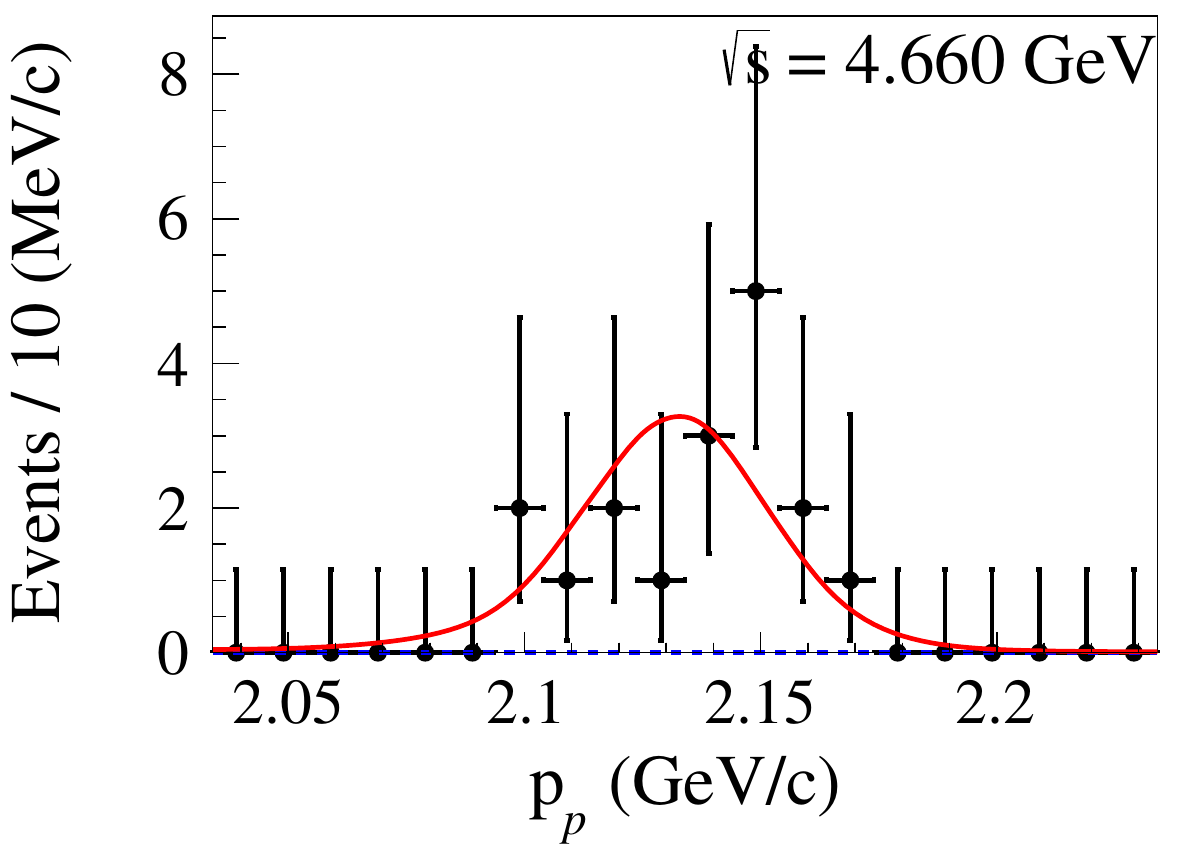}
    \includegraphics[width=.245\linewidth]{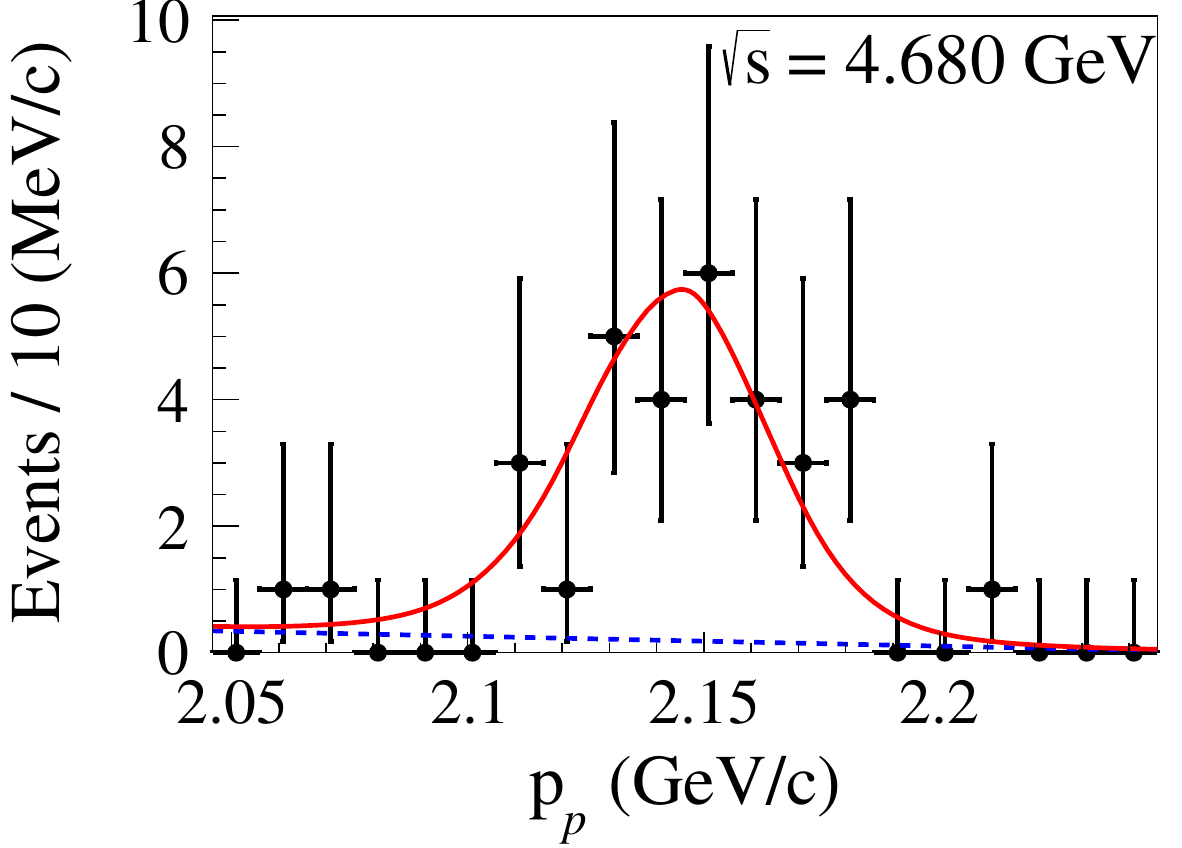}\\
    \includegraphics[width=.245\linewidth]{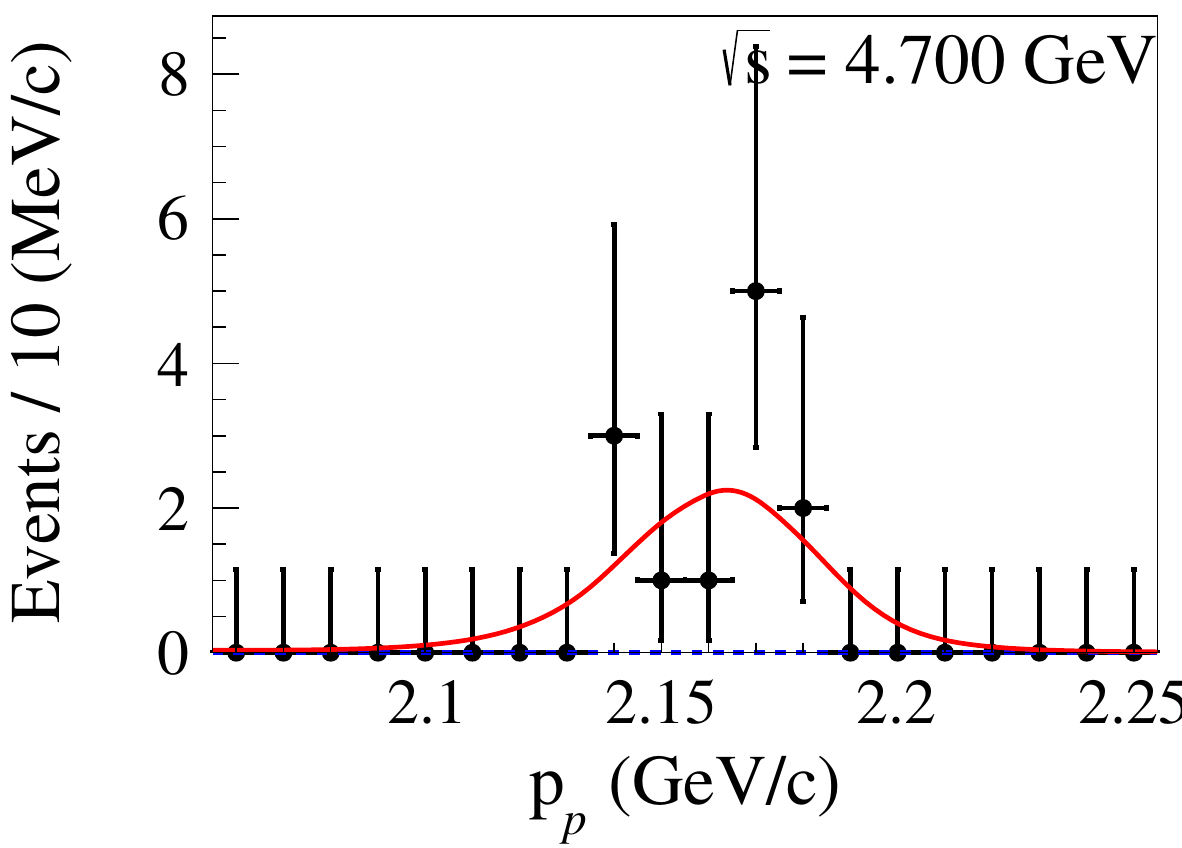}
    \includegraphics[width=.245\linewidth]{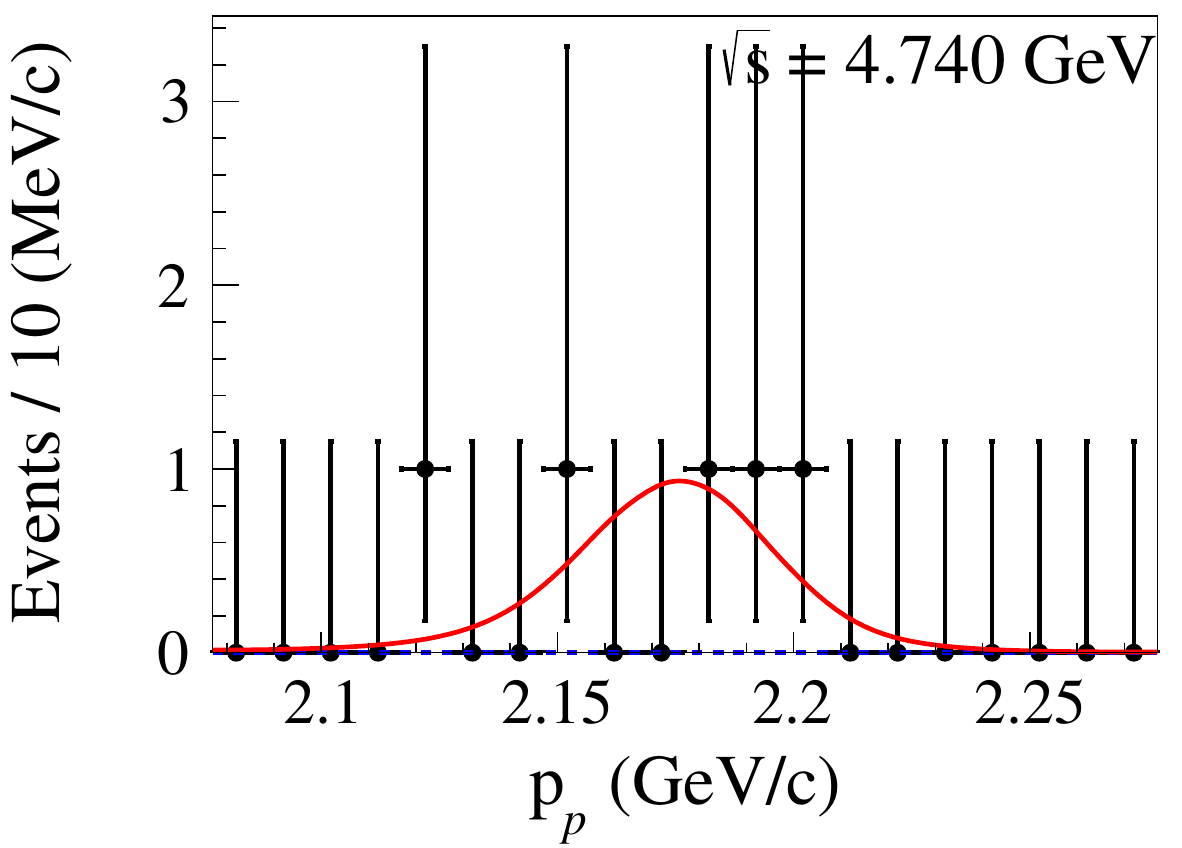}
    \includegraphics[width=.245\linewidth]{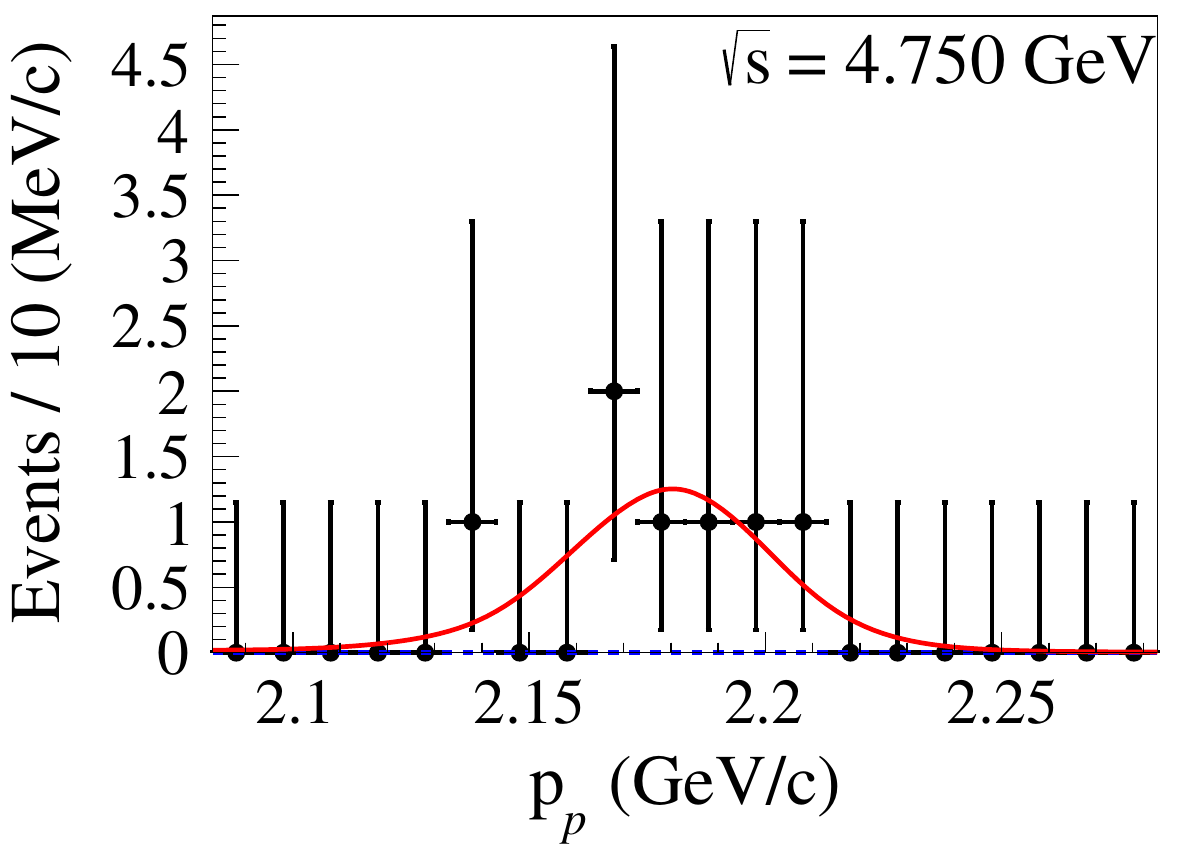}    
    \includegraphics[width=.245\linewidth]{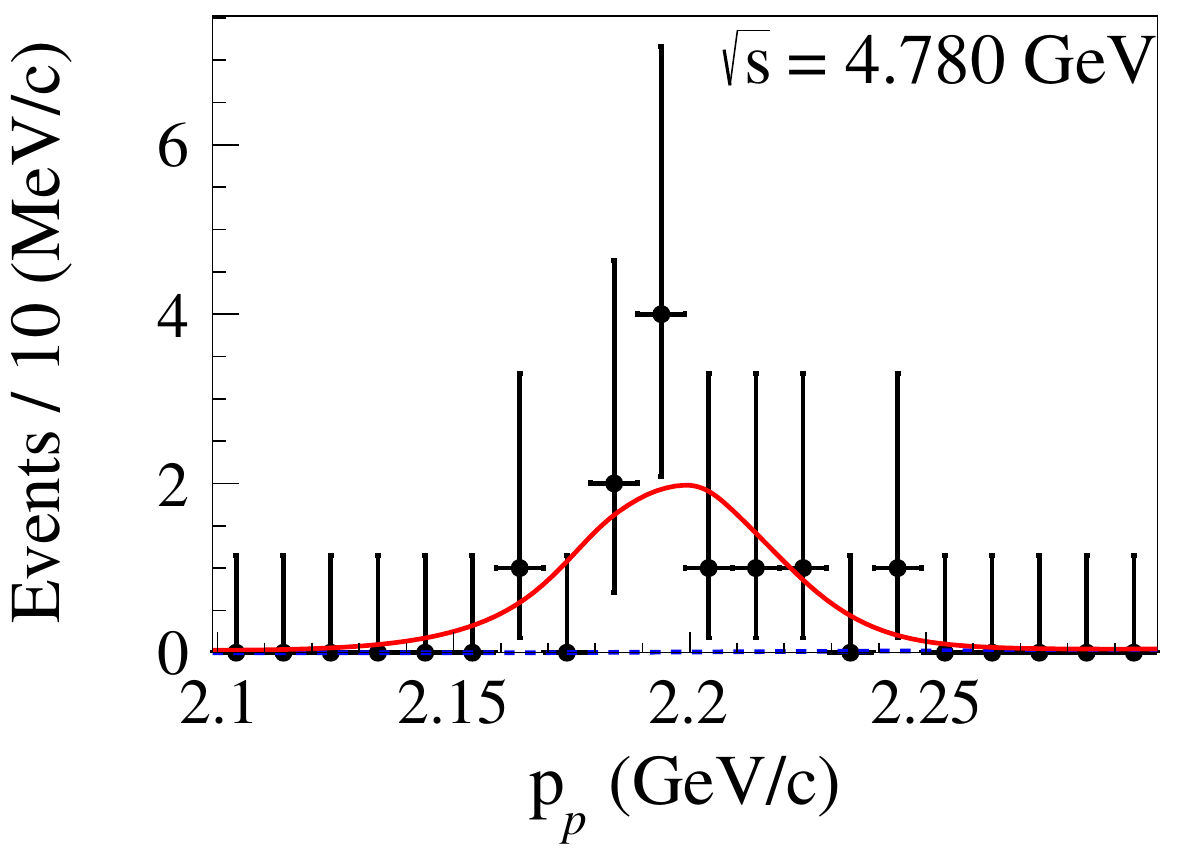}\\
    \includegraphics[width=.245\linewidth]{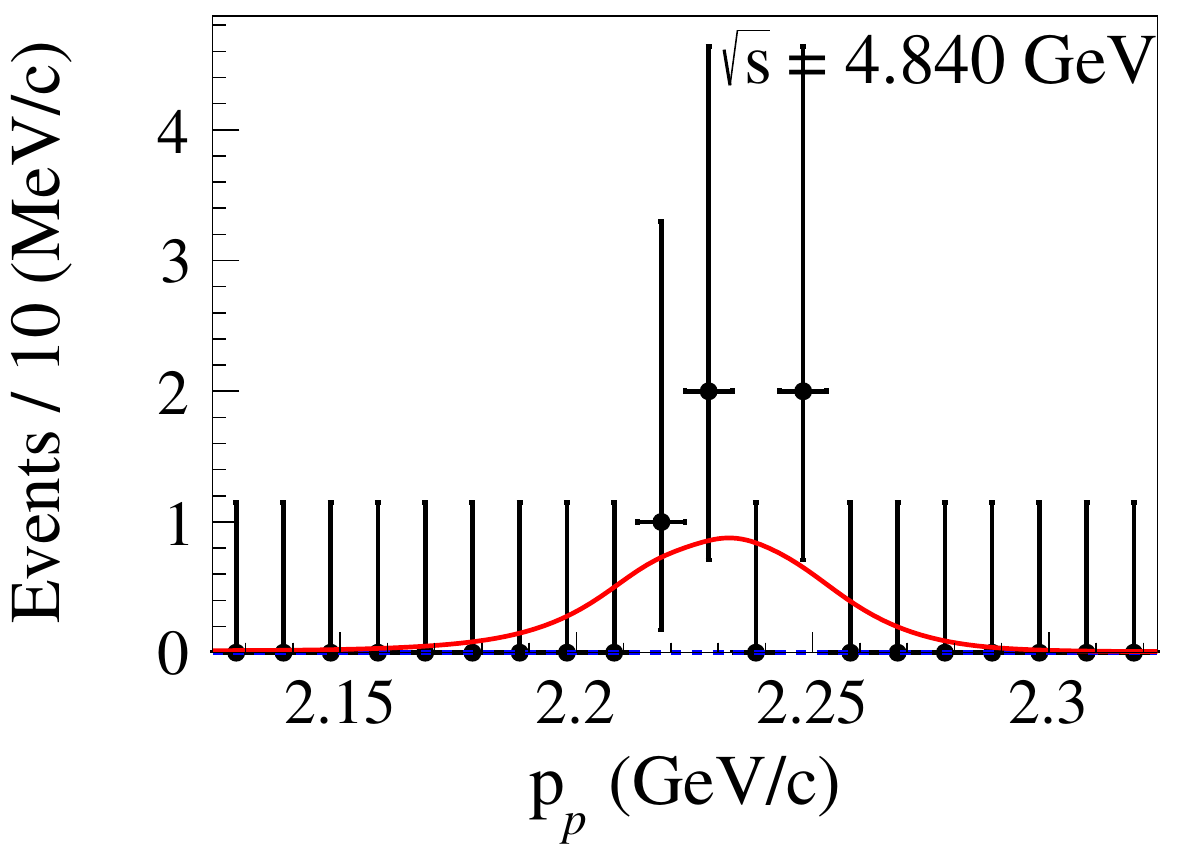}
    \includegraphics[width=.245\linewidth]{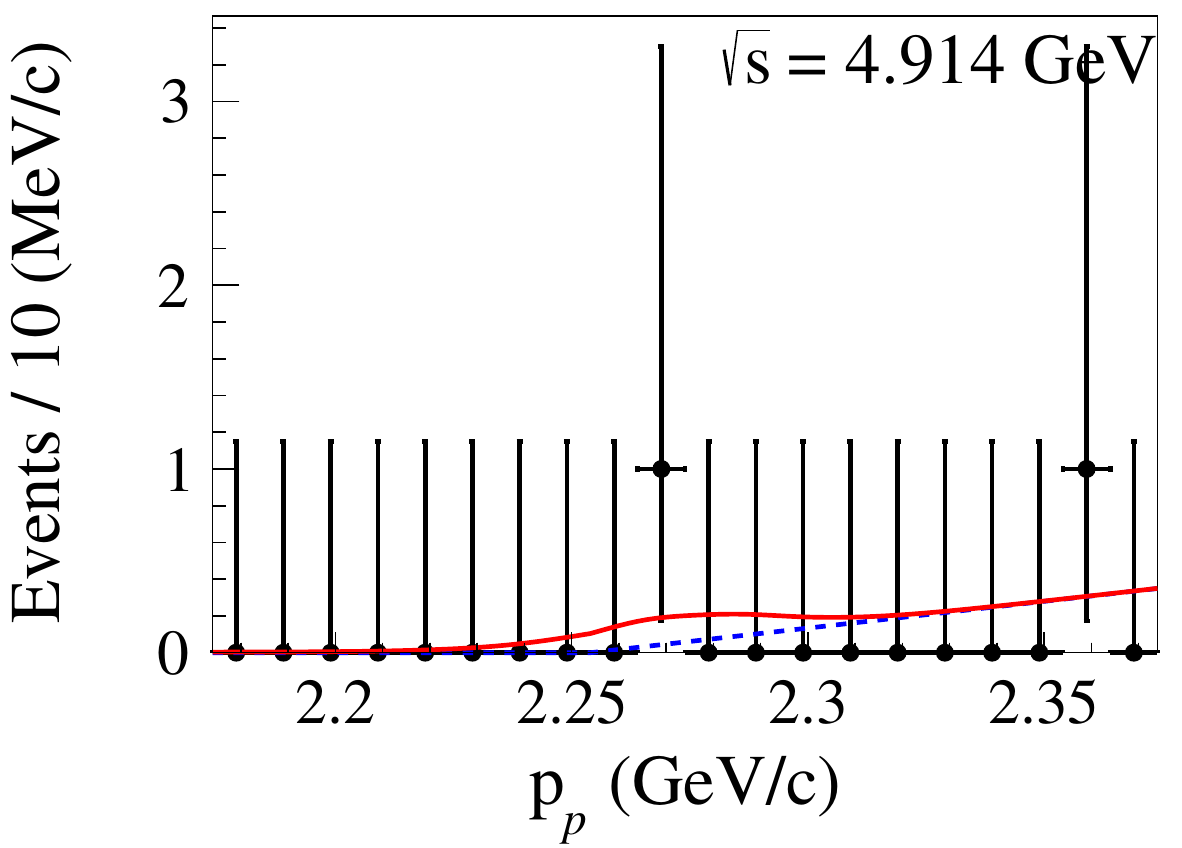}
    \includegraphics[width=.245\linewidth]{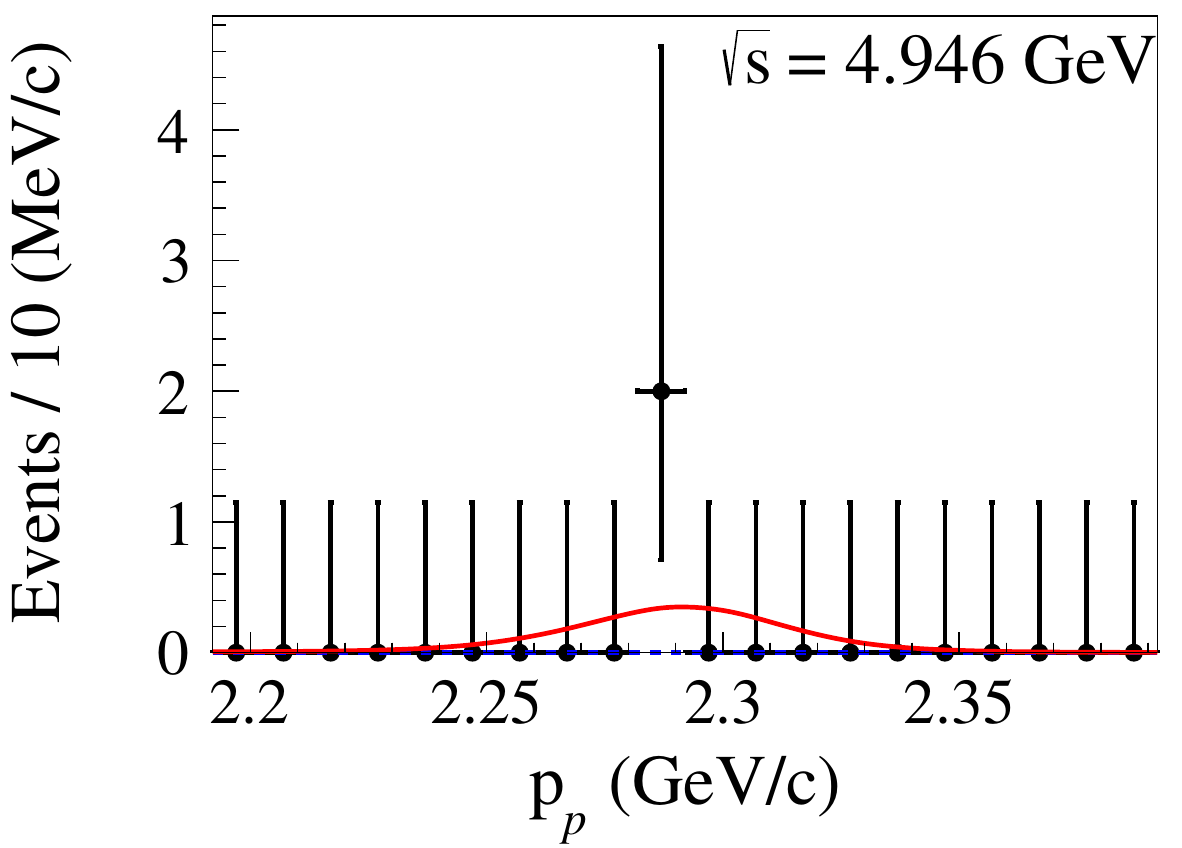}
    \caption{Fits to the proton momentum distributions at $\sqrt{s}=4.390-4.946$ GeV. The dots with error bars are data. The red solid lines represent the total fit, and the blue dashed lines stand for the backgrounds.}
    \label{extract_signal_yield:AD}
\end{figure*}

To extract the signal yields for the $e^+e^-\to p\bar p$ reaction at each energy point, an extended maximum likelihood fit to the proton momentum spectrum is performed as shown in Figs.~\ref{extract_signal_yield} and~\ref{extract_signal_yield:AD}. In the fit, the signal shape for energy points with luminosity $L>300\,\rm pb^{-1}$ is modeled by the simulated MC shape convolved with a Gaussian function, thereby accounting for the mass resolution difference between data and MC simulation. The parameters of the Gaussian function are left free. While for other data samples with luminosity $L<300\,\rm pb^{-1}$ the parameters of the Gaussian function are fixed to the values from the fits to neighboring high-statistics energy points. The background shapes are modeled with a linear function. In cases where the fit of background event yields zero, the background contribution is set to zero. At $\sqrt{s}=3.773$ GeV, an additional peaking background from ISR to the lower lying $\psi(3686)$ resonance is present. This background is modeled using a simulated MC distribution convolved with a Gaussian function. The signal yields at each energy point are summarized in Table~\ref{tab:table1:BCS_num}.  

\subsection{ISR correction and detection efficiency}
The ISR process is a universal effect in which one or more photons are emitted by an electron or a positron before annihilation. The emitted photons carry away part of the energy from the initial electron or positron, thereby reducing the effective $e^+e^-$ c.m.~energy. This effect is considered by an iterative MC weighting method~\cite{Sun:2020ehv}. 
Initially, the initial ISR correction factor $(1+\delta)_1$ and efficiency $\rm \epsilon_1=N^{sel}/N^{gen}$ are obtained from the MC samples generated with a flat cross section lineshape, where $\rm N^{sel}$ is the number of selected signal MC events after applying weight factor, $\rm N^{gen}$ is the total number of the generated signal MC events. Then, the preliminary dressed cross section is obtained by
\begin{equation}
\rm \sigma^{dressed}_{1} =\frac{N^{obs}}{{\cal{L}}\cdot(1 + \delta)_1\cdot\epsilon_1},
\end{equation}
where $\rm N^{obs}$ denotes the signal yields and $\cal{L}$ is the integrated luminosity. In subsequent iterations, the same MC samples with ISR effect are weighted according to the dressed cross section lineshape from the previous iteration,
and the ISR correction factor and detection efficiency are then redetermined. The weighting factor $\rm W_{\it i}$ for the $i^{\rm th}$ event is defined as:
\begin{equation}
\rm
W_{\it i}=\frac{\sigma^{dr}_{\it i}(\it m_{\it j})}{\sigma^{dr}(\it E_{\rm cms})}\times\frac{\sigma^{input}_{\it i}(\it E_{\rm cms})}{\sigma^{input}(\it m_{\it j})},
\end{equation}
where $\rm \sigma^{dr}_{\it i}(\it m_{\it j})$ and $\rm \sigma^{dr}(\it E_{\rm cms})$ are the dressed cross sections at the invariant mass of $\it m_{\it j}$ and $\it E_{\rm cms}$. Then the ISR correction factor for the $ (i+1)^{\rm th}$ iteration $\rm (1+\delta)_{\it i\rm +1}$ can be weighted as
\begin{equation}
\rm 
(1+\delta)_{\it i\rm +1}=(1+\delta)_{1}\times\frac{\sum_{\it i=\rm1}^{N^{total}}W_{\it i}}{N^{total}}.
\end{equation}
The detection efficiency for the $(i+1)^{\rm th}$ iteration can also be estimated in terms of weights by
\begin{equation}
\label{eff_weight}
\rm 
\epsilon_{\it i\rm +1}=\frac{
\sum_{\it i\rm =1}^{N^{selected}}W_{\it i}
}{
\sum_{\it i\rm =1}^{N^{total}}W_{\it i}
}.
\end{equation}
The final results of ISR correction factors and detection efficiency based on the iteration procedure are summarized in Table \ref{tab:table1:BCS_num}.

\subsection{Determination of Born cross section and effective form factor}
The Born cross section for the $e^+e^-\to p\bar p$ reaction is calculated by
\begin{equation}
\rm 
\sigma^{B}(s) =\frac{\sigma^{dressed}(s)}{f_{\rm VP}}=\frac{N^{\rm obs}}{{\cal{L}}\cdot(1 + \delta)\cdot\frac{1}{|1 - \prod|^{2}}\cdot\epsilon},
\end{equation}
where $\rm f_{VP} = 1/|1-\Pi|^2$ is the vacuum polarization (VP) correction factor as calculated in Ref.~\cite{Jegerlehner:2011ti}. 
The dressed cross sections are determined based on the previous iterative procedure until the values converge, which happens when the difference between the values of $(1 + \delta)\times\epsilon$ of the last two iterations is less than 0.1\%~\cite{Sun:2020ehv}. The Born cross section is then calculated according to the extracted ISR factor and detection efficiency.

Under the assumption that the dominant process for the $e^+e^-\to p\bar{p}$ reaction is the one-photon exchange, the proton effective form factor is defined as~\cite{Wang:2022zyc},
\begin{equation}
        G_{\rm eff}(s) =
        \sqrt{\frac{3s\tau\sigma^B}{2\pi\alpha^2C\beta(2\tau+1)}},
\end{equation}
where $s$ is the square of the c.m.~energy, $\alpha$ is the fine structure constant, the variable $\beta = \sqrt{1-1/\tau}$ is the velocity of $p$ in the laboratory frame, $\tau = s/4m_{p}^2$, and $C$ is the Coulomb factor $C$~\cite{Wang:2022zyc, Baldini:2007qg, Arbuzov:2011ff}. The Coulomb factor parameterizes the electromagnetic interaction between the outgoing baryon and anti-baryon. For neutral baryons, $C=1$, for point-like charged fermions, $C=\frac{\pi\alpha}{\beta}\cdot\frac{\sqrt{1-\beta^2}}{1-e^{-\frac{\pi\alpha}{\beta}\sqrt{1-\beta^2}}}$
\cite{Sommerfeld:1931qaf, Tzara:1970ne,Sakharov:1948plh}. 
The numerical results of the measured Born cross section and the proton effective form factor $G_{\rm eff}(s)$ for each energy points are listed in Table~\ref{tab:table1:BCS_num}. 

\begin{table*}
\renewcommand{\arraystretch}{1.15}
\caption{\label{tab:table1:BCS_num}Numerical results for the Born cross sections of $e^+e^-\to p\bar p$ at different energy points. $\sqrt{s}$ is the c.m. energy, ${\cal{L}}$ is the integrated luminosity, $1/|1 - \prod|^{2}$ is the VP correction factor, $\epsilon\cdot(1 + \delta)$ is the ISR correction factor times the detection efficiency, $N_{\rm obs}$ is the signal yield, $\sigma^{B}$ is the Born cross section, and $G_{\rm eff}(s)$ is the effective from factor. The first and second uncertainties for $\sigma^{B}$ and $G_{\rm eff}(s)$ are statistical and systematic, respectively.}
\begin{ruledtabular}
\begin{tabular}{lllcclr}
\multicolumn{1}{c}{$\sqrt{s}$ (GeV)} &\multicolumn{1}{c}{${\cal{L}}$ (pb$^{-1})$} &\multicolumn{1}{c}{$N_{\rm obs}$ }&\multicolumn{1}{c}{$\epsilon(1+\delta)\,(\%)$} &\multicolumn{1}{c}{\,\,$1/|1 - \Pi|^{2}$} &\multicolumn{1}{c}{$\sigma^{B}$ (fb)} &\multicolumn{1}{c}{$G_{\rm eff}(s)$ $\times 10^{-3}$}\\
\hline
$ 3.510 $  & $ 405.4 $   & $ 388.1^{+21.1}_{-20.5}$   &  $ 41.52 $ &  $1.04$ & $2208    ^{+120}      _{-117}      \pm  119 $  & $18.1 	^{+0.5} _{-0.5} \pm 0.5 $ \\  
$ 3.581 $  & $ 85.7 $    & $ 69.6 ^{+9.0}_{-8.5}$     &  $ 49.55 $ &  $1.04$ & $1588    ^{+204}      _{-193}      \pm  86 $   & $15.9 	^{+1.0} _{-0.9} \pm 0.4 $ \\       
$ 3.650 $  & $ 410.0 $     & $ 276.7 ^{+17.9}_{-17.3}$  &  $ 55.00 $ &  $1.02$ & $1204    ^{+78}      _{-75}      \pm  65 $   & $14.9 	^{+0.5} _{-0.5} \pm 0.4 $ \\      
$ 3.768 $  & $ 415.8 $ & $ 117.3 ^{+14.8}_{-13.3}$  &  $ 59.67 $ &  $1.05$ & $449    ^{+56}      _{-51}      \pm  24 $        & $9.3 	^{+0.6} _{-0.5} \pm 0.2 $ \\    
$ 3.773 $  & $ 2916.9 $ & $ 729.8 ^{+28.9}_{-28.3}$  &  $ 63.84 $ &  $1.06$ & $371    ^{+15}      _{-14}      \pm  20 $       & $8.5 	^{+0.2} _{-0.2} \pm 0.2 $ \\    
$ 3.780 $  & $ 410.0 $ & $ 106.5 ^{+12.1}_{-11.4}$  &  $ 51.25 $ &  $1.06$ & $475    ^{+54}      _{-51}      \pm  25 $        & $9.6 	^{+0.5} _{-0.5} \pm 0.2 $ \\    
$ 3.867 $  & $ 108.9 $   & $ 38.0 ^{+6.5}_{-5.8}$     &  $ 52.30 $ &  $1.05$ & $635    ^{+109}      _{-98}      \pm  34 $     & $10.9 	^{+0.9} _{-0.8} \pm 0.3 $ \\     
$ 3.871 $  & $ 110.3 $   & $ 29.0 ^{+5.7}_{-5.1}$     &  $ 52.31 $ &  $1.05$ & $479    ^{+94}      _{-83}      \pm  26 $      & $9.5 	^{+0.9} _{-0.8} \pm 0.2 $ \\    
$ 3.896 $  & $ 52.6 $    & $ 12.5 ^{+4.1}_{-3.4}$     &  $ 52.54 $ &  $1.05$ & $431    ^{+141}      _{-118}      \pm  23 $    & $9.1 	^{+1.5} _{-1.2} \pm 0.2 $ \\      
$ 4.008 $  & $ 482.0 $     & $ 102.0 ^{+10.4}_{-9.8}$   &  $ 52.31 $ &  $1.04$ & $388    ^{+40}      _{-37}      \pm  21 $    & $8.9 	^{+0.5} _{-0.4} \pm 0.2 $ \\    
$ 4.128 $  & $ 401.5 $   & $ 30.6 ^{+6.2}_{-5.5}$     &  $ 54.39 $ &  $1.05$ & $133    ^{+27}      _{-24}      \pm  7 $       & $5.4 	^{+0.5} _{-0.5} \pm 0.1 $ \\    
$ 4.157 $  & $ 408.7 $   & $ 47.2 ^{+7.7}_{-7.0}$     &  $ 54.46 $ &  $1.05$ & $201    ^{+33}      _{-30}      \pm  11 $      & $6.7 	^{+0.5} _{-0.5} \pm 0.2 $ \\    
$ 4.178  $  & $ 3189.0 $    & $ 378.4 ^{+21.1}_{-20.6}$  &  $ 52.42 $ &  $1.05$& $215    ^{+12}      _{-12}      \pm  12 $    & $6.8 	^{+0.2} _{-0.2} \pm 0.2 $ \\    
$ 4.188  $  & $ 526.7 $   & $ 56.0 ^{+8.7}_{-8.1}$     &  $ 52.50 $ &  $1.06$& $192    ^{+30}      _{-28}      \pm  10 $      & $6.5 	^{+0.5} _{-0.5} \pm 0.2 $ \\    
$ 4.199 $  & $ 526.0 $     & $ 50.7 ^{+8.7}_{-8.2}$     &  $ 53.58 $ &  $1.06$ & $170    ^{+29}      _{-28}      \pm  9 $     & $6.2 	^{+0.5} _{-0.5} \pm 0.2 $ \\    
$ 4.209  $  & $ 517.1 $   & $ 69.9 ^{+8.9}_{-8.3}$     &  $ 53.49 $ &  $1.06$& $239    ^{+31}      _{-28}      \pm  13 $      & $7.4 	^{+0.5} _{-0.4} \pm 0.2 $ \\    
$ 4.219  $  & $ 514.6 $   & $ 60.3 ^{+8.6}_{-7.9}$     &  $ 53.90 $ &  $1.06$& $206    ^{+29}      _{-27}      \pm  11 $      & $6.9 	^{+0.5} _{-0.4} \pm 0.2 $ \\     
$ 4.226 $  & $ 1100.9 $ & $ 139.6 ^{+15.4}_{-11.8}$  &  $ 53.46 $ &  $1.06$ & $225    ^{+25}      _{-19}      \pm  12 $       & $7.2 	^{+0.4} _{-0.3} \pm 0.2 $ \\      
$ 4.236  $  & $ 530.3 $   & $ 57.1 ^{+8.4}_{-7.2}$     &  $ 54.13 $ &  $1.06$& $188    ^{+28}      _{-24}      \pm  10 $      & $6.6 	^{+0.5} _{-0.4} \pm 0.2 $ \\     
$ 4.244  $  & $ 538.1 $   & $ 51.5 ^{+7.9}_{-7.3}$     &  $ 53.89 $ &  $1.06$& $168    ^{+26}      _{-24}      \pm  9 $       & $6.3 	^{+0.5} _{-0.4} \pm 0.2 $ \\      
$ 4.258 $  & $ 828.4 $   & $ 78.8 ^{+10.3}_{-9.6}$    &  $ 53.77 $ &  $1.05$ & $168    ^{+22}      _{-21}      \pm  9 $       & $6.3 	^{+0.4} _{-0.4} \pm 0.2 $ \\      
$ 4.267  $  & $ 531.1 $   & $ 39.0 ^{+6.8}_{-6.1}$     &  $ 55.42 $ &  $1.05$& $126    ^{+22}      _{-20}      \pm  7 $       & $5.5 	^{+0.5} _{-0.4} \pm 0.1 $ \\      
$ 4.278  $  & $ 175.7 $   & $ 7.2 ^{+3.3}_{-2.7}$      &  $ 54.70 $ &  $1.05$& $71    ^{+33}      _{-27}      \pm  4 $        & $4.2 	^{+1.0} _{-0.8} \pm 0.1 $ \\     
$ 4.288 $  & $ 502.4 $   & $ 33.4 ^{+6.4}_{-5.7}$     &  $ 56.14 $ &  $1.05$ & $113    ^{+22}      _{-19}      \pm  6 $       & $5.3 	^{+0.5} _{-0.5} \pm 0.1 $ \\     
$4.312 $  & $ 501.2 $   & $ 23.5 ^{+6.0}_{-5.3}$     &  $ 55.54 $ &  $1.05$  & $80    ^{+20}      _{-18}      \pm  4 $        & $4.5 	^{+0.6} _{-0.5} \pm 0.1 $ \\     
$4.337 $  & $ 505.0 $     & $ 35.3 ^{+6.6}_{-6.0}$     &  $ 56.24 $ &  $1.05$  & $118    ^{+22}      _{-20}      \pm  6 $     & $5.5 	^{+0.5} _{-0.5} \pm 0.1 $ \\     
$4.358 $  & $ 544.0 $     & $ 39.6 ^{+7.2}_{-6.5}$     &  $ 53.74 $ &  $1.05$  & $129    ^{+23}      _{-21}      \pm  7 $     & $5.7 	^{+0.5} _{-0.5} \pm 0.1 $ \\      
$4.377 $  & $ 522.7 $   & $ 30.7 ^{+7.0}_{-6.4}$     &  $ 55.73 $ &  $1.05$  & $100    ^{+23}      _{-21}      \pm  5 $       & $5.1 	^{+0.6} _{-0.5} \pm 0.1 $ \\     
$4.387 $  & $ 55.6 $   & $ 7.4 ^{+3.4}_{-2.6}$      &  $ 53.67 $ &  $1.05$  & $238    ^{+107}      _{-84}      \pm  13 $      & $7.8 	^{+1.8} _{-1.4} \pm 0.2 $ \\      
$4.396 $  & $ 507.8 $   & $ 9.8 ^{+4.0}_{-3.4}$      &  $ 54.63 $ &  $1.05$  & $34    ^{+14}      _{-11}      \pm  2 $        & $2.9 	^{+0.6} _{-0.5} \pm 0.1 $ \\     
$ 4.416 $  & $ 1090.7 $  & $ 75.2 ^{+9.9}_{-9.3}$     &  $ 56.92 $ &  $1.05$ & $115    ^{+15}      _{-14}      \pm  6 $       & $5.6 	^{+0.4} _{-0.3} \pm 0.1 $ \\  
$ 4.436 $  & $ 569.9 $   & $ 23.4 ^{+6.8}_{-2.8}$     &  $ 58.51 $ &  $1.05$ & $67    ^{+19}      _{-8}      \pm  4 $         & $4.1 	^{+0.9} _{-0.5} \pm 0.1 $ \\     
$ 4.461 $  & $ 111.1 $  & $ 8.2 ^{+3.7}_{-3.0}$      &  $ 56.24 $ &  $1.05$ & $124    ^{+56}      _{-46}      \pm  7 $        & $5.8 	^{+1.3} _{-1.1} \pm 0.1 $ \\    
$ 4.527 $  & $ 112.1 $  & $ 5.3 ^{+3.6}_{-2.9}$      &  $ 57.57 $ &  $1.05$ & $77    ^{+52}      _{-43}      \pm  4 $         & $4.7 	^{+1.6} _{-1.3} \pm 0.1 $ \\     
$ 4.600 $  & $ 586.9 $   & $ 7.8 ^{+5.8}_{-5.2}$      &  $ 57.59 $ &  $1.05$ & $22    ^{+16}      _{-15}      \pm  1 $        & $2.5 	^{+0.9} _{-0.8} \pm 0.1 $ \\    
$ 4.612 $  & $ 103.8 $  & $ 5.1 ^{+3.2}_{-2.5}$      &  $ 53.65 $ &  $1.05$ & $102    ^{+54}      _{-43}      \pm  6 $        & $4.3 	^{+1.9} _{-1.6} \pm 0.1 $ \\    
$ 4.628   $  & $ 521.53 $  & $ 27.6 ^{+7.9}_{-7.5}$     & $ 53.49 $ &  $1.05$& $94    ^{+27}      _{-26}      \pm  5 $        & $4.8 	^{+0.8} _{-0.7} \pm 0.1 $ \\    
$ 4.641 $  & $ 552.4 $   & $ 23.1 ^{+7.1}_{-6.6}$     &  $ 53.28 $ &  $1.05$ & $74    ^{+23}      _{-21}      \pm  4 $        & $4.2 	^{+0.8} _{-0.7} \pm 0.1 $ \\    
$ 4.661 $  & $ 529.6 $   & $ 16.3 ^{+7.4}_{-6.9}$     &  $ 54.85 $ &  $1.05$ & $53    ^{+24}      _{-23}      \pm  3 $        & $3.8 	^{+0.9} _{-0.9} \pm 0.1 $ \\    
$ 4.682 $  & $ 1167.4 $ & $ 26.0 ^{+11.9}_{-11.2}$   &  $ 54.68 $ &  $1.05$ & $39    ^{+18}      _{-17}      \pm  2 $         & $1.8 	^{+1.4} _{-1.4} \pm 0.0 $ \\    
$ 4.699 $  & $ 535.5 $  & $ 10.8 ^{+7.6}_{-7.2}$     &  $ 54.27 $ &  $1.05$ & $35    ^{+25}      _{-24}      \pm  2 $         & $2.2 	^{+1.5} _{-1.4} \pm 0.1 $ \\    
$ 4.740 $  & $ 163.9 $  & $ 3.7 ^{+4.3}_{-3.7}$      &  $ 62.19 $ &  $1.05$ & $34    ^{+40}      _{-34}      \pm  2 $         & $1.5 	^{+4.7} _{-3.9} \pm 0.0 $ \\  
$ 4.750 $  & $ 366.6 $  & $ 7.0 ^{+3.0}_{-2.3}$      &  $ 62.86 $ &  $1.05$ & $29    ^{+12}      _{-10}      \pm  1 $         & $1.3 	^{+1.4} _{-1.1} \pm 0.0 $ \\  
$ 4.781 $  & $ 511.5 $  & $ 11.3 ^{+8.8}_{-8.2}$     &  $ 64.28 $ &  $1.06$ & $33    ^{+25}      _{-24}      \pm  2 $         & $2.5 	^{+1.7} _{-1.6} \pm 0.1 $ \\    
$ 4.843 $  & $ 525.2 $  & $ 5.0 ^{+2.6}_{-1.9}$     &  $ 64.57 $ &  $1.06$ & $14    ^{+7}      _{-5}   \pm  1 $               & $2.0    ^{+0.5}_{-0.4}  \pm 0.1$\\    
$ 4.918 $  & $ 207.8 $  & $ 0.7 ^{+1.4}_{-0.7}$     &  $ 65.25 $ &  $1.06$ & $5    ^{+10}      _{-5}   \pm  0 $               & $1.2    ^{+1.2}_{-0.6}  \pm 0.1$\\    
$ 4.951 $  & $ 159.3 $  & $ 2.0 ^{+1.8}_{-1.1}$     &  $ 65.94 $ &  $1.06$ & $18    ^{+16}      _{-10}   \pm  1 $             & $2.3    ^{+1.0}_{-0.7}  \pm 0.1$\\    
\end{tabular}
\end{ruledtabular}
\end{table*}

\section{Measurement of EMFFs and angular distribution}
\begin{figure}[h]
    \centering
    \includegraphics[width=0.99\linewidth]{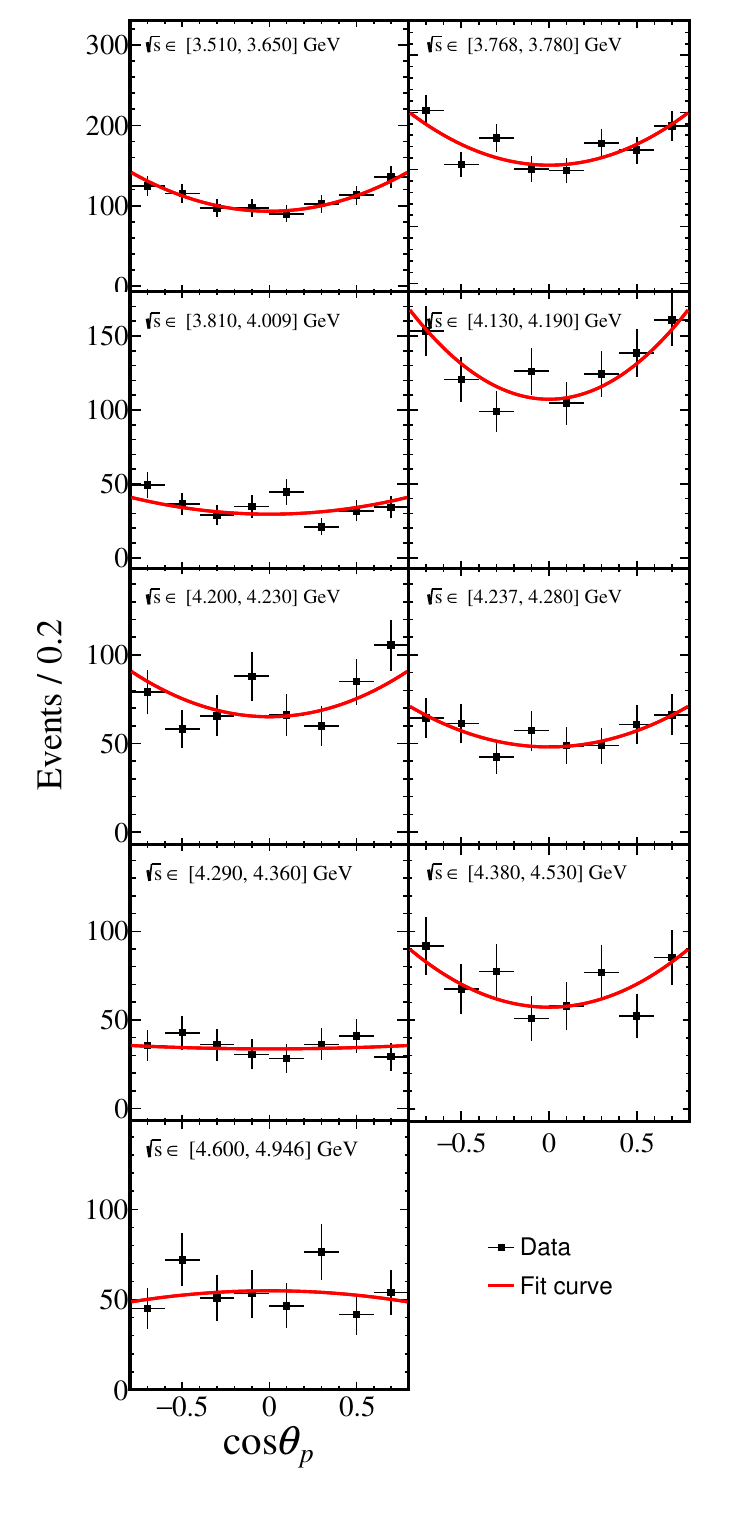}
    \caption{Fits to the proton angular distributions in different energy ranges.}
    \label{fit_dn:AD}
\end{figure}

The moduli of the EMFFs $|G_E|$ and $|G_M|$, or equivalently their ratio $R=|G_E/G_M|$ and $|G_M|$ in the $e^+e^-\to p\bar p$ reaction, can be determined from a fit to the proton angular distribution by combining data samples from several energy points to increase statistics. The combination of data samples strikes a balance between statistical precision and energy resolution, ensuring sufficient signal yield ($N_{obs}>300$) while minimizing variations in efficiency and production dynamics across the interval. The central value of each combined energy range is used to represent that interval.
The proton angular distribution ($\theta_p$) is fitted using the following expression:
\begin{equation}
\begin{split}
      \mathcal{W} = \mathcal{W}(\xi,\boldsymbol{\Omega}) 
      &= \frac{{\cal L}\epsilon\pi\alpha^2\beta C}{2s}|G_M|^2\Bigg [(1+{\rm cos}^2\theta)\\
       &+\frac{1}{\tau}R^2(1-{\rm cos}^2\theta)\Bigg ],
\end{split}
\label{angular:dis}
\end{equation}
which is described by scalar $\xi = \theta$ and the vector of decay parameters $\boldsymbol{\Omega}$ = $(R, G_{M})$. The angular distribution parameter is $\eta=(\tau - R^{2})/(\tau + R^{2})$~\cite{Pacetti:2014jai}.
To determine $\boldsymbol{\Omega}$, an unbinned maximum likelihood function $\mathcal{L}$ is implemented and constructed from the probability density function, ${\cal{P}}({\xi}_i,\boldsymbol{\Omega})$, for the event $i$ characterized by the measured angles ${\xi}_i$
\begin{align}
    \label{L}
    \mathcal{L} =\prod_{i=1}^{N}\mathcal{P}(\xi_{i},\mathbf{\Omega})
    =\prod_{i=1}^{N}\frac{\mathcal{W}(\xi_{i},\mathbf{\Omega})\epsilon (\xi_{i})}{\cal{N}(\boldsymbol{\Omega})},
\end{align}
where $\cal{N}$ is the number of data events after applying all selection criteria, ${\cal{W}}({{\xi}}_i, {\boldsymbol{\Omega}})$ is given by Eq.~(\ref{angular:dis}), $\epsilon({\xi}_i)$ is the detection efficiency, and $\cal{N}(\boldsymbol{\Omega}) = \int {\cal{W}}({{\xi}}_i, {\boldsymbol{\Omega}})\epsilon({\xi}_i)d{\xi}$ is the normalization factor. The negative value of the logarithm of $\mathcal{L}$ is minimized by using the MINUIT package from the CERN library~\cite{James:1975dr} and is defined as
\begin{align}
  \label{S}
\mathit{S} = -\mathrm{ln}\mathcal{L} =-  \sum_{i=1}^{N}\mathrm{ln}\frac{\mathcal{W}(\xi_{i},\mathbf{\Omega})}{\cal{N}(\boldsymbol{\Omega})}.
\end{align}
No background term is included in the fit function, as background contamination in
the final sample is negligible. However, background is considered as a source of systematic uncertainty.
Figure~\ref{fit_dn:AD} shows the fitted angular distributions for several sub-samples across different c.m.~energy ranges.
The numerical results of the ratio $R=|G_E/G_M|$, the modulus of the magnetic form factor $|G_M|$ and $\eta$ in the fit are summarized in Table~\ref{tab:fiteta}. At higher energies, the $1/\tau$ suppression of $G_E$ reduces the sensitivity to $R$, making $|G_E|$ and $|G_M|$ more difficult to disentangle and resulting in larger statistical uncertainties. 
\begin{table*}[!htbp]
\renewcommand{\arraystretch}{1.15}
  \caption{Numerical results of the ratio $R$, modulus of the magnetic FF $|G_M|$ and the angular asymmetry factor and variation of $(q^{2})^2G_M(q^2)/\mu_{p}$ at different energy ranges with statistical uncertainties only.}
  \begin{ruledtabular}
            \begin{tabular}{c c c c c} 
            $\sqrt{s}$ (GeV) & $ |G_E/G_M|$&$|G_M|\,(\times 10^{-2})$&$\eta$ &$(q^{2})^2G_M(q^2)/\mu_{p}\,({\rm GeV^{4}})$	 \\
            \hline
            3.510-3.650      &{$0.58 \pm 0.54 \pm 0.03$} & $  1.47\pm0.09 \pm 0.05$&{0.83 $\pm 0.29 \pm 0.09$} & $ 0.87 \pm 0.05 \pm 0.03$	  \\
            3.768-3.780      &{$0.89 \pm 0.32 \pm 0.06$} & $  0.86\pm0.04 \pm 0.03$&{0.66 $\pm 0.19 \pm 0.13$} & $ 0.62 \pm 0.03 \pm 0.02$   \\
            3.867-4.009      &{$1.03 \pm 0.91 \pm 0.06$} & $  0.89\pm0.12 \pm 0.03$&{0.61 $\pm 0.56 \pm 0.13$} & $ 0.74 \pm 0.10 \pm 0.03$   \\
            4.130-4.190      &{$0.52 \pm 0.87 \pm 0.03$} & $  0.82\pm0.05 \pm 0.03$&{0.89 $\pm 0.33 \pm 0.14$} & $ 0.87 \pm 0.06 \pm 0.03$   \\
            4.200-4.230      &{$1.08 \pm 0.75 \pm 0.06$} & $  0.78\pm0.08 \pm 0.03$&{0.63 $\pm 0.43 \pm 0.14$} & $ 0.88 \pm 0.08 \pm 0.03$   \\
            4.237-4.280      &{$0.84 \pm 0.98 \pm 0.05$} & $  0.71\pm0.07 \pm 0.03$&{0.76 $\pm 0.49 \pm 0.15$} & $ 0.83 \pm 0.09 \pm 0.04$   \\
            4.290-4.360      &{$2.00 \pm 0.98 \pm 0.11$} & $  0.53\pm0.10 \pm 0.02$&{0.14 $\pm 0.48 \pm 0.14$} & $ 0.67 \pm 0.12 \pm 0.03$	  \\
            4.380-4.530      &{$0.52 \pm 2.62 \pm 0.05$} & $  0.78\pm0.09 \pm 0.03$&{0.91 $\pm 0.88 \pm 0.27$} & $ 1.10 \pm 0.12 \pm 0.04$	  \\
            4.600-4.946      &{$2.99 \pm 1.29 \pm 0.15$} & $  0.42\pm0.09 \pm 0.02$&{-0.17 $\pm 0.42 \pm 0.16$}& $ 0.73 \pm 0.15 \pm 0.03$ 	  \\ 
   \end{tabular}
  \label{tab:fiteta}
  \end{ruledtabular}
\end{table*}
\section{Systematic uncertainty on cross section measurement}
Systematic uncertainties in the measurement of the cross section stem from the integrated luminosity, proton tracking and PID, the line-shape structure, the requirements on the open angle $\rm \theta_{opening}$ and the $E/p$ ratio, and the momentum distribution fit procedure. The uncertainty arising from vacuum polarization is negligible. Among these, the systematic uncertainties associated with the integrated luminosity and proton tracking and PID are considered correlated, while the other systematic uncertainties are assumed to be uncorrelated.

\subsection{Luminosity}
The integrated luminosities at all energy points are determined by analyzing Bhabha scattering events, with uncertainties of 1.0\% below 4.0 GeV, 0.7\% from 4.0 to 4.6 GeV and 0.6\% above 4.6 GeV ~\cite{BESIII:2022dxl,BESIII:2022ulv}.
 
\subsection{Proton tracking and PID}
The uncertainties due to tracking and PID for the selection of proton and antiproton are evaluated by using the control sample of $e^+e^-\to p\bar{p}\pi^+\pi^-$. Here, the systematic uncertainty in tracking for both proton and antiproton is estimated to be 2\% for each at $\sqrt{s}=4.230, 4.260, 4.360, 4.420$ and \SI{4.600}{GeV}. While the systematic uncertainty in PID is assigned as $1\%$ for each. The total systematic uncertainties for the tracking and PID are calculated by adding up the uncertainties of proton and antiproton, with the values of $4.0\%$ and $2.0\%$, respectively. 

\subsection{Input line shape}
The ISR factor and the detection efficiency are determined based on the line shape of the cross section. The resulting uncertainty stems from the statistical uncertainty of the measured cross section and is estimated by varying the central value of the measured cross section within its $\pm1\sigma$ statistical uncertainty . Subsequently, the $(1 + \delta)\epsilon$ values for each energy point are recalculated. This process is repeated 3000 times, and a Gaussian function is employed to fit the distribution of $(1 + \delta)\epsilon$ based on the 3000 repetitions. The standard deviation of the Gaussian function is regarded as the systematic uncertainty due to the input line shape. This uncertainty is found to be negligible. 

\subsection{Fit method}
The uncertainties in the fit of the momentum distribution stem from the signal and background shapes. The former is examined by varying the nominal fit strategy from the signal MC shape convolved with a Gaussian function to the signal MC shape only, and the resulting difference is found to be negligible. Regarding the latter one, the uncertainty is assessed based on several energy points with larger statistics at $\sqrt{s} =$ 3.510, 3.773, 4.178, 4.628, 4.641 and 4.661 GeV, aiming to avoid the influence of statistical fluctuations. The resulting difference by alternately changing the nominal fit function from a first-order to a second-order polynomial is found to have an uncertainty of $1.6\%$. 

\subsection{Requirements on $\rm \theta_{opening}$ and $E/p$ ratio}
Two Barlow tests are performed to assess whether the uncertainties due to the requirements on $\rm \theta_{opening}$ and $E/p$ ratio are statistically significant or consistent with statistical fluctuations. The Barlow test for the requirements on $\rm \theta_{opening}$ is examined by alternatively varying the $\rm \theta_{opening}$ requirements from the nominal one to the values: 3.090, 3.095, 3.100, 3.105, 3.110 rad. And the Barlow test for the requirements on $E/p$ ratio is evaluated by alternatively changing the requirement on the ratio itself from the nominal one to the values: 0.40, 0.45, 0.50, 0.55 and 0.60 with the control sample of $e^+e^-\to p\bar p$ at $\sqrt{s} =$ \SI{3.773}{GeV}.
The tests confirm that the observed variations are consistent with statistical fluctuations; therefore, no associated systematic uncertainty is assigned.

\subsection{Total systematic uncertainty}
The systematic uncertainties on the measured cross section for the reaction \( e^+e^-\to p\bar p \) are evaluated under the assumption that all sources are independent. A summary of the individual systematic uncertainties and their respective values is provided in Table~\ref{tab:sys}.

\begin{table}[h]
    \centering
        \caption{
        Systematic uncertainties and their sources for each energy point on the Born cross section measurement.}
    \begin{tabular}{lc}\hline\hline
        Source & \% \\\hline
        Luminosity&1.0\\
        Tracking & 4.0\\
        PID &2.0\\
        Input line shape&Negligible\\
        Fit method  &1.6 \\
        Total&4.9
         \\\hline\hline
    \end{tabular}
    \label{tab:sys}
\end{table}

\section{Systematic uncertainty on measurement of EMFFs and angular distribution}
Systematic uncertainties on the measurements of the angular distribution in the $e^+e^-\to p\bar p$ process arise from the proton tracking and PID, requirements on $\rm \rm \theta_{opening}$ and $E/p$ ratio, background, fit method, respectively. 

\subsection{Proton tracking and PID}
The differences between the data and MC simulation for tracking and PID efficiencies are evaluated with the same procedure as before. Then we apply a correction to the MC simulation with the efficiency difference described above, and repeat the fit. The differences between the new and nominal values are taken as the systematic uncertainties resulting from the tracking and PID, which are 2.2\% and 3.1\%, respectively.

\subsection{Requirements on $E/p$ ratio and opening angle}
As mentioned above, the efficiency differences between data and MC simulation for the requirements on the $E/p$ ratio and opening angle are obtained by varying alternatively the different requirements. Subsequently, a correction to the MC simulation is carried out based on the resulting efficiency difference, and the fit to the angular distribution is repeated. The differences between the new and nominal values are regarded as the systematic uncertainties resulting from the requirements on the $E/p$ ratio and opening angle, with the values of 3.0\% and 5.0\%. 

\subsection{Background}
To evaluate a systematic uncertainty from background source that is not considered in the nominal fit, we repeated the fitting procedure by incorporating contributions from the sideband region and continuum processes into the log-likelihood function. The resulting differences from the nominal values were found to be negligible.

\subsection{Fit method}
The reliability of the fit method is verified by conducting an input and output check based on 500 pseudo experiments, using the helicity amplitude formula from Ref.~\cite{BESIII:2018cnd}, wherein the proton angular distribution is fitted. The mean values of polarization and decay-asymmetric parameters measured in this analysis are employed as input in the formula, and the number of events in each generated MC sample is 100 times that of the data sample. The differences between the input and output results are found to be negligible. 

\subsection{Total systematic uncertainties}
Assuming all sources are independent, the total systematic uncertainty is calculated by adding the individual contributions in quadrature. For \(|G_M|\) and \(|G_E/G_M|\), the systematic uncertainties are \(3.7\%\) and \(5.8\%\), respectively.

\section{Summary} 
In summary, with a data sample corresponding to an integrated luminosity of 26~fb$^{-1}$, collected by the BESIII detector at the BEPCII collider, we performed a measurement of the exclusive Born cross section and the effective form factor for the $e^+e^- \to p\bar p$ reaction at 47 c.m.~energy points ranging from 3.510 to 4.946~GeV. The results are summarized in Table~\ref{tab:table1:BCS_num}. For the first time, we also measure the modulus of the ratio between the electric and magnetic proton EMFFs $|G_{E}/G_{M}|$ together with $|G_{M}|$, the modulus of the single magnetic FF, by analyzing the proton polar angle distribution with high precision even at large timelike momentum transfer $q^2$ values. The results are summarized in Table~\ref{tab:fiteta}. These results provide valuable experimental insight into the dynamical mechanisms underlying the baryon-pair production of charmonium(-like) states. They may also serve as important input for theoretical models. 

\section{acknowledgement}
The BESIII Collaboration thanks the staff of BEPCII (https://cstr.cn/31109.02.BEPC) and the IHEP computing center for their strong support. This work is supported in part by National Key R\&D Program of China under Contracts Nos. 2025YFA1613900, 2023YFA1606000, 2023YFA1606704; National Natural Science Foundation of China (NSFC) under Contracts Nos. 12075107,
11635010, 11935015, 11935016, 11935018, 12025502, 12035009, 12035013, 12061131003, 12192260, 12192261, 12192262, 12192263, 12192264, 12192265, 12221005, 12225509, 12235017, 12342502, 12361141819; 
the Fundamental Research Funds for the Central Universities No. lzujbky-2025-ytA05,  No. lzujbky-2025-it06,  No. lzujbky-2024-jdzx06;
the Natural Science Foundation of Gansu Province No. 22JR5RA389, No.25JRRA799;
the ‘111 Center’ under Grant No. B20063;
the Chinese Academy of Sciences (CAS) Large-Scale Scientific Facility Program; the Strategic Priority Research Program of Chinese Academy of Sciences under Contract No. XDA0480600; CAS under Contract No. YSBR-101; 100 Talents Program of CAS; The Institute of Nuclear and Particle Physics (INPAC) and Shanghai Key Laboratory for Particle Physics and Cosmology; ERC under Contract No. 758462; German Research Foundation DFG under Contract No. FOR5327; Istituto Nazionale di Fisica Nucleare, Italy; Knut and Alice Wallenberg Foundation under Contracts Nos. 2021.0174, 2021.0299, 2023.0315; Ministry of Development of Turkey under Contract No. DPT2006K-120470; National Research Foundation of Korea under Contract No. NRF-2022R1A2C1092335; National Science and Technology fund of Mongolia; Polish National Science Centre under Contract No. 2024/53/B/ST2/00975; STFC (United Kingdom); Swedish Research Council under Contract No. 2019.04595; U. S. Department of Energy under Contract No. DE-FG02-05ER41374.
\bibliography{ref}
\onecolumngrid
\begin{center}
M.~Ablikim$^{1}$\BESIIIorcid{0000-0002-3935-619X},
M.~N.~Achasov$^{4,c}$\BESIIIorcid{0000-0002-9400-8622},
P.~Adlarson$^{81}$\BESIIIorcid{0000-0001-6280-3851},
X.~C.~Ai$^{87}$\BESIIIorcid{0000-0003-3856-2415},
C.~S.~Akondi$^{31A,31B}$\BESIIIorcid{0000-0001-6303-5217},
R.~Aliberti$^{39}$\BESIIIorcid{0000-0003-3500-4012},
A.~Amoroso$^{80A,80C}$\BESIIIorcid{0000-0002-3095-8610},
Q.~An$^{77,64,\dagger}$,
Y.~H.~An$^{87}$\BESIIIorcid{0009-0008-3419-0849},
Y.~Bai$^{62}$\BESIIIorcid{0000-0001-6593-5665}
% ...
\\
\vspace{0.2cm}
(BESIII Collaboration)\\
\vspace{0.2cm} {\it
$^{1}$ Institute of High Energy Physics, Beijing 100049, People's Republic of China\\
% ...
$^{4}$ Budker Institute of Nuclear Physics SB RAS (BINP), Novosibirsk 630090, Russia\\
% ...
$^{31}$ INFN Sezione di Ferrara, (A)INFN Sezione di Ferrara, I-44122, Ferrara, Italy; (B)University of Ferrara, I-44122, Ferrara, Italy\\
% ...
$^{39}$ Johannes Gutenberg University of Mainz, Johann-Joachim-Becher-Weg 45, D-55099 Mainz, Germany\\
% ...
$^{62}$ Southeast University, Nanjing 211100, People's Republic of China\\
% ...
$^{64}$ State Key Laboratory of Particle Detection and Electronics, Beijing 100049, Hefei 230026, People's Republic of China\\
% ...
$^{77}$ University of Science and Technology of China, Hefei 230026, People's Republic of China\\
% ...
$^{80}$ University of Turin and INFN, (A)University of Turin, I-10125, Turin, Italy; (B)University of Eastern Piedmont, I-15121, Alessandria, Italy; (C)INFN, I-10125, Turin, Italy\\
$^{81}$ Uppsala University, Box 516, SE-75120 Uppsala, Sweden\\
% ...
$^{87}$ Zhengzhou University, Zhengzhou 450001, People's Republic of China\\

\vspace{0.2cm}
$^{\dagger}$ Deceased\\
% ...
$^{c}$ Also at the Novosibirsk State University, Novosibirsk, 630090, Russia\\
% ...

}
 
\end{center}
\end{document}